\documentclass[11pt]{article}

\usepackage[table]{xcolor}
\usepackage{amsmath}
\usepackage{graphicx,psfrag,epsf}
\usepackage{enumerate}
\usepackage[numbers, sort&compress]{natbib}
\usepackage[hyphens]{url} 
\usepackage{ulem} 
\usepackage{standalone}

\newcommand{\blind}{0}

\usepackage{makecell}
\usepackage{amsthm}
\usepackage{bm}
\usepackage{bbm}
\usepackage{mathtools}
\usepackage{collcell}
\usepackage{dsfont}
\usepackage{rotating}
\usepackage{wrapfig}
\usepackage[utf8]{inputenc}
\usepackage{tikz}
\usepackage{lettrine}
\usepackage{dblfloatfix}
\usepackage[english]{babel} 
\usepackage{amsfonts}
\usepackage{animate}
\usepackage{subfig}
\graphicspath{ {fig/} }
\usepackage[hyperfootnotes=false]{hyperref}

\usepackage{arydshln}
\usepackage{lipsum}
\usetikzlibrary{arrows,decorations.markings}
\usepackage{standalone}
\usepackage{tablefootnote}
\usepackage{animate}

\newcommand\blfootnote[1]{%
	\begingroup
	\renewcommand\thefootnote{}\footnote{#1}%
	\addtocounter{footnote}{-1}%
	\endgroup
}

\definecolor{gray}{rgb}{0.5,0.5,0.5}
\definecolor{red}{rgb}{0.8,0,0}
\definecolor{dred}{rgb}{0.5,0,0}
\definecolor{blue}{rgb}{0,0.1,1}
\definecolor{dblue}{rgb}{0,0.1,0.6}
\definecolor{cyan}{rgb}{0,0.5,.5}
\definecolor{dcyan}{rgb}{0,0.3,.3}
\definecolor{mpurple}{rgb}{.7,0,.9}

\definecolor{b}{rgb}{0,0,.8}	
\definecolor{g}{rgb}{0,.6,0}	
\definecolor{n}{rgb}{0,0,0}	
\definecolor{h}{rgb}{0.4,0.2,0.2}	
\definecolor{v}{rgb}{0.2,0.6,0}

\newcommand{\A}{{\mathbb A}}
\newcommand{\B}{{\mathbb B}}
\newcommand{\C}{{\mathbb C}}
\newcommand{\D}{{\mathbb D}}
\newcommand{\E}{{\mathbb E}}
\newcommand{\F}{{\mathbb F}}
\newcommand{\G}{{\mathbb G}}
\renewcommand{\H}{{\mathbb H}}
\newcommand{\I}{{\mathbb I}}
\newcommand{\J}{{\mathbb J}}
\newcommand{\K}{{\mathbb K}}
\renewcommand{\L}{{\mathbb L}}
\newcommand{\M}{{\mathbb M}}
\newcommand{\N}{{\mathbb N}}
\renewcommand{\O}{{\mathbb O}}
\renewcommand{\P}{{\mathbb P}}
\newcommand{\Q}{{\mathbb Q}}
\newcommand{\R}{{\mathbb R}}
\newcommand{\T}{{\mathbb T}}
\newcommand{\U}{{\mathbb U}}
\newcommand{\V}{{\mathbb V}}
\newcommand{\W}{{\mathbb W}}
\newcommand{\X}{{\mathbb X}}
\newcommand{\Y}{{\mathbb Y}}
\newcommand{\Z}{{\mathbb Z}}

\renewcommand{\AA}{{\mathcal{A}}}
\newcommand{\BB}{{\mathcal{B}}}
\newcommand{\CC}{{\mathcal{C}}}
\newcommand{\DD}{{\mathcal{D}}}
\newcommand{\EE}{{\mathcal{E}}}
\newcommand{\FF}{{\mathcal{F}}}
\newcommand{\CG}{{\mathcal{G}}}
\newcommand{\HH}{{\mathcal{H}}}
\newcommand{\II}{{\mathcal{I}}}
\newcommand{\JJ}{{\mathcal{J}}}
\newcommand{\KK}{{\mathcal{K}}}
\newcommand{\LL}{{\mathcal{L}}}
\newcommand{\MM}{{\mathcal{M}}}
\newcommand{\NN}{{\mathcal{N}}}
\newcommand{\OO}{{\mathcal{O}}}
\newcommand{\PP}{{\mathcal{P}}}
\newcommand{\QQ}{{\mathcal{Q}}}
\newcommand{\RR}{{\mathcal{R}}}
\renewcommand{\SS}{{\mathcal{S}}}
\newcommand{\TT}{{\mathcal{T}}}
\newcommand{\UU}{{\mathcal{U}}}
\newcommand{\VV}{{\mathcal{V}}}
\newcommand{\WW}{{\mathcal{W}}}
\newcommand{\XX}{{\mathcal{X}}}
\newcommand{\YY}{{\mathcal{Y}}}
\newcommand{\ZZ}{{\mathcal{Z}}}

\newcommand{\bsa}{\boldsymbol a}
\newcommand{\bsb}{\boldsymbol b}
\newcommand{\bsc}{\boldsymbol c}
\newcommand{\bsd}{\boldsymbol d}
\newcommand{\bse}{\boldsymbol e}
\newcommand{\bsf}{\boldsymbol f}
\newcommand{\bsg}{\boldsymbol g}
\newcommand{\bsh}{\boldsymbol h}
\newcommand{\bsi}{\boldsymbol i}
\newcommand{\bsj}{\boldsymbol j}
\newcommand{\bsk}{\boldsymbol k}
\newcommand{\bsl}{\boldsymbol l}
\newcommand{\bsm}{\boldsymbol m}
\newcommand{\bsn}{\boldsymbol n}
\newcommand{\bso}{\boldsymbol o}
\newcommand{\bsp}{\boldsymbol p}
\newcommand{\bsq}{\boldsymbol q}
\newcommand{\bsr}{\boldsymbol r}
\newcommand{\bss}{\boldsymbol s}
\newcommand{\bst}{\boldsymbol t}
\newcommand{\bsu}{\boldsymbol u}
\newcommand{\bsv}{\boldsymbol v}
\newcommand{\bsw}{\boldsymbol w}
\newcommand{\bsx}{\boldsymbol x}
\newcommand{\bsy}{\boldsymbol y}
\newcommand{\bsz}{\boldsymbol z}
\newcommand{\bsA}{\boldsymbol A}
\newcommand{\bsB}{\boldsymbol B}
\newcommand{\bsC}{\boldsymbol C}
\newcommand{\bsD}{\boldsymbol D}
\newcommand{\bsE}{\boldsymbol E}
\newcommand{\bsF}{\boldsymbol F}
\newcommand{\bsG}{\boldsymbol G}
\newcommand{\bsH}{\boldsymbol H}
\newcommand{\bsI}{\boldsymbol I}
\newcommand{\bsJ}{\boldsymbol J}
\newcommand{\bsK}{\boldsymbol K}
\newcommand{\bsL}{\boldsymbol L}
\newcommand{\bsM}{\boldsymbol M}
\newcommand{\bsN}{\boldsymbol N}
\newcommand{\bsO}{\boldsymbol O}
\newcommand{\bsP}{\boldsymbol P}
\newcommand{\bsQ}{\boldsymbol Q}
\newcommand{\bsR}{\boldsymbol R}
\newcommand{\bsS}{\boldsymbol S}
\newcommand{\bsT}{\boldsymbol T}
\newcommand{\bsU}{\boldsymbol U}
\newcommand{\bsV}{\boldsymbol V}
\newcommand{\bsW}{\boldsymbol W}
\newcommand{\bsX}{\boldsymbol X}
\newcommand{\bsY}{\boldsymbol Y}
\newcommand{\bsZ}{\boldsymbol Z}
\newcommand{\bsone}{\boldsymbol 1}
\newcommand{\bstwo}{\boldsymbol 2}
\newcommand{\bsthree}{\boldsymbol 3}
\newcommand{\bsfour}{\boldsymbol 4}
\newcommand{\bsfive}{\boldsymbol 5}
\newcommand{\bssix}{\boldsymbol 6}
\newcommand{\bsseven}{\boldsymbol 7}
\newcommand{\bseight}{\boldsymbol 8}
\newcommand{\bsnine}{\boldsymbol 9}
\newcommand{\bszero}{\boldsymbol 0}
\newcommand{\bsnought}{\boldsymbol 0}
\newcommand{\bsnaught}{\boldsymbol 0}
\newcommand{\bsnull}{\boldsymbol 0}
\newcommand{\bsnil}{\boldsymbol 0}
\newcommand{\bsNull}{\text{\textbf{O}}}

\newcommand{\bPP}{\boldsymbol{\mathcal{P}}}

\newcommand{\bsalpha}{\boldsymbol \alpha}
\newcommand{\bsbeta}{\boldsymbol \beta}
\newcommand{\bslambda}{\boldsymbol \lambda}
\newcommand{\bsmu}{\boldsymbol \mu}
\newcommand{\bseps}{\boldsymbol \varepsilon}
\newcommand{\bstheta}{\boldsymbol \theta}
\newcommand{\bssigma}{\boldsymbol \sigma}
\newcommand{\bszeta}{\boldsymbol \zeta}
\newcommand{\bsbarphi}{\boldsymbol \varphi}
\newcommand{\bstau}{\boldsymbol \tau}
\newcommand{\bsxi}{\boldsymbol \xi}

\newcommand{\bsSamma}{\boldsymbol \Gamma}
\newcommand{\bsPhi}{\boldsymbol \Phi}
\newcommand{\bsSigma}{\boldsymbol \Sigma}
\newcommand{\bsPi}{\boldsymbol \Pi}

\newcommand{\Om}{{\Omega}}
\newcommand{\Ome}{{\Omega}}
\newcommand{\Omeg}{{\Omega}}
\newcommand{\eps}{{\varepsilon}}
\newcommand{\epsi}{{\epsilon}}
\newcommand{\kap}{{\kappa}}

\DeclareMathOperator*{\klimsup}{K-lim\,sup}
\DeclareMathOperator*{\kliminf}{K-lim\,inf}
\DeclareMathOperator*{\klim}{K-lim}
\DeclareMathOperator*{\argmin}{arg\,min}
\DeclareMathOperator*{\argmax}{arg\,max}
\DeclareMathOperator*{\epi}{epi}
\DeclareMathOperator{\lev}{lev}
\DeclareMathOperator{\gr}{graph}
\DeclareMathOperator{\conv}{conv}	
\DeclareMathOperator{\inter}{int}		
\DeclareMathOperator{\cl}{cl}
\DeclareMathOperator{\bdy}{bdy}
\DeclareMathOperator{\relint}{relint}	
\DeclareMathOperator{\relbdy}{relbdy}	
\DeclareMathOperator{\dist}{dist}
\DeclareMathOperator{\LSC}{LSC}
\DeclareMathOperator{\EPI}{EPI}
\DeclareMathOperator{\id}{id}
\DeclareMathOperator{\ind}{\boldsymbol 1 }
\DeclareMathOperator{\gdc}{gdc}
\DeclareMathOperator{\modulo}{mod}
\DeclareMathOperator{\logit}{logit}
\DeclareMathOperator{\spt}{spt}
\DeclareMathOperator{\sign}{sgn}
\DeclareMathOperator{\sgn}{sgn}
\newcommand{\abs}{|\cdot|}

\DeclareMathOperator{\var}{\V ar}
\DeclareMathOperator{\cov}{\C ov}
\DeclareMathOperator{\cor}{\C or}
\DeclareMathOperator{\corr}{\C orr}
\DeclareMathOperator{\as}{\text{a.s.}}

\DeclareMathOperator{\Diag}{Diag}
\DeclareMathOperator{\diag}{diag}

\newcommand{\MC}{\text{MC}} 
\newcommand{\MLE}{\text{MLE}} 
\newcommand{\MCMLE}{\text{MC-MLE}} 

\newcommand{\ov}\overline
\newcommand{\what}{\widehat}
\newcommand{\wtilde}{\widetilde}
\newcommand{\ow}{\text{ otherwise}}
\newcommand{\rig}\right
\newcommand{\lef}\left
\newcommand{\nf}\normalfont

\usepackage{forest} 
\usepackage{multirow}

\usepackage{subcaption} 

\usepackage{longtable}

\newcommand{\MSP}{\text{MSP}} 
\newcommand{\MVP}{\text{MVP}} 

\newcommand{\MAE}{\text{MAE}} 
\newcommand{\MMAE}{\text{MMAE}} 
\newcommand{\bsMAE}{\text{\textbf{MAE}}} 
\newcommand{\bsMMAE}{\text{\textbf{MMAE}}} 
\newcommand{\RMSE}{\text{RMSE}} 
\newcommand{\MRMSE}{\text{MRMSE}} 
\newcommand{\bsRMSE}{\text{\textbf{RMSE}}} 
\newcommand{\bsMRMSE}{\text{\textbf{MRMSE}}} 

\newcommand{\EXAA}{\text{EXAA}} 

\newcommand{\bsYY}{\boldsymbol \YY}
\newcommand{\bsDDelta}{\boldsymbol \varDelta}
\newcommand{\bsDelta}{\boldsymbol \Delta}

\newcommand{\Sup}{\text{S}} 
\newcommand{\Dem}{\text{D}}

 \newcommand{\FZ}[1]{\color{BrickRed} #1 \color{BrickRed}} 
 \long\def\FZC#1{{\color{BrickRed}{[FZ: #1]}\color{BrickRed}}}

 \newcommand{\PG}[1]{\color{black} #1 \color{black}} 
 \long\def\PGC#1{{\color{black}{[PG: #1]}\color{black}}}

 \newcommand{\specialcell}[2][c]{%
 \begin{tabular}[#1]{@{}c@{}}#2\end{tabular}}

 \setcitestyle{numbers} 
  \renewcommand{\citet}[1]{(\citepauthor{#1},~\citepyear{#1})}

\usepackage{float}
\floatstyle{plaintop}
\restylefloat{table}

\definecolor{greenpos}{RGB}{219, 255, 218}
\definecolor{redneg}{RGB}{255, 218, 218}
\definecolor{yellowsignif}{RGB}{250, 255, 184}
\definecolor{blueinsignif}{RGB}{184, 215, 255}

\usetikzlibrary{fit,positioning, tikzmark, calc}

\usepackage{etoolbox}
\makeatletter
\def\tikz@valign{c}
\tikzset{
  enforce alignment/.code={
    \csname if#1\endcsname
      \def\tikz@text@width
        {\pgfkeysvalueof{/pgf/minimum width}-2*(\pgfkeysvalueof{/pgf/inner xsep})}%
    \else
      \let\tikz@text@width\pgfutil@empty
    \fi},
  enforce alignment/.default=true,
  valign/.is choice,
  valign/top/.code=\def\tikz@valign{t},
  valign/center/.code=\def\tikz@valign{c},
  valign/bottom/.code=\def\tikz@valign{b},
  valign height/.initial=%
    \pgfkeysvalueof{/pgf/minimum height}-2*(\pgfkeysvalueof{/pgf/inner ysep})}
\patchcmd\tikz@fig@continue{\tikz@node@transformations}{%
  \pgfmathsetlength\pgf@x{\pgfkeysvalueof{/tikz/valign height}}%
  \pgf@y\ht\pgfnodeparttextbox
  \advance\pgf@y\dp\pgfnodeparttextbox
  \ifdim\pgf@y<\pgf@x
  \if\tikz@valign b%
    \advance\pgf@x-\dp\pgfnodeparttextbox
    \setbox\pgfnodeparttextbox\vbox to \pgf@x{\vfill\hbox{\unhbox\pgfnodeparttextbox}}%
  \else\if\tikz@valign t%
    \setbox\pgfnodeparttextbox\vbox to \pgf@x{\hbox{\unhbox\pgfnodeparttextbox}\vfill}%
  \fi\fi\fi
  \tikz@node@transformations}{}{}
\makeatother

\tikzset{
diagonal fill/.style 2 args={fill=#2, path picture={
\fill[#1, sharp corners] (path picture bounding box.south west) -|
                         (path picture bounding box.north east) -- cycle;}},
reversed diagonal fill/.style 2 args={fill=#2, path picture={
\fill[#1, sharp corners] (path picture bounding box.north west) |- 
                         (path picture bounding box.south east) -- cycle;}}
}

\begin{document}

	\def\spacingset#1{\renewcommand{\baselinestretch}%
		{#1}\small\normalsize} \spacingset{1}

	
	\if0\blind
	{
		\title{\bf From day-ahead to mid and long-term horizons with econometric electricity price forecasting models}
		\author{Paul Ghelasi\footnote{Corresponding author.} \footnote{Chair of Environmental Economics, esp. Economics of Renewable Energy. House of Energy Markets and Finance. University of Duisburg-Essen, Essen, Germany. Email addresses:  paul.ghelasi@uni-due.de (Paul Ghelasi), florian.ziel@uni-due.de (Florian Ziel).} , Florian Ziel\footnotemark[2] \\
		University of Duisburg-Essen, Germany}
		\maketitle
	} \fi

	
	\if1\blind
	{
		
		\bigskip
		
		\begin{center}
			{\LARGE\bf From day-ahead to mid and long-term horizons with econometric electricity price forecasting models}
		\end{center}
	
	} \fi


	\begin{abstract}
		The recent energy crisis starting in 2021 led to record-high gas, coal, carbon and power prices, with electricity reaching up to 40 times the pre-crisis average. This had dramatic consequences for operational and risk management prompting the need for robust econometric models for mid to long-term electricity price forecasting. After a comprehensive literature analysis, we identify key challenges and address them with novel approaches: 
		1) Fundamental information is incorporated by constraining coefficients with bounds derived from fundamental models offering interpretability; 
		2) Short-term regressors such as load and renewables can be used in long-term forecasts by incorporating their seasonal expectations to stabilize the model; 
		3) Unit root behavior of power prices, induced by fuel prices, can be managed by estimating same-day relationships and projecting them forward.
		We develop interpretable models for a range of forecasting horizons from one day to one year ahead, providing guidelines on robust modeling frameworks and key explanatory variables for each horizon. 
		Our study, focused on Europe's largest energy market, Germany, analyzes hourly electricity prices using regularized regression methods and generalized additive models.
	\end{abstract}
	
	\noindent%
	\vfill
	
	\spacingset{1.45} 

\newpage

\section{Introduction}
\subsection{Context and motivation}
\begin{figure}[htb]
    \centering
	\includegraphics[width=1\textwidth]{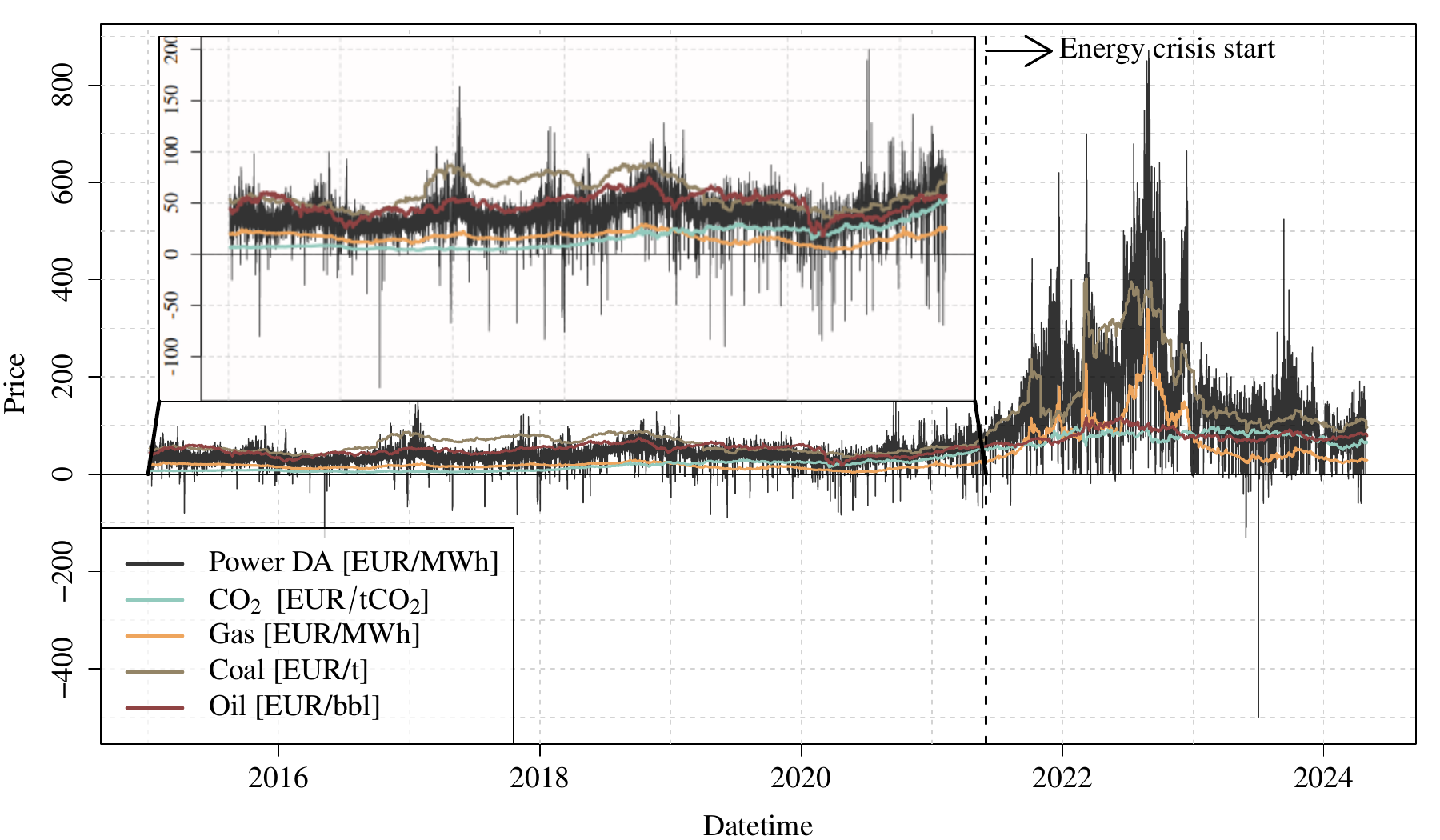}
	\caption{Price times series of German day-ahead power prices and front-month futures of $\textrm{CO}_2$ (as European Union Allowances), gas, coal, oil.}
	\label{fig:time_series}
\end{figure}

The global energy crisis that began in mid-2021 led to record-high european power prices by 2022 prompting regulators to reconsider energy policies and many investment projects to be reevaluated amid spiked future uncertainty (Figure \ref{fig:time_series}) \citep{emiliozzi2023european, adolfsen2024gas, neri2023energy}. 
Although prices have somewhat stabilized since then, they remain on average around twice the pre-crisis levels and exhibit much higher volatility \citep{segarra2024electricity}. 
This unprecedented situation raises critical questions: To what extent can existing power price forecasting models still be applied in the current environment? Can they be extended to forecast even further into the future, and if so, how far? These questions are vital also from a practical perspective as the crisis has heightened interest in mid to long-term forecasts among multiple market actors. Governments need such models for policy making reasons such as inflation containment and revision of subsidy schemes \citep{fabra2023reforming, kotek2023can, neri2023energy, adolfsen2024gas}. Industrial customers, producers, investors and asset managers are also highly interested in longer term forecasts for power purchase agreement (PPA) contracting, hedging, and valuation purposes \citep{segarra2024electricity}. Accurate forecasts are essential as they can reduce perceived investment risk and enhance the liquidity of futures markets.

\begin{figure}[htb]
    \centering
	\includegraphics[width=1\textwidth]{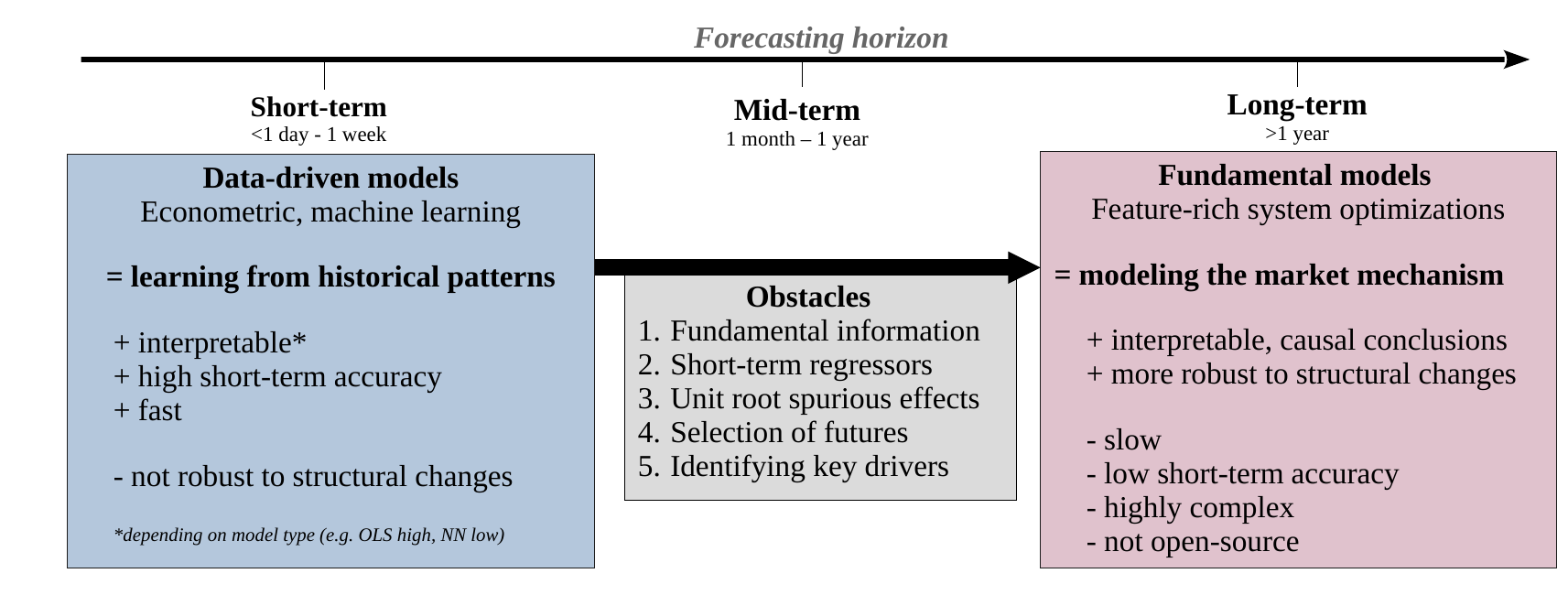}
	\caption{Sketch of relations between data-driven and fundamental models.}
	\label{fig:sketch}
\end{figure}

Electricity price forecasting methods can be broadly classified into two categories: data-driven and fundamental models \citep{weron2014electricity, ziel2016electricity}. Figure \ref{fig:sketch} illustrates the relationships between these two types of models, highlighting their strengths and weaknesses.
Data-driven models, such as econometric, machine-learning, or artificial intelligence (AI) models, utilize historical data to forecast power prices and are primarily employed for short-term forecasting \citep{cuaresma2004forecasting, weron2014electricity, uniejewski2019understanding, ziel2018probabilistic, narajewski2020econometric}. They have high short-term forecasting accuracy mainly due to autoregressive terms, as they can capture strategic and speculative behavior  \citep{bello2016medium}. However, they might have difficulty in incorporating market mechanisms and dealing with structural changes \citep{marcos2019electricity}. In contrast, fundamental models replicate the market mechanisms that lead to price formation. They take into account the expected market conditions and power plant fleet within a geographic zone, deriving power prices from intersections of supply and demand curves or as shadow prices of the demand constraint \citep{kallabis2016plunge, beran2019modelling, pape2016fundamentals}. These models are typically used for long-term forecasting, as they require detailed information on the power plant fleet, including technical parameters, fuel costs, availabilities and complex production schedules \citep{ringkjob2018review}. They are mostly not open source and require significant computational resources and technical expertise to implement. They are not as accurate in the short-term as data-driven models as they tend to underestimate volatility \citep{beran2021multi}. However, they are highly interpretable and causal conclusions can be drawn from them, which are essential properties for correct model assessment in the context of long-term forecasting.

This study explores the potential of extending data-driven models to long-term forecasting. Since interpretability and model transparency are crucial for such horizons we restricted our analysis to regularized linear regression models. In doing so we encountered several key obstacles, each addressed with a novel combination of approaches:
\begin{enumerate}
	\item \textbf{Integrating fundamental structures}: By constraining coefficients, we enhance model interpretability and stability.
	\item \textbf{Inclusion of load and renewables in the long-term}: We generate long-term forecasts for load and RES (renewable energy sources), which prove to be crucial for model stabilization.
	\item \textbf{Addressing unit roots from commodity markets}: Spurious effects becomes significant after one month; we resolve it by estimating same-day relationships and then forecasting.
	\item \textbf{Selecting optimal futures contracts for carbon and fuel prices}: Our analysis reveals that only the most recent contracts are essential for precise forecasting.
	\item \textbf{Identifying the importance of key drivers}: We thoroughly explore model combinations and offer detailed guidelines tailored for different forecasting horizons.
\end{enumerate}


\subsection{Literature review}
While there is an abundance of literature on short-term electricity price forecasting, research on mid to long-term forecasting remains relatively scarce, especially for an hourly resolution and for the energy crisis period between 2021-2023. \citep{ziel2018probabilistic} compiled an exhaustive list of papers covering econometric short, mid and long-term power price forecasting published between 2011 and 2017. Among these, only four papers, including the cited one, addressed long-term forecasting, while thirteen papers focused on mid-term forecasting. The authors refer to forecasting horizons from one month to one year ahead as mid-term, while long-term are horizons beyond one year. We will be using this definition as well.

Since then, more papers in mid to long-term forecasting of power prices have been published, which are systemically summarized in Tables \ref{tab:papers1} and \ref{tab:papers2}. 
The papers were found using the Scopus search engine with queries as in \citep{ziel2018probabilistic}, which can be found in the Appendix \ref{sec:appendix_scopus}. We report five key aspects of each paper: the market, the forecasting horizon, the period covered, the regressors used, and the forecasting method. The most common forecasting methods are autoregressive models, neural networks, and support vector machines. The most common regressors are autoregressive terms, with many papers not including external regressors at all. The markets covered are mainly European and US markets, with the most commonly ones being the PJM and ISO-NE regions in the US, Germany, and the Iberian Electricity Market (MIBEL). The forecasting horizons range from one day to five years, with most papers focusing on horizons between one week and one month ahead. The periods covered range from 2002 to 2023, with most papers focusing on the years 2016-2021. Even recent papers tend to use data before the recent energy crisis.
\begin{table}[htpb]
	\setlength{\tabcolsep}{1pt}
	\centering
	\scalebox{0.6}{
		\setlength\extrarowheight{-3pt}	
		\begin{tabular}{|>{\centering\arraybackslash}p{4cm}|>{\centering\arraybackslash}p{2cm}|>{\centering\arraybackslash}p{2cm}|>{\centering\arraybackslash}p{2cm}|>{\centering\arraybackslash}p{7cm}|>{\centering\arraybackslash}p{7cm}|}
			\hline
			\textbf{Paper} &\textbf{Market} & \textbf{Horizon} & \textbf{ Period} & \textbf{Regressors} & \textbf{Method} \\
			\hline
			\citeauthor{abroun2024predicting}, \citeyear{abroun2024predicting}, \citep{abroun2024predicting} \tablefootnote[1]{The authors use monthly average prices.} & Bulgaria, Greece, Hungary, Romania & 1-7 months & 2021-2022 & Autoregressive & Nonlinear regression, neural networks, support vector regression (SVR) \\
			\hline
			\citeauthor{niafenderi2023rolling}, \citeyear{niafenderi2023rolling}, \citep{niafenderi2023rolling} & PJM (US) & 1 day, 1 week & 2006 & Autoregressive & LSTM, convolutional neural networks \\
			\hline
			\citeauthor{foroni2023low}, \citeyear{foroni2023low}, \citep{foroni2023low} & Germany, Italy & 1-28 days & 2013-2020 & Autoregressive, industrial production index, manufacturing performance indices, oil and gas prices & MIDAS regression \\
			\hline
			\citeauthor{lee2023temporal}, \citeyear{lee2023temporal}, \citep{lee2023temporal} & PJM (US) & 1 day, 1 week & 2006 & Autoregressive & Temporal convolutional neural networks \\
			\hline
			\citeauthor{gomez2023electricity}, \citeyear{gomez2023electricity}, \citep{gomez2023electricity} & PJM (US), Australia & 2, 10, and 30 days & 2020-2021 & Autoregressive & LSTM, ensemble empirical mode decomposition \\
			\hline
			\citeauthor{sgarlato2022role}, \citeyear{sgarlato2022role}, \citep{sgarlato2022role} & Germany & 1-10 days & 2020-2021 & Autoregressive, coal, $\textrm{CO}_2$, gas, meteorological data, weekday dummies  & Lasso \\
			\hline
			\citeauthor{wagner2022short}, \citeyear{wagner2022short}, \citep{wagner2022short} & Germany & 1 day, 4 years & 2015-2019 & Autoregressive, temperature, renewables infeed (for day-ahead only), weekdays, holidays & Neural networks \\
			\hline
			\citeauthor{gabrielli2022data}, \citeyear{gabrielli2022data}, \citep{gabrielli2022data} \tablefootnote[2]{The authors use yearly average prices.} & UK, Germany, Sweden, Denmark & 1 year & 2015-2019 & Autoregressive, power demand, generation, imports, renewables infeed, production from conventional plants, commodity prices & Linear regression, gaussian process regression, neural networks \\
			\hline
			\citeauthor{karmakar2022long}, \citeyear{karmakar2022long}, \citep{karmakar2022long} & Germany & 1-17 weeks & 2013-2014 & Autoregressive & Robust Bayes, exponential smoothing state space models, neural networks, ARMAX \\
			\hline
			\citeauthor{iqbal2022optimized}, \citeyear{iqbal2022optimized}, \citep{iqbal2022optimized} & Australia & 1,2 weeks & 2020-2021 & Autoregressive, load, infeed from renewables & Gated recurrent unit, bagged trees ensemble \\
			\hline
			\citeauthor{irfan2022week}, \citeyear{irfan2022week}, \citep{irfan2022week} & ISO-NE (US) & 1 week & 2017-2020 & Autoregressive, load, temperature, weather & DenseNet neural networks,  \\
			\hline
			\citeauthor{ciarreta2022jumps}, \citeyear{ciarreta2022jumps}, \citep{ciarreta2022jumps} & Germany, Austria & 1-7 days & 2016-2018 & Autoregressive, weekday dummies, jump dummies & Lasso \\
			\hline
			\citeauthor{aslam2021towards}, \citeyear{aslam2021towards}, \citep{aslam2021towards} & ISO-NE (US) & 1-3 days, 1 week & 2012-2020 & Autoregressive, load, temperature, weather & Convolutional neural networks, XGBoost, Adaboost \\
			\hline
			\citeauthor{matsumoto2021one}, \citeyear{matsumoto2021one}, \citep{matsumoto2021one} \tablefootnote[3]{The authors use weekly average prices.} & Japan & 1 week & 2016-2020 & Autoregressive, temperature & Quantile regression, GARCH \\
			\hline
			\citeauthor{teixeira2021extreme}, \citeyear{teixeira2021extreme}, \citep{teixeira2021extreme} \footnotemark[1] & MIBEL (Iberian Electricity Market)  & 1 month & 2019-2020 & Autoregressive & Artificial neural networks, extreme learning machines \\
			\hline
			\citeauthor{imani2021forecasting}, \citeyear{imani2021forecasting}, \citep{imani2021forecasting} & Italy  & 1 hour, 1 day, 1 week & 2020 & Autoregressive, Load, natural gas & Support vector machines, tree-based methods, multilayer perceptron, gaussian process regression\\
			\hline
			\citeauthor{spiliotis2021forecasting}, \citeyear{spiliotis2021forecasting}, \citep{spiliotis2021forecasting} & Belgium  & 1 week & 2012-2016 & Autoregressive, load, generation capacity, hour, year & Exponential smoothing, linear regression, multi-layer perceptron, random forest \\
			\hline
			\citeauthor{ribeiro2020electricity}, \citeyear{ribeiro2020electricity}, \citep{ribeiro2020electricity}, \citep{ribeiro2020electricity2} & Brazil & 1-3 months & 2015-2020 & Autoregressive, supply, demand & Coyote optimization, empirical mode decomposition, extreme learning machine, gaussian process, SVR, gradient boosting machine \\
			\hline
		\end{tabular}
	}
	\caption{Papers covering mid and long-term forecasting of electricity prices from 2018 to 2024. Continued on Table \ref{tab:papers2}.}
	\label{tab:papers1}
\end{table}
\begin{table}[htbp]
	\setlength{\tabcolsep}{1pt}
	\centering
	\scalebox{0.6}{
		\setlength\extrarowheight{-3pt}	
		\begin{tabular}{|>{\centering\arraybackslash}p{4cm}|>{\centering\arraybackslash}p{2cm}|>{\centering\arraybackslash}p{2cm}|>{\centering\arraybackslash}p{2cm}|>{\centering\arraybackslash}p{7cm}|>{\centering\arraybackslash}p{7cm}|}
			\hline
			\textbf{Paper} &\textbf{Market} & \textbf{Horizon} & \textbf{ Period} & \textbf{Regressors} & \textbf{Method} \\
			\hline
			\citeauthor{monteiro2020strategy}, \citeyear{monteiro2020strategy}, \citep{monteiro2020strategy} \footnotemark[1] & Iberian Electricity Market (MIBEL) & 1-6 months & 2018-2019 & Autoregressive, delivery month, time to maturity, baseload physical power futures prices, spot price & Multilayer perceptron \\
			\hline
			\citeauthor{yeardley2020efficient}, \citeyear{yeardley2020efficient}, \citep{yeardley2020efficient} & UK & 4 weeks & 2017-2018 & Autoregressive, demand, generation from conventional plants, renewables infeed & Gaussian process, clustering \\
			\hline
			\citeauthor{vcesnavivcius2020lithuanian}, \citeyear{vcesnavivcius2020lithuanian}, \citep{vcesnavivcius2020lithuanian} \footnotemark[1] & Lithuania & 1 month & 2020 &  Autoregressive & ARIMA, SARIMA \\
			\hline
			\citeauthor{hammaddcnn}, \citeyear{hammaddcnn}, \citep{hammaddcnn} & NYISO (US) & 1 day, 1 week & 2016 & Autoregressive & LDA, random forest, deep convoluted neural networks\\
			\hline
			\citeauthor{windler2019one}, \citeyear{windler2019one}, \citep{windler2019one} & Germany, Austria& 1-29 days  & 2016 & Autoregressive, hour/ day/ week/ month/ year dummies & Weighted nearest neighbors, exponential smoothing state space model, box-cox transformation, ARMA errors, trend and seasonal components (TBATS), deep neural networks \\
			\hline
			\citeauthor{marcos2019electricity}, \citeyear{marcos2019electricity}, \citep{marcos2019electricity} & Iberian Electricity Market (MIBEL) & 1 day, 1 week & 2016 & Autoregressive, fundamentally estimated market price, load, renewables infeed, seasonality dummies & Linear programming, artificial neural networks \\
			\hline
			\citeauthor{razak2019hybrid}, \citeyear{razak2019hybrid}, \citep{razak2019hybrid} \footnotemark[1] & Ontario (Canada) & 1 month & 2012 & Autoregressive & Least square support vector machine, bacterial foraging optimization \\
			\hline
			\citeauthor{ferreira2019linear}, \citeyear{ferreira2019linear}, \citep{ferreira2019linear} \footnotemark[1] & Iberian Electricity Market (MIBEL) & 1 month & 2017 & Autoregressive, electricity consumption/ imports/ exports, heating degree days, cooling degree days, macroeconomic indices, oil, renewables infeed & Linear regression \\
			\hline
			\citeauthor{steinert2019short}, \citeyear{steinert2019short}, \citep{steinert2019short} & Germany  & 28 days & 2016-2017 & Autoregressive, base and peak daily/weekly/weekend/month futures, weekday dummies, periodic splines & Lasso \\
			\hline
			\citeauthor{ali2019short}, \citeyear{ali2019short}, \citep{ali2019short} & NYISO (US) & 1 week & 2014 & Autoregressive, weather data, load & SVM, KNN, decision trees \\
			\hline
			\citeauthor{yousefi2019long}, \citeyear{yousefi2019long}, \citep{yousefi2019long} \footnotemark[1] & California (US) & 3, 5 years & 2012-2017 & Autoregressive, gas consumption, coal consumption, power generation/ imports, GDP, renewables infeed & LSTM, SARIMA \\ 
			\hline
			\citeauthor{ubrani2019lstm}, \citeyear{ubrani2019lstm}, \citep{ubrani2019lstm} & India & 1 day, 1 week & 2017 & Autoregressive & LSTM, gated recurrent unit \\
			\hline
			\citeauthor{ma2018month}, \citeyear{ma2018month}, \citep{ma2018month}  \footnotemark[1] & ERCOT (US)& 1 month  & 2016 & Autoregressive, prices of gas, nuclear, coal, renewables infeed, load, weather, power import/export, calendar days & Support vector machines \\
			\hline
			\citeauthor{campos2018short}, \citeyear{campos2018short}, \citep{campos2018short} & Spain, PJM (US) & 1 day - 1 week & 2002, 2006 & Autoregressive & Wavelet transform, hybrid particle swarm optimization, adaptive neuro-fuzzy inference systems \\
			\hline
			\citeauthor{razak2018optimization}, \citeyear{razak2018optimization}, \citep{razak2018optimization} \footnotemark[1] & Ontario (Canada) & 1 month & 2009-2010 & Autoregressive & SVM, LSSVM, generic algorithm \\
			\hline
			\citeauthor{marin2018forecasting}, \citeyear{marin2018forecasting}, \citep{marin2018forecasting} \footnotemark[1] & Colombia & 1 month & 2014-2016 & Autoregressive, hydro, availability & ARMAX, artifical neural networks \\
			\hline
			\citeauthor{ziel2018probabilistic}, \citeyear{ziel2018probabilistic}, \citep{ziel2018probabilistic} & Germany & 3 years & 2015-2016 & Autoregressive, temperature, renewables infeed, production from conventional plants, electricity load, auction data, day/week/season/holiday dummies & Lasso, supply and demand curve modeling \\
			\hline
		\end{tabular}
	}
	\caption{Papers covering mid and long-term forecasting of electricity prices from 2018 to 2024. Continued from Table \ref{tab:papers1}.}
	\label{tab:papers2}
\end{table}

Some results of the query included false positives which we removed. For example \citep{najafi2023application} forecast day-ahead prices but refer to them as mid-term forecasts, and \citep{aruldoss2021week} refer to them as weekly forecasts. Other papers do not focus on market power prices, such as \citep{yang2022electricity} who analyze company-level prices, and others  such as \citep{jkedrzejewski2021importance} and \citep{marcjasz2019importance} forecast day-ahead prices but include long-term components as regressors.
The majority of papers focus on mid-term forecasting with forecasting horizons under a year, many under a month. Many studies model average monthly, weekly or yearly prices. None of the papers tackle forecasting power prices on an hourly basis around the 2022 energy crisis and the period thereafter.

\subsection{Obstacles when forecasting beyond day-ahead}
To illustrate the obstacles encountered with econometric models when forecasting far into the future, consider the following established day-ahead power price model, which we will refer to as the \textbf{expert model} \citep{chai2023forecasting, bille2023forecasting, marcjasz2023distributional, maciejowska2020pca, fezzi2020size, serafin2019averaging, ziel2018day, uniejewski2017variance}:
\begin{align}
	\label{eq:expert}
	\text{Price}_{t+h} = & \beta_0 + \underbrace{\beta_1 \text{Mon}_{t+h} + \beta_2 \text{Fri}_{t+h} + \beta_3 \text{Sat}_{t+h} + \beta_4 \text{Sun}_{t+h}}_{\text{Weekly seasonality}}  \\
	& + \underbrace{\beta_5 \text{Winter}_{t+h} + \beta_6 \text{Spring}_{t+h} + \beta_7 \text{Summer}_{t+h}}_{\text{Annual seasonality}} \nonumber \\ 
	& + \underbrace{\beta_8 \text{Price}_{t} + \beta_9 \text{Price}_{t-1} + \beta_{10} \text{Price}_{t-6} + \beta_{11} \text{PriceLastHour}_{t}}_{\text{Autoregressive terms}}  \nonumber  \\
	&+ \underbrace{\beta_{12} \what{\text{RES}}_{t+1} + \beta_{13} \what{\text{Load}}_{t+1}}_{\text{Power market forecasts}} 
	+ \underbrace{\beta_{14} \textrm{CO}_{2t} + \beta_{15} \text{Gas}_{t} + \beta_{16} \text{Coal}_{t} + \beta_{17} \text{Oil}_{t}}_{\text{Commodity prices}} + \epsilon_t \nonumber 
\end{align}
where $h$ is the forecasting horizon and $h=1$ corresponds to forecasting one day ahead.

\begin{figure}[htb]
	\centering
	\includegraphics[width=0.49\textwidth]{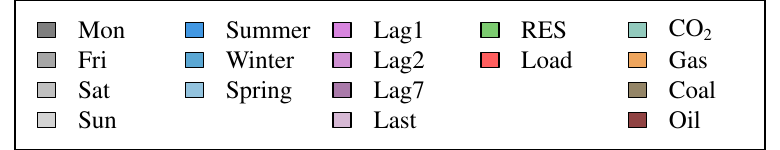}
	\linebreak
    \subfloat[Coefficients 2015-2019]{
        \includegraphics[width=0.49\textwidth]{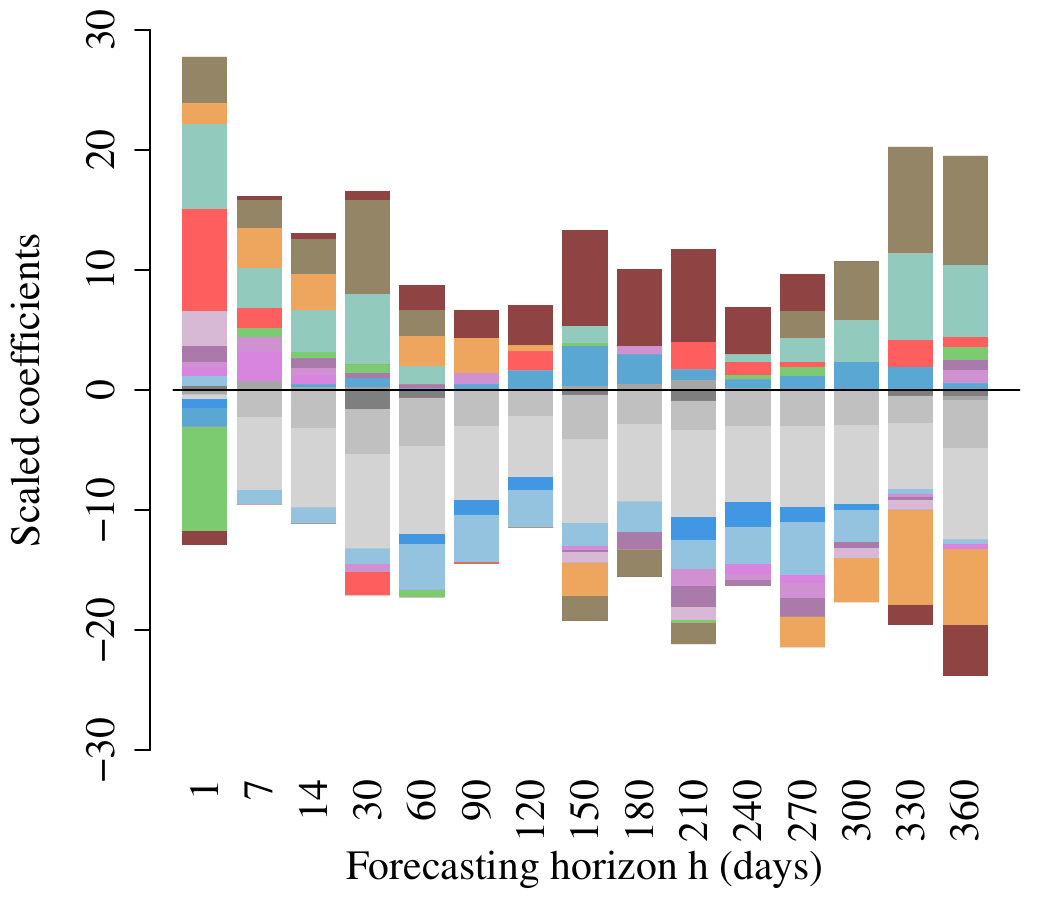}}
    \subfloat[Coefficients 2020-2024]{
        \includegraphics[width=0.49\textwidth]{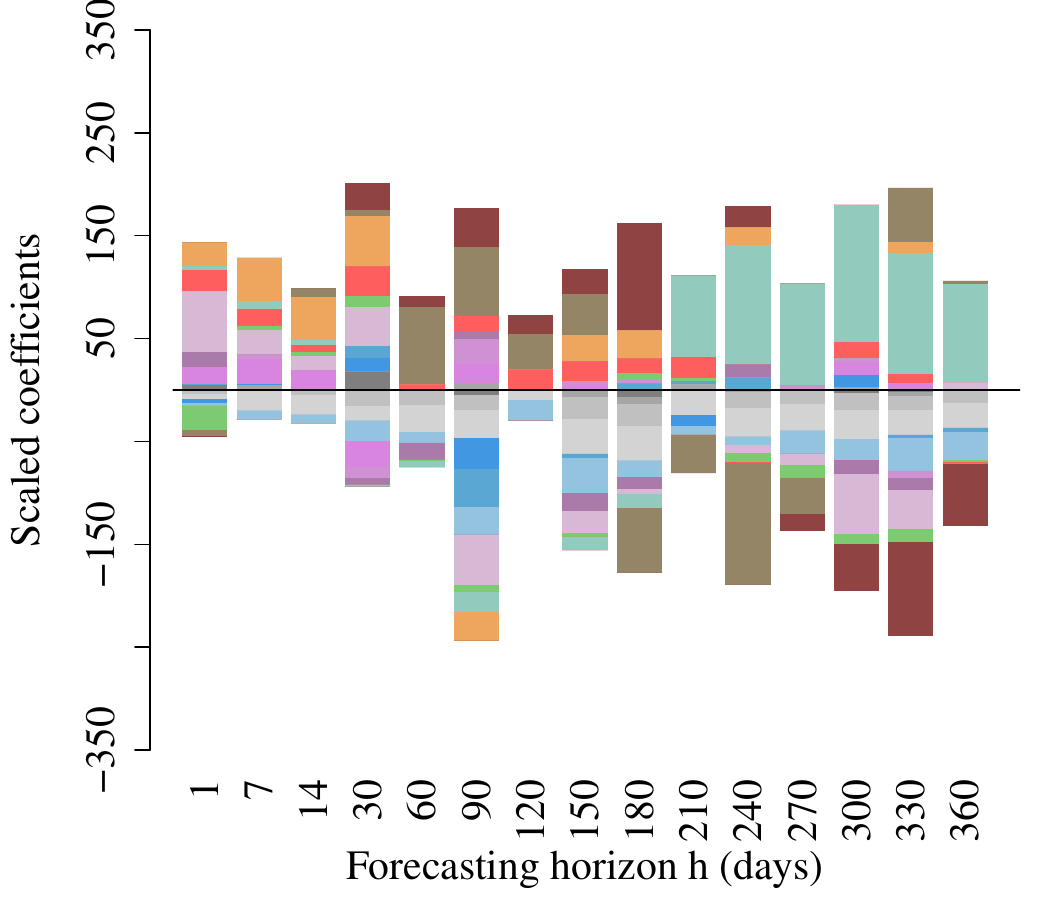}}
    \caption{Scaled coefficients of the \textbf{expert model} (\ref{eq:expert}) for hour 9, for forecasting horizons from 1 to 360 days ahead, estimated using elastic net. Data from two periods were used to fit each model, before and after the energy crisis. Similar results were observed for all other hours.}
    \label{fig:expert_noconstr_coefs}
\end{figure}

Figure \ref{fig:expert_noconstr_coefs} depicts the estimated coefficients of this model across various forecasting horizons, ranging from one day to 360 days ahead, before and after the energy crisis. The estimates exhibit high instability with increasing forecasting horizon $h$, particularly following the energy crisis. The effects of renewables infeed, load, and autoregressive terms initially diminish rapidly, in line with expectations. However, they sometimes resurface at higher horizons without a discernible pattern and with different signs. The coefficient estimates of the commodity futures prices are also sometimes negative, which is not in line with energy economics theory, since these represent fuel prices of power plants. All of these can be traced back to the obstacles listed in the beginning.

While some of these issues are addressed in a few papers (see Tables \ref{tab:papers1} and \ref{tab:papers2}), they are not considered as a whole. The provided solutions are often very specific to the pre-crisis period or the considered forecasting horizon, making it questionable if they remain robust when varying these two factors. 

For incorporating fundamental information, \citep{gonzalez2011forecasting}, \citep{marcos2019electricity}, \citep{beran2021multi}, and \citep{gabrielli2022data} use a hybrid approach by including the price returned by a fundamental model as an explanatory variable into the econometric model. We incorporate fundamentally-derived information by setting coefficient bounds, thus stabilizing the model and reducing spurious effects.

Regarding the inclusion of short-term regressors, \citep{ziel2018probabilistic} simulate wind and solar generation averages for long-term forecasts. \citep{gabrielli2022data} use yearly wind and solar data. \citep{wagner2022short} include renewables for short-term but not for long-term forecasts. Other studies that model horizons up to a month use renewables as lagged regressors. We provide a method to incorporate variables such as renewables by generating seasonal forecasts and projecting them into the future.

Studies such as \citep{nowotarski2016importance}, \citep{marcjasz2019importance}, and \citep{jkedrzejewski2021importance} examine long-term seasonal components of power prices for day-ahead forecasts, but do not address unit root behavior drivers and their implications for mid to long-term forecasting. We show that unit root behavior is explained by commodity prices, though this diminishes with increasing horizons. We also show that after a certain threshold the best approach is to estimate the current unlagged relationship between power prices and their regressors, and projecting it into the future.

This paper is organized as follows. In the next section \ref{sec:data}, we provide a brief overview of the data used and present the basic model and study design. In section \ref{sec:models}, we tackle each of the challenges presented above and present modifications of the basic model to address them. In section \ref{sec:results} we report on the accuracy of the new proposed models and compared them to some benchmarks. We offer a guideline of which input variables are important and what methodology performs best for which forecasting horizon. Lastly, in section \ref{sec:conclusion} we conclude and present directions for further research.

\section{Data and design}
\label{sec:data}

\subsection{Data}
In this study we make use of the following data pertaining to Europe's largest energy market, Germany: day-ahead power prices, infeed from solar, onshore wind and offshore wind plants, and load data. This data is in hourly resolution and is freely available on the ENTSOE Transparency platform. We also use daily closing prices for futures of the commodities gas, coal, oil and european emission allowances (EUA), further referred to as $\textrm{CO}_2$, which were provided by the information platform Refinitiv Eikon. The data spans 2015 to 2024. We apply standard clock-change adjustment to the hourly data (interpolation in spring, averaging in fall).

\begin{figure}
    \centering
      \includegraphics[width=1\textwidth]{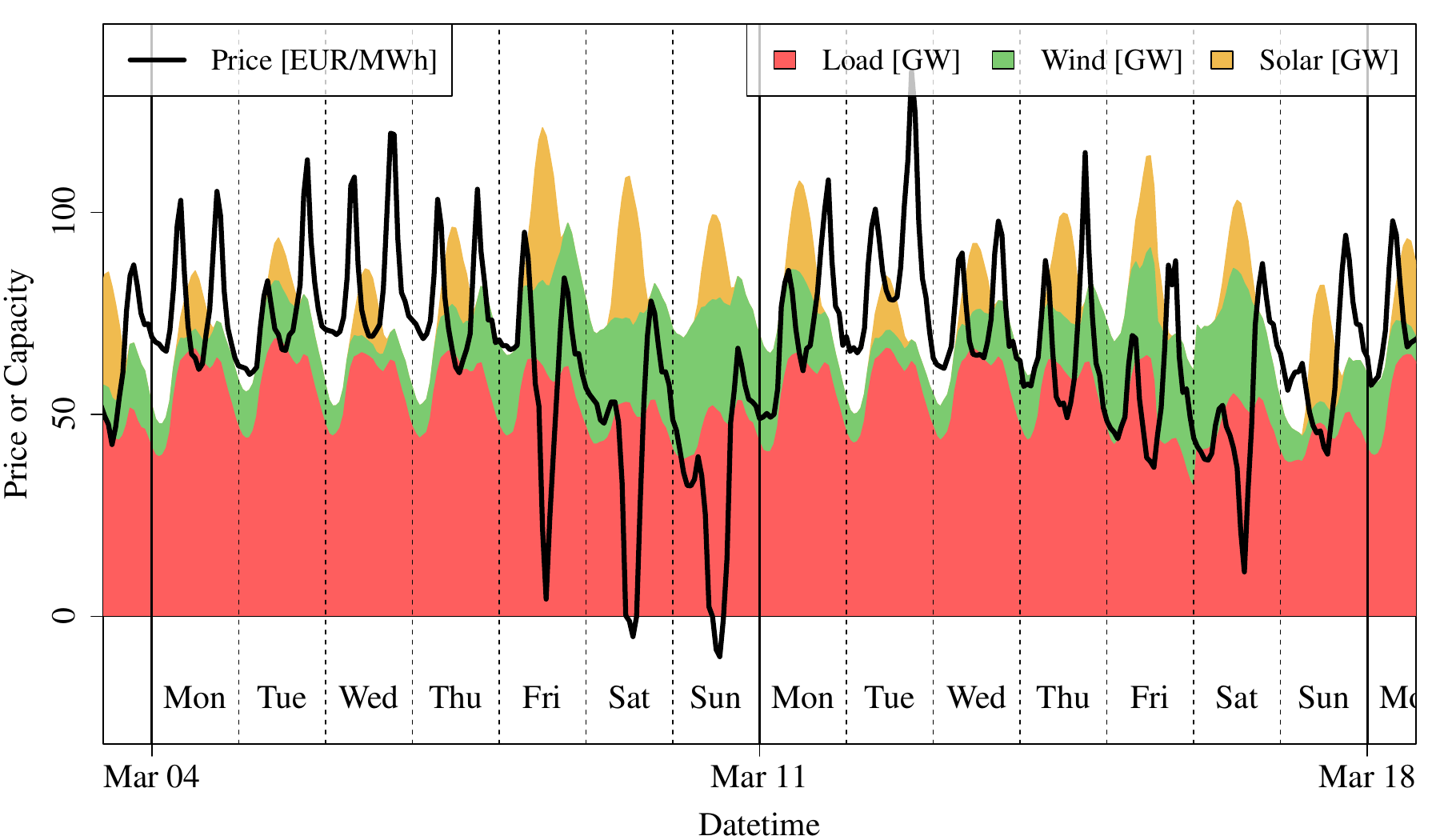}
    \caption{Power price seasonalities, interactions with load, RES for 2 weeks of 2024.}
	\label{fig:price_res_load}
\end{figure}

\begin{figure}[hbtp]
    \centering
	\scalebox{0.5}{
	\includegraphics[width=0.5\textwidth]{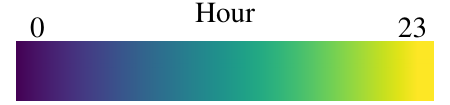}}
	\linebreak
    \subfloat[Power price means by year and hour]{
        \includegraphics[width=0.49\textwidth]{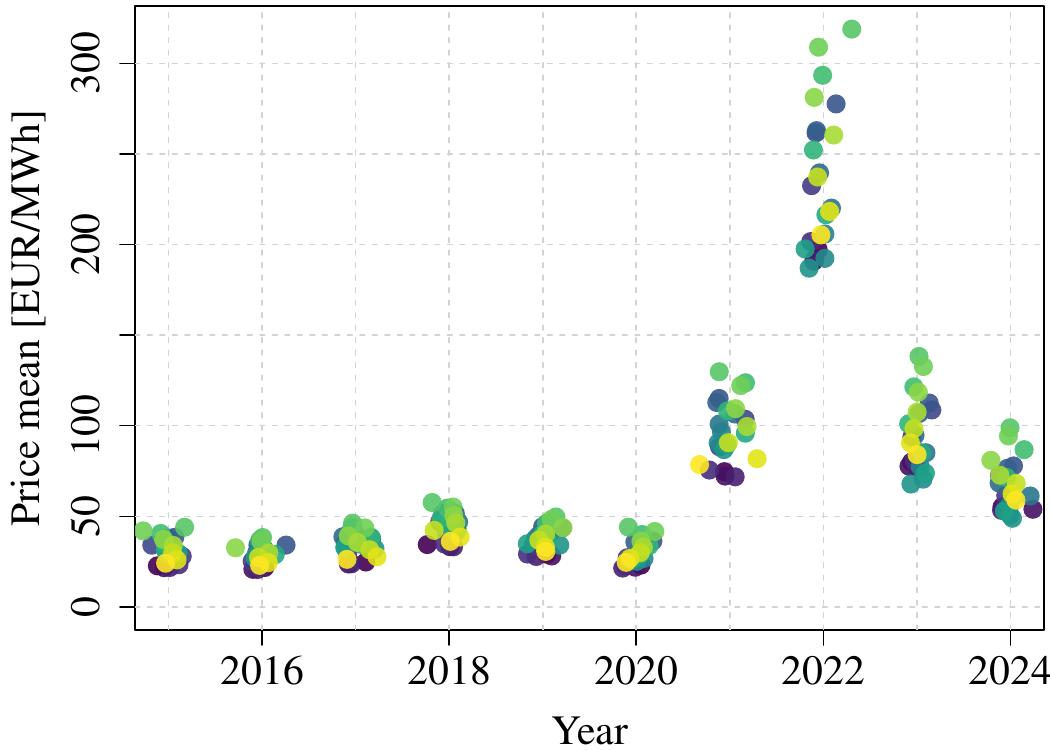}}
    \subfloat[Power price standard deviations by year and hour]{
        \includegraphics[width=0.49\textwidth]{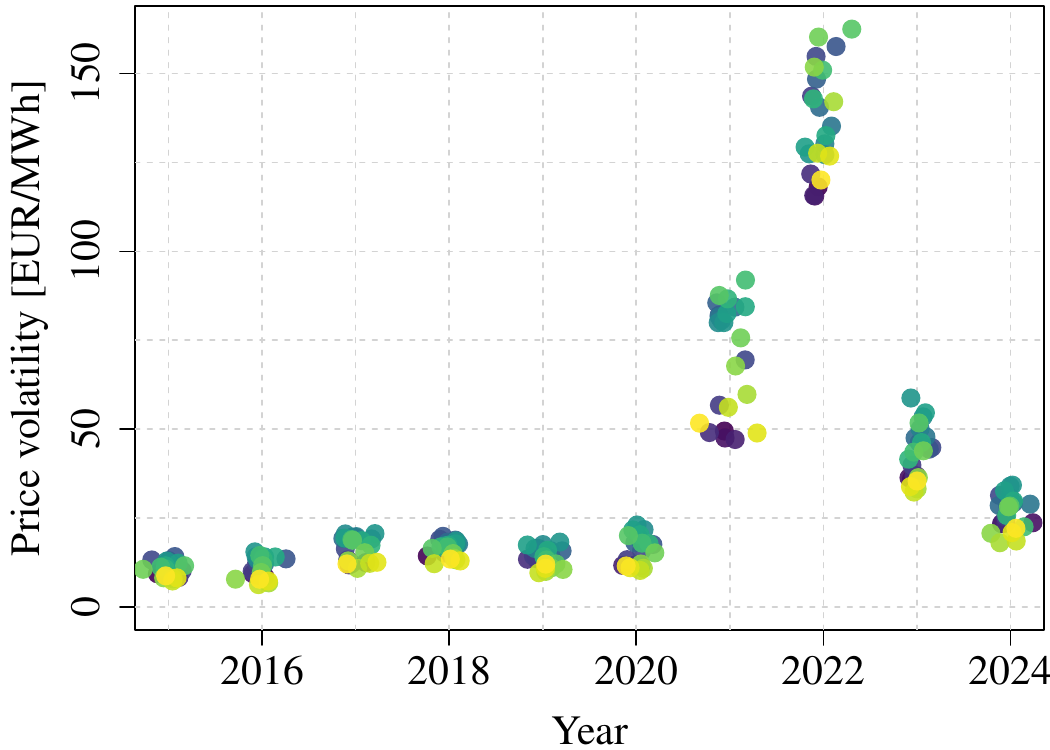}}
	\linebreak
	\subfloat[Power price means by week and hour]{
        \includegraphics[width=0.49\textwidth]{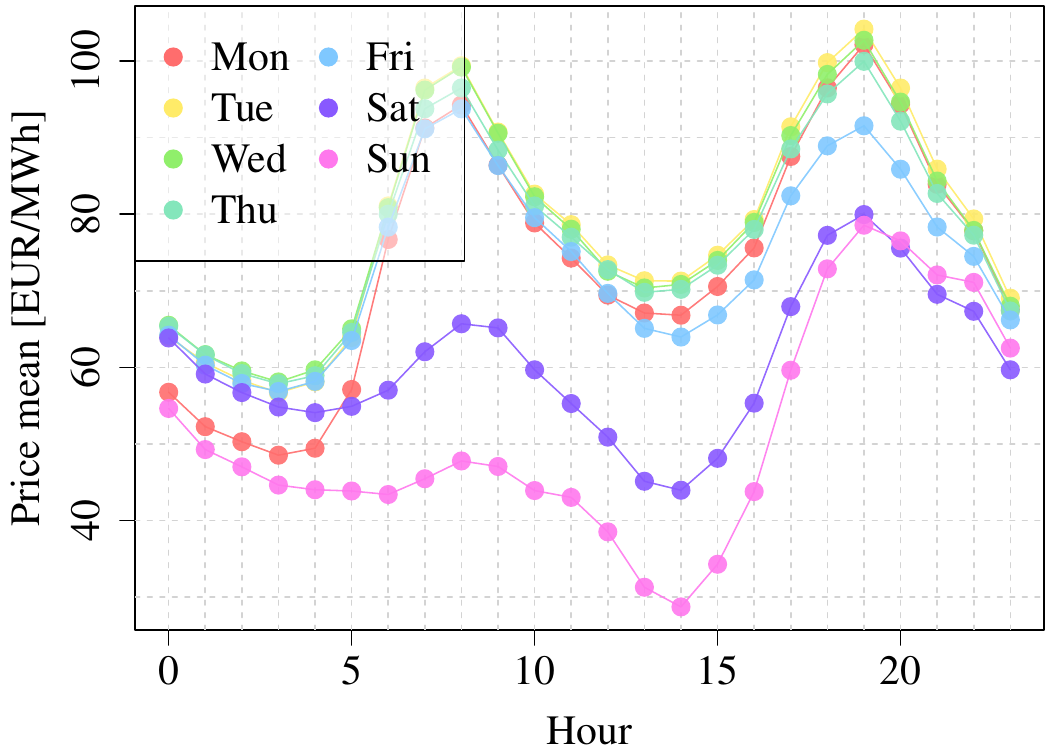}}
    \subfloat[Power price means by season and year]{
        \includegraphics[width=0.49\textwidth]{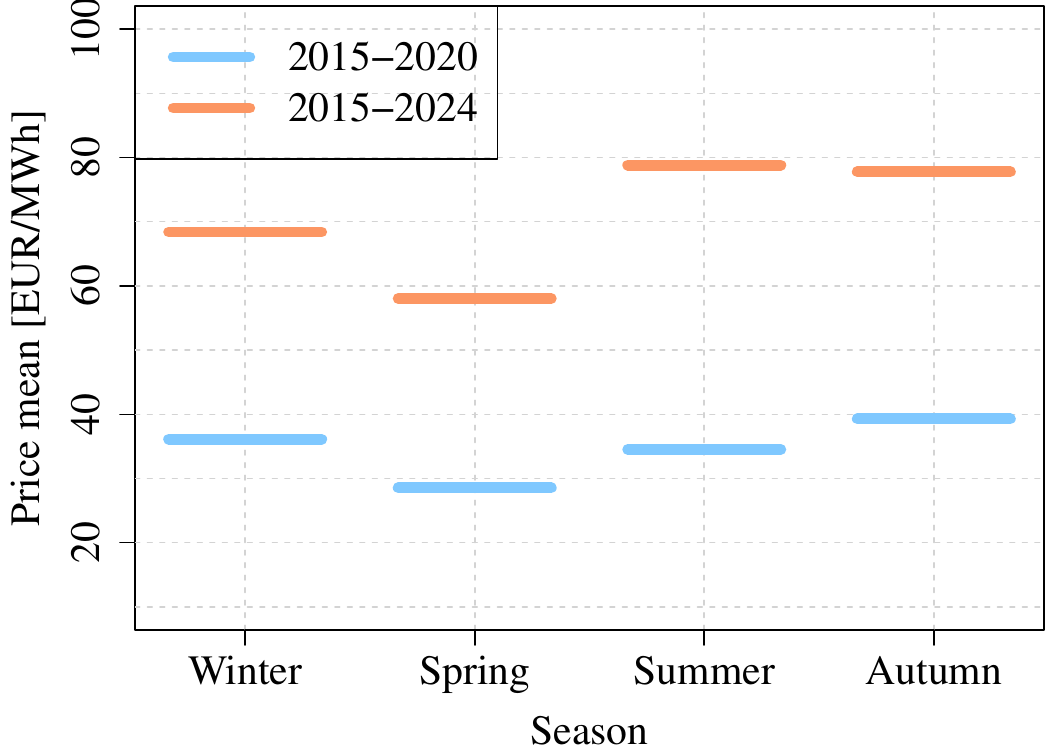}}
	\linebreak
	\centering
		\includegraphics[width=0.7\textwidth]{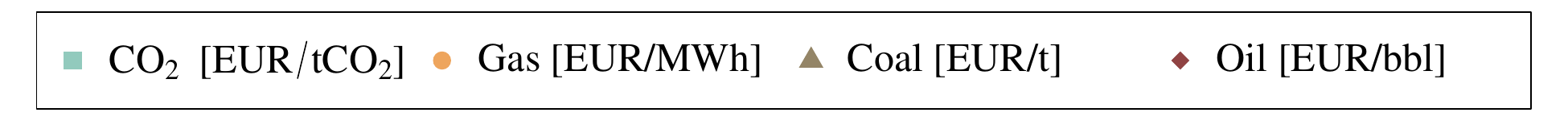}
	\linebreak
	\subfloat[Commodity futures means by year]{
		\includegraphics[width=0.49\textwidth]{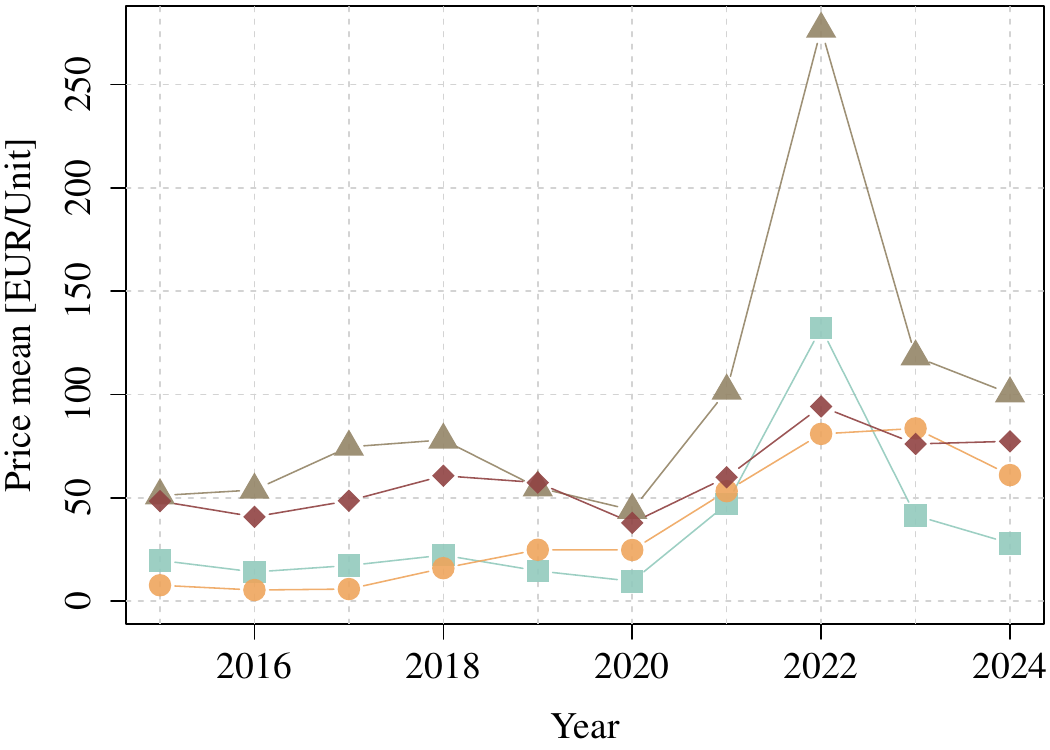}}
	\subfloat[Commodity futures standard deviations by year]{
		\includegraphics[width=0.49\textwidth]{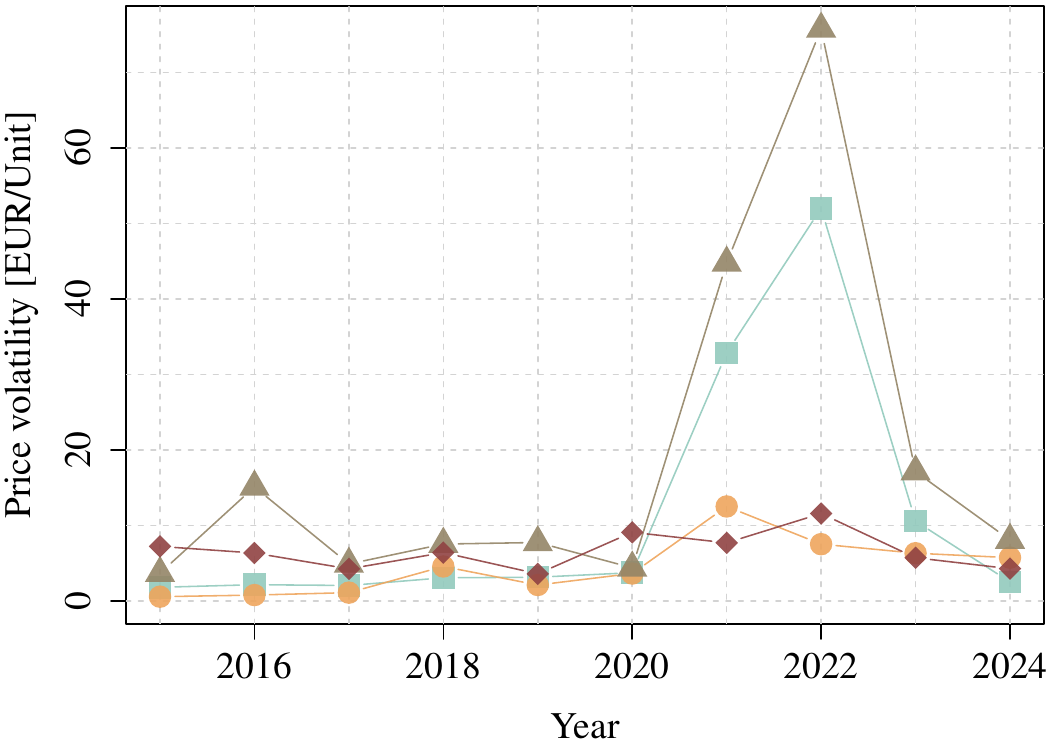}}	
    \caption{Descriptive statistics (means and standard deviations) of German day-ahead power prices (a)-(d) and of front-month futures commodities: $\textrm{CO}_2$, gas, coal, oil (e)-(f).}
    \label{fig:price_stats}
\end{figure}

Figure \ref{fig:price_res_load} shows power prices, renewables infeed broken down by solar, onshore wind and offshore wind, and load (demand) data for two weeks of 2024. Typical power price characteristics can be observed: first, the weekly seasonality as a result of lower load during weekends, second the daily seasonality with lower prices during mid-day as a result of solar infeed, and third the strong impact of renewables infeed on power prices, also known as the \emph{merit order effect of renewables}. This states that when renewables infeed, especially infeed from wind, is high, then power prices become lower often reaching negative values. 

Figure \ref{fig:price_stats} illustrates mean and standard deviation statistics of power and commodity prices highlighting some of the same characteristics.
It is evident that, until the energy crisis started in late 2021, power prices remained relatively stable around a mean of 40 EUR/MWh. With the onset of the energy crisis they surged to over 800 EUR/MWh by 2022 before subsiding to a mean slightly below 100 EUR/MWh since 2023. Nevertheless, the volatility remains at over 25 EUR/MWh which is notably higher than pre-crisis levels. A similar pattern is observed for the futures prices of $\textrm{CO}_2$, gas, coal, and oil. As these commodities are used as fuels in the generation of electricity, their prices are key drivers of power prices. Indeed, their long-term co-movement with the power prices is evident in the time series, especially for gas and coal (see Figure \ref{fig:time_series}).

The recent energy crisis which started in late 2021, its timeline and effects are well documented \citep{di2022natural, medzhidova2022return, szpilko2022european, emiliozzi2023european}. In summary, it was triggered by perceived scarcity of gas supply in Europe mainly as a result of Russia's reduction of gas supply amid rising geopolitical tensions which ultimatively led to the currently ongoing Russia-Ukraine conflict. Nevertheless, the energy crisis comes in the more complex context of the post-pandemic recovery and the unfolding of the European Green Deal. 


\subsection{Basic model and study design}

Model structures for short-term power price forecasting are well-researched and regressors with high explanatory power have been identified in the literature \citep{chai2023forecasting, bille2023forecasting, marcjasz2023distributional, maciejowska2020pca, fezzi2020size, ziel2018day, uniejewski2017variance}. These models are referred to as \emph{expert models} as they are current state-of-the art models with solid theoretic background and empirical backup \citep{weron2019electricity}. For day-ahead forecasts variables with high explanatory power include: 
\begin{enumerate}
	\item Deterministic calendar variables such as dummy variables and smooth periodic functions to capture daily, weekly and annual seasonalities,
	\item Autoregressive terms with the most important lags being days 1, 2 and 7, but complex models use autoregressive terms up to 30 days,
	\item Cross-period and non-linear effects: these are prices of other past hours, such as the last price of the previous day, or the highest and lowest prices of the previous day,
	\item Day-ahead forecasts for load and infeed from renewable energy sources,
	\item Futures prices of $\textrm{CO}_2$ emission allowances, gas, coal, and oil. 
\end{enumerate}

The \textbf{expert model} (\ref{eq:expert}) is a representative expert model with all of these regressors, except for the highest and lowest prices of the previous day. 
The problems with the \textbf{expert model} become apparent when looking at Figure \ref{fig:expert_noconstr_coefs}. The coefficients are as expected for the day-ahead forecast, but become unstable with increasing horizons and many apparently spurious effects can be observed. Hence the performance of the \textbf{expert model} is expected to decline drastically with increasing forecasting horizons.

Throughout this study we consider variations of the \textbf{expert model} (\ref{eq:expert}). For every considered model we conduct a rolling window forecasting study with a window size of $365 \times 3$. Here the window size refers to the number of rows of the regressor matrix. 
The window is rolled forward by one day after each forecast for an evaluation set of 6 years from April 2018 to April 2024. For each day 24 models are used, one for each hour of the day. Hence, at every rolling window step 24 models and sets of coefficients are estimated. 
We fit the models on normalized\footnotemark[4]\footnotetext[4]{By normalization we refer to substracting the sample mean and dividing by the sample standard deviation of a variable in order to get zero mean and unit variance.} data using elastic net as implemented in the glmnet R package \citep{glmnet1}. The elastic net penalty parameter $\alpha$ is set to 0.5, which is a balanced mix of LASSO and ridge regression. The regularization parameter $\lambda$ is chosen using the Bayesian information criterion (BIC) from an exponential grid of $\lambda$. The models are evaluated using the root mean squared error (RMSE) and the mean absolute error (MAE) as performance measures.

\section{Model extensions}
\label{sec:models}
In this section we will thoroughly discuss each of the challenges associated with forecasting power prices beyond day ahead (see Figure \ref{fig:sketch}) and inspect ways of addressing these. 

\subsection{Fundamental information} 

In the absence of restrictions which would force the \textbf{expert model} to adhere to energy economic principles, the estimated coefficients become highly unstable (see Figure \ref{fig:expert_noconstr_coefs}). Fundamental information refers to the underlying economic and physical principles that drive power prices. In the context of power markets this includes how the day-ahead market mechanism works to form the power price. The simplest model of a power market is the \emph{supply stack model}, also referred to as the \emph{merit order model}. According to this model the marginal costs of power plants are ordered from lowest to highest generating the merit order curve and the power price is determined as the intersection of this curve with the inelastic load. In the following we will show that the expert model has an implicit merit-order model representation and hence that its coefficients are closely related to some technical parameters of the power plants. Thus coefficients can be constrained to be consistent with fundamental theory and as a result provide much needed stability and interpretability.

\subsubsection{Merit order representation of econometric models}

To illustrate how the \textbf{expert model} relates to the \emph{merit order}, we consider a market where power is produced only by a single gas power plant. The simplified variable costs of this plant depend on the gas price, the $\textrm{CO}_2$ price, and on the characteristics of the power plant. They are calculated as:
\begin{align}
	\text{VarCost}_{t}  \left [ \frac{\text{EUR}}{\text{MWh}_{el}} \right ] & = 
	\frac{\text{Gas}_{t} \left [ \frac{\text{EUR}}{\text{MWh}_{th}} \right ]}{\eta_{Gas} \left [ \frac{\text{MWh}_{el}}{\text{MWh}_{th}} \right ] } + 
	\frac{\varepsilon_{Gas} \left [ \frac{tCO_2}{\text{MWh}_{th}} \right ] \text{CO}_{2t} \left [ \frac{\text{EUR}}{tCO_2} \right ] }{\eta_{Gas} \left [ \frac{\text{MWh}_{el}}{\text{MWh}_{th}} \right ] } + \textrm{OtherCost}_t
	\label{eq:varcost}
\end{align}
where $\eta_{Gas}$ is the efficiency and $\varepsilon_{Gas}$ is the $\textrm{CO}_2$ intensity factor of the power plant. According to the merit order model, the power price is equal to the marginal costs of the price-setting technology. In this case, the only technology is gas, hence $\textrm{Price}_t = \textrm{VarCost}_t$ together with (\ref{eq:varcost}) results in 
\begin{align}
	\textrm{Price}_t =  \frac{1}{\eta_{Gas}} \textrm{Gas}_t + \frac{\varepsilon_{Gas}}{\eta_{Gas}} \textrm{CO}_{2t} + \varepsilon_t 
	\label{eq:price_model_simple_tech}
\end{align}
Now consider a simplified econometric model for the power price with only two explanatory variables, namely gas and $\textrm{CO}_2$ prices:
\begin{align}
	\textrm{Price}_t =  \beta_{Gas} \textrm{Gas}_t + \beta_{\textrm{CO}_{2}} \textrm{CO}_{2t} + \varepsilon_t  
	\label{eq:price_model_simple}
\end{align}
Since this is equivalent to (\ref{eq:price_model_simple_tech}), the corresponding coefficients $\beta_{Gas} = \frac{1}{\eta_{Gas}}$ and $\beta_{\textrm{CO}_{2}} = \frac{\varepsilon_{Gas}}{\eta_{Gas}} $ are equal to the inverse of the power plant efficiency, also known as the \emph{heat rate}, and the ratio of $\textrm{CO}_2$ intensity to efficiency respectively.

\begin{figure}[htb]
	\centering
    \subfloat[$\textrm{Price}_t \sim \beta^A_1 \textrm{Gas}_t + \beta^A_2 \textrm{CO}_{2t}$]{
        \includegraphics[width=0.33\textwidth]{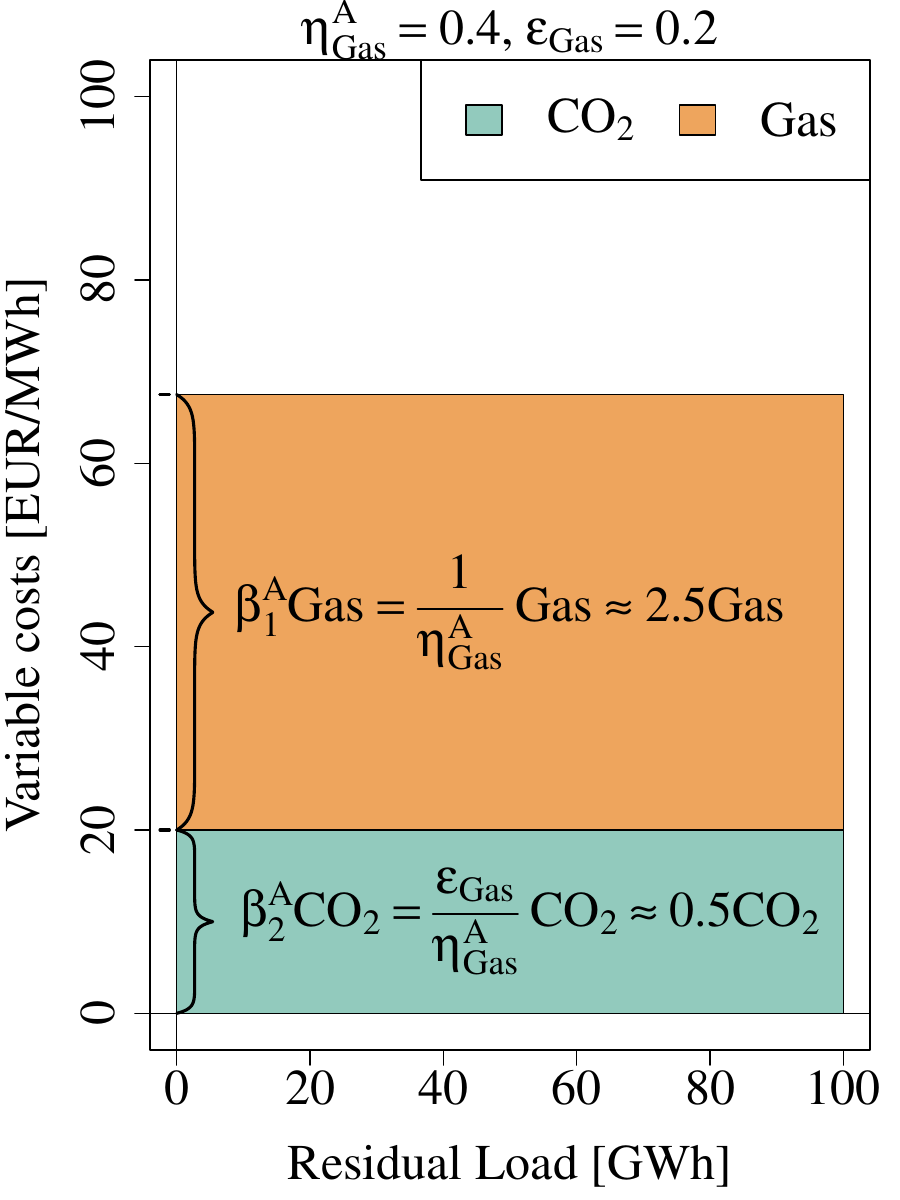}}
    \subfloat[$\textrm{Price}_t \sim \beta^B_1 \textrm{Gas}_t + \beta^B_2 \textrm{CO}_{2t}$]{
        \includegraphics[width=0.33\textwidth]{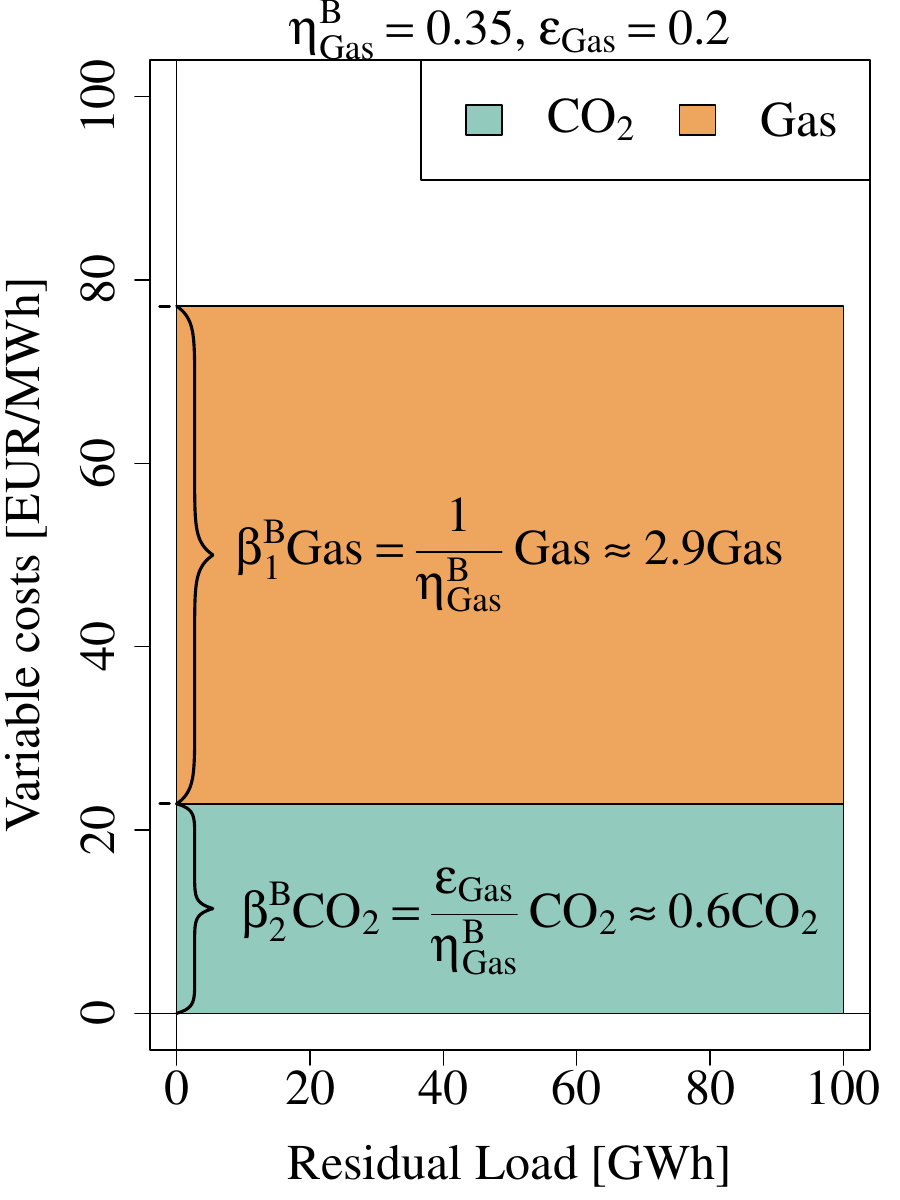}}
	\subfloat[$\textrm{Price}_t \sim \beta^C_1 \textrm{Coal}_t + \beta^C_2 \textrm{CO}_{2t}$]{
		\includegraphics[width=0.33\textwidth]{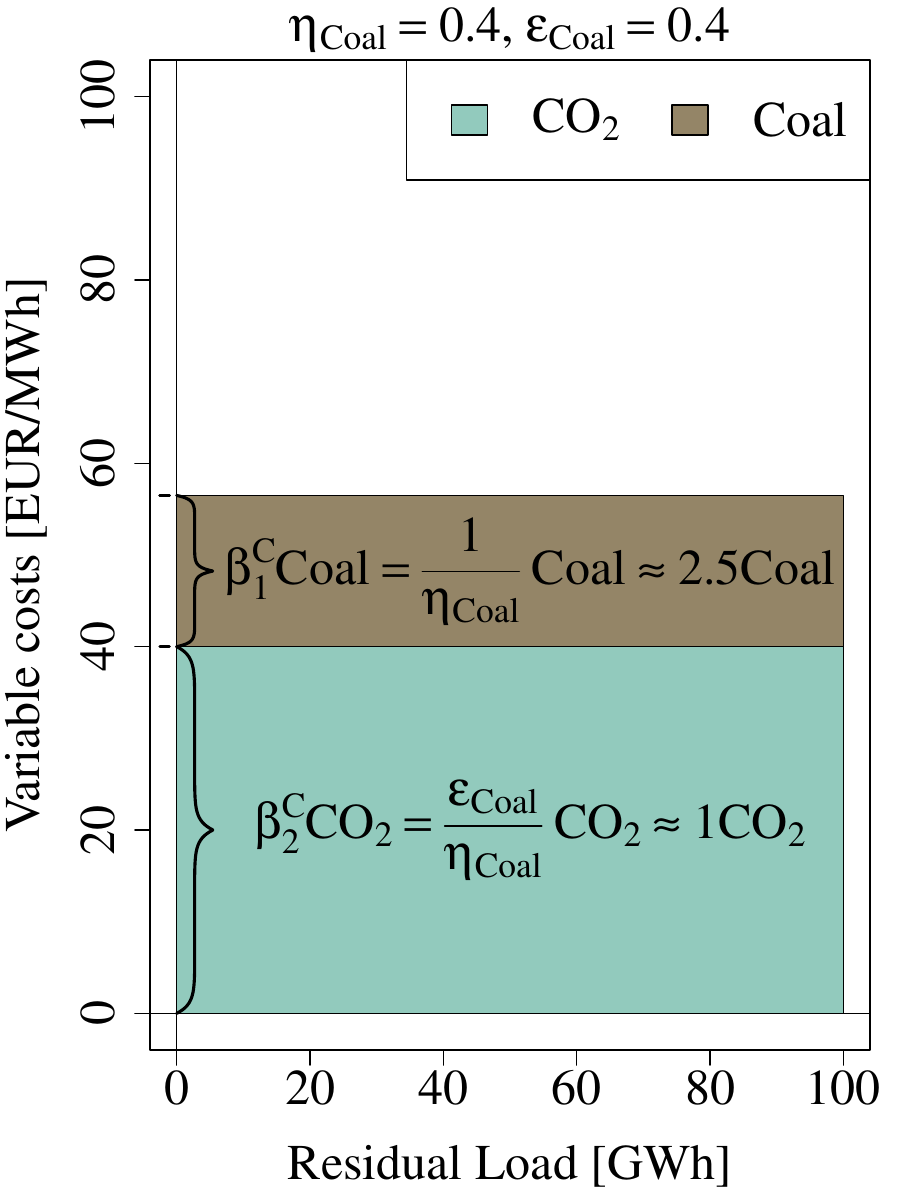}}
    \caption{Merit order representation of each price equation, where $\eta$ is the power plant efficiency and $\epsilon$ the $\textrm{CO}_2$ intensity factor.  Illustrative examples of single-technology markets for: (a) efficient gas, (b) inefficient gas, and (c) coal. Coefficients follow from (\ref{eq:varcost})-(\ref{eq:price_model_simple_tech}).}
    \label{fig:mo}
\end{figure}

Figure \ref{fig:mo} shows the merit order representation of the same illustrative example for an efficient gas power plant, an inefficient one, and for a coal power plant. Here the components of the power price are shown as colored areas. Note that for the more inefficient gas plant shown in the middle, the gas price component is higher than for the efficient plant, since more gas is needed to produce the same amount of electricity. For the coal plant on the right the $\textrm{CO}_2$ component is higher than for the gas plants, since coal produces more $\textrm{CO}_2$ per MWh as illustrated by the higher $\textrm{CO}_2$ intensity factor. 
 
Now consider a market where all of the three technologies from Figure \ref{fig:mo} are simultaneously active. We consider the simplified power price model:
\begin{align}
	\textrm{Price}_t =  \beta_{Gas} \textrm{Gas}_t + \beta_{Coal} \textrm{Coal}_t + \beta_{\textrm{CO}_{2}} \textrm{CO}_{2t} + \varepsilon_t.
	\label{eq:price_model_simple_ext}
\end{align}
Similar to (\ref{eq:price_model_simple_tech}), $\beta_{Gas}$ will correspond to the average heat rate of the gas power plants weighted by the corresponding capacities, and $\beta_{\textrm{CO}_{2}}$ will correspond to the weighted average $\textrm{CO}_2$-to-efficiency ratio of all plants. However, since these are averages, $\beta_{Gas}$ will never exceed the highest individual heat rate nor will $\beta_{\textrm{CO}_{2}}$ exceed the highest individual $\textrm{CO}_2$-to-efficiency ratio. Hence, using the notations from Figure \ref{fig:mo}, we can derive upper bounds for these coefficients:
\begin{align}
	&\beta_{Gas} \le \max (\beta_1^A, \beta_1^B) \nonumber \\
	\label{eq:coefs_upper_bounds_example}
	&\beta_{Coal} \le \max (\beta_1^C) \\
	&\beta_{\textrm{CO}_{2}} \le \max (\beta_2^A, \beta_2^B, \beta_2^C) \nonumber
\end{align}

\subsubsection{Coefficient constraints}
Doing such calculations for the \textbf{expert model} (\ref{eq:expert}) requires knowledge of the power plant characteristics, which were summarized by \citep{beran2019modelling} and are also openly published by ENTOSE in the European Resource Adequacy Assessment (ERAA) report \footnotemark[5]\footnotetext[5]{Link: https://www.entsoe.eu/outlooks/eraa/2023/eraa-downloads/}. We summarize these characteristics in Table \ref{tab:fundamentals}. Note that the units of the coefficients must be chosen in such a way that the unit of the variable multiplied by the unit of the coefficient equals the unit of the power price (EUR/MWh). Now, upper bounds for the fuel coefficients can be derived similar to (\ref{eq:coefs_upper_bounds_example}). The resulting lower and upper bounds for the coefficients are shown in Table \ref{tab:bounds} and the corresponding calculations can be found in the Appendix \ref{sec:appendix_bounds}. In addition to the fuel prices we also restrict RES to only have a negative effect on the power price, since the marginal costs of renewables are close to zero and hence they push the power price downwards according to the merit order model. Similarly, we restrict the load coefficient to be positive, since increasing demand should increase the equilibrium price according to standard economic theory. Autoregressive effects were also restricted to be positive. This is because we noticed that fuel and autoregressive coefficients can be replaced by one another. For example, when fuel coefficient is unconstrained, the autoregressive coefficient is small. When the fuel coefficient is constrained, the autoregressive coefficient becomes as relevant as the previously unconstrained fuel coefficient. This is expected since the time series strongly co-move in the long term. To mitigate this we impose the positivity constraint, which is plausible in this context since autoregressive terms should not be relevant too far out into the future once seasonalities are accounted for.

\begin{table}[ht]
	\centering
	\scalebox{0.9}{
		\begin{tabular}{|l|cc|cc|c|cc|}
			\hline
			\textbf{Type} & \multicolumn{2}{c|}{\textbf{Units}} & \multicolumn{5}{c|}{\textbf{Power Plant Characteristics}} \\
			\hline
			 & \textbf{Variable} & \textbf{Coefficient} & \multicolumn{2}{c|}{\textbf{Efficiency}} & \textbf{CO2 factor} & \multicolumn{2}{c|}{\textbf{Ratio}} \\
			& \multicolumn{2}{c|}{} & \multicolumn{2}{c|}{$\eta$   $\left[ \frac{MWh_{el}}{MWh_{th}} \right]$} & \multicolumn{1}{c|}{ $\varepsilon$  $\left[ \frac{tCO_2}{MWh_{th}} \right]$} & \multicolumn{2}{c|}{ $\frac{\varepsilon}{\eta}$  $\left[ \frac{tCO_2}{MWh_{el}} \right]$} \\
			& \multicolumn{2}{c|}{} & \textbf{Old} & \textbf{New} & & \textbf{Old} & \textbf{New} \\
			\hline 
			Power 	& $\frac{\text{EUR}}{\text{MWh}_{el}}$ & - & -  & - & - & - & - \\
			Lignite & $\frac{\text{EUR}}{\text{t}}$ & - & 0.3  & 0.43 & 0.4  & 1.33 & 0.99 \\
			Coal    & $\frac{\text{EUR}}{\text{t}}$ & $\frac{\text{t}}{\text{MWh}_{el}}$ & 0.35 & 0.46 & 0.3  & 0.86 & 0.65 \\
			Gas     & $\frac{\text{EUR}}{\text{MWh}_{th}}$ & $\frac{\text{MWh}_{th}}{\text{MWh}_{el}}$ & 0.25 & 0.4  & 0.2  & 0.8  & 0.5 \\
			Oil     & $\frac{\text{EUR}}{\text{bbl}}$ & $\frac{\text{bbl}}{\text{MWh}_{el}}$ & 0.24 & 0.44 & - & -  & - \\
			$\textrm{CO}_2$ 	& $\frac{\text{EUR}}{\text{tCO}_2}$ & $\frac{\text{tCO}_2}{\text{MWh}_{el}}$ & - & - & - & -  & - \\
			\hline
			\hline
			Coal conversion & \multicolumn{7}{l|}{$1 \ \text{t coal} = 8.141 \ \text{MWh}_{th}$} \\
			Oil conversion & \multicolumn{7}{l|}{$1,000 \ \text{bbl oil} = 1.700 \ \text{MWh}_{th}$} \\
			\hline
		\end{tabular}
	}
	\caption{Units of variables and coefficients used in the \textbf{expert model} (\ref{eq:expert}). Power plant characteristics including efficiency and CO2-intensity factors for old and new power plants. Conversion factors for coal and oil. $\text{MWh}_{el}$ and $\text{MWh}_{th}$ stand for output and input power, respectively.}
	\label{tab:fundamentals}
\end{table}

\begin{table}[ht]
	\centering
	\scalebox{0.8}{
		\begin{tabular}{|l|ccccccc|}
			\hline
			& \textbf{Lags} & \textbf{Load} & \textbf{RES} & \textbf{$\textrm{CO}_2$} & \textbf{Gas} & \textbf{Coal} & \textbf{Oil}\\ 
			\hline
			\textbf{Lower bound} & $0$      & $0$      & $-\infty$ & $0$    & $0$ & $0$     & $0$\\
			\textbf{Upper bound} & $\infty$ & $\infty$ & $0$       & $1.33$ & $4$ & $0.123$ & $0.588$\\
			\hline
		\end{tabular}
	}
	\caption{Fundamentally-derived bounds for the coefficinets of selected regressors. All other coefficients are unbounded. The derivations can be found in the Appendix \ref{sec:appendix_bounds}. }
	\label{tab:bounds}
\end{table}

\begin{figure}[hbt]
    \centering
	\includegraphics[width=0.49\textwidth]{expert_win3_coefs_legend.pdf}
	\linebreak
    \subfloat[Coefficients 2015-2019]{
        \includegraphics[width=0.49\textwidth]{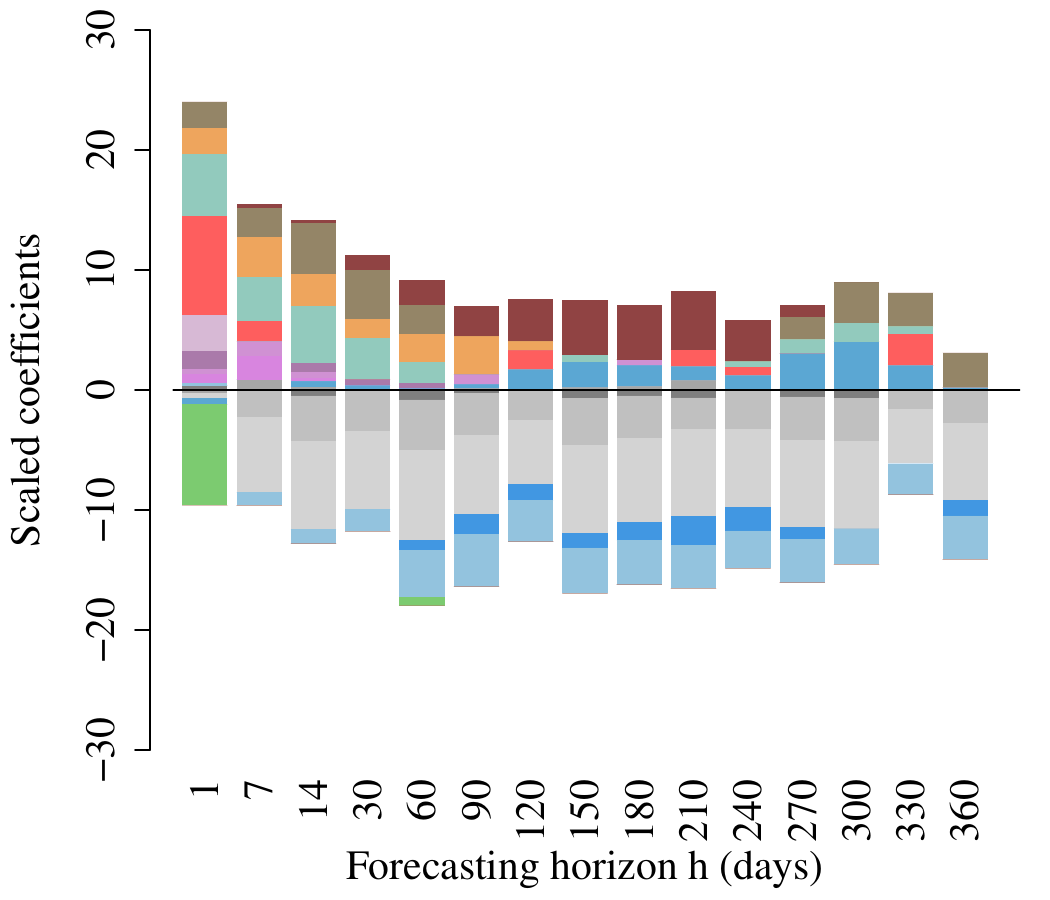}}
    \subfloat[Coefficients 2020-2024]{
        \includegraphics[width=0.49\textwidth]{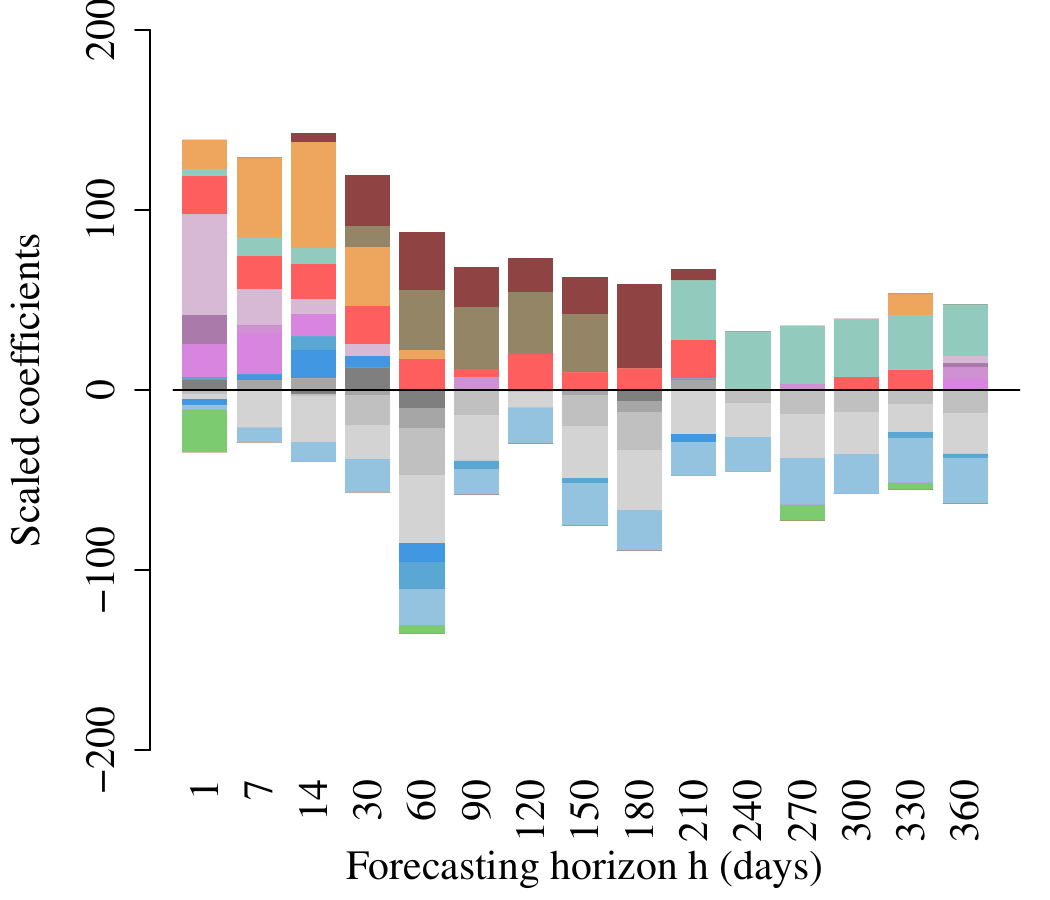}}
    \caption{Scaled coefficients of the constrained model, i.e. \textbf{expert model} (\ref{eq:expert}) with coefficient constraints in Table \ref{tab:bounds} for hour 9. Similar results were observed for all other hours.}
    \label{fig:expert_constr_coefs}
\end{figure}

Figure \ref{fig:expert_constr_coefs} shows the estimated coefficients of the \textbf{constrained expert model} \ref{eq:expert} using these bounds. Much more stable coefficients can be observed. This allows for some interpretation. First, the coefficients of short-term regressors such as renewables, load and autoregressive terms rapidly diminish to zero with increasing forecasting horizon. Nevertheless, these terms  reappear at higher horizons without clear patterns, hinting towards remaining spurious effects. Second, the weekend and seasonal effects become relevant with increasing horizons. This stems from the fact that seasonal effects are mainly driven by load and renewables, for which there are no accurate forecasts. Third,  
gas and EUA seem to be most relevant for short forecasting horizons, while coal and oil become more relevant towards middle horizons. Towards the end EUA seems to be the most relevant fuel variable. However, it is unclear if these are causal effects or spurious correlations. We will tackle this aspect in the corresponding section below.

\subsubsection{Portfolio effects}
\begin{wrapfigure}{r}{8cm}
	\includegraphics[width=8cm]{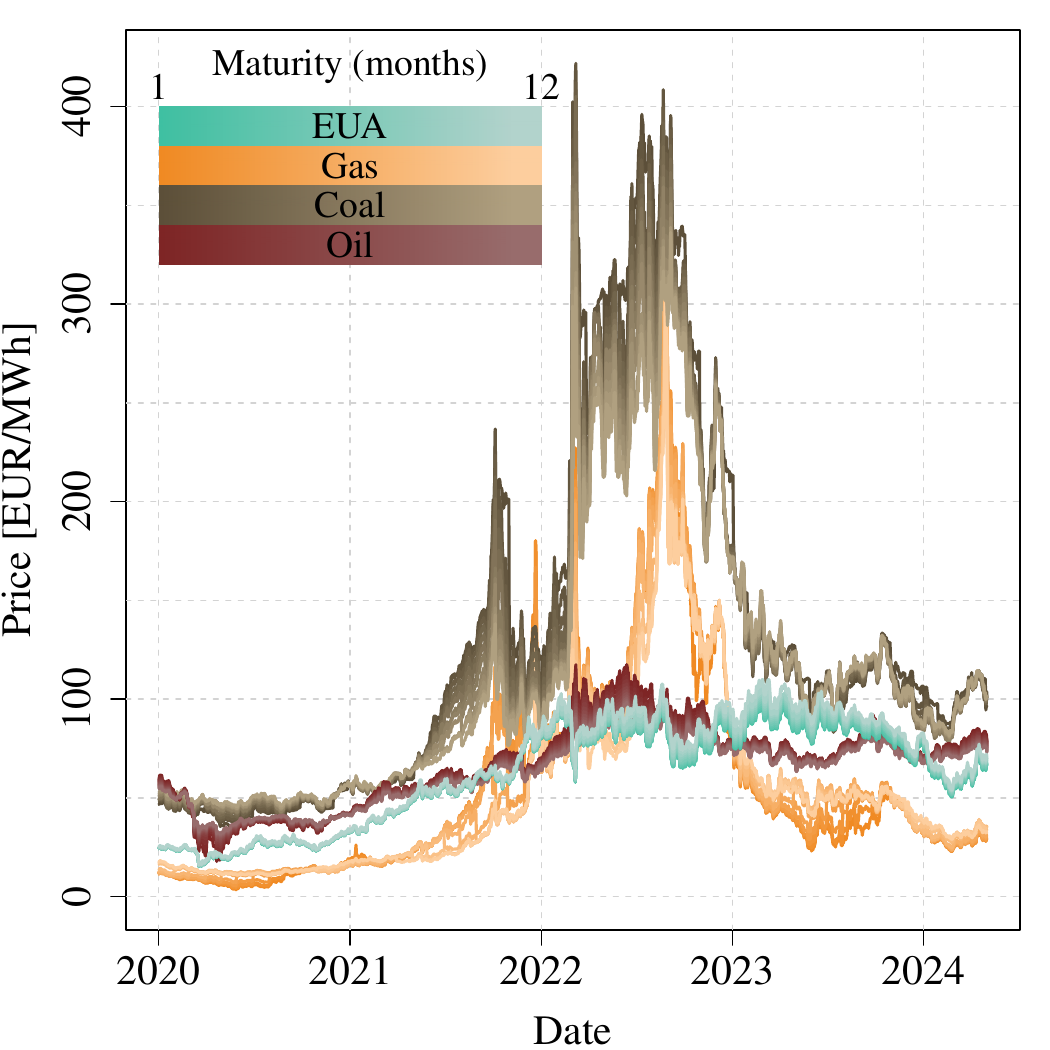}
	\caption{Prices of 1 to 12-month futures.}
	\label{fig:fuels_maturities}
\end{wrapfigure}

In the \textbf{constrained expert model} we used the front-month futures of the commodities $\textrm{CO}_2$, gas, coal, and oil. However, there are multiple futures available each day with maturities ranging from of one day up to several years. These are the prices that can be locked in for selling or buying the commodity at fixed time points in the future. They are highly correlated as shown in Figure \ref{fig:fuels_maturities} and theoretically represent the market participants' expectations of future prices under a risk-neutral measure. Sometimes the longer maturity futures are traded at higher prices than the shorter maturities or vice-versa leading to upward (contagos) or downwards (backwardation) sloping futures curves. This information is expected to be relevant for power price forecasting, as market participants often hedge their fuel costs by entering into futures contracts thus forming a portfolio \citep{caldana2017electricity, steinert2019short}. 
We define $D0$ to be the delivery period corresponding to the forecasting horizon $h$ such that the model
\begin{align}
	\label{eq:expert_portfolio}
	\text{Price}_{t+h} &= \sum_{i=0}^{12} \beta_i \overline{\text{Gas}}_{t-30i}^{(D0)} + \varepsilon_t 
\end{align}
represents the power price $h$ days ahead explained by average gas future prices of the last 12 months with delivery in $h$ days from today. 
We can extend the model to include also the trailing $D-1$ and leading $D+1$ months to delivery. Intuitively we would expect the future with a maturity corresponding to the forecasting horizon to be most relevant. However, when we extend the \textbf{constrained expert model} by including the terms in (\ref{eq:expert_portfolio}) with additional $D-1$ and $D+1$, we observe that the $D0$ future is not always the most relevant, but rather the most recent and liquid one. For example, it seems that when forecasting one month ahead, the current front-month future $D0$ is the most relevant. When forecasting 4 months ahead, it seems that the current trailing month future $D-1$, which is the 3-month future, is the most relevant, because it is closer to the present and hence more liquid and more informative. The lag structure used and the results are included in the Appendix in Tables \ref{tab:portfolio_lag_struct} and \ref{tab:portfolio} respectively. This observation aligns with the concept of an efficient market, where all available information is rapidly incorporated into the latest prices resembling a Markov chain dynamic. This represents some evidence in favor of always using front-month futures for every forecasting horizon. Furthermore, the expectations implied by the shape of the futures curve rarely happened. For example, at the start of the energy crisis in late 2021, the 12-month coal and gas futures were traded at a discount compared to the front-month futures, which implied that the market expected prices to decrease in the future. However, this was not the case, as prices increased dramatically over the next year. The effect was observed in early 2023. This gives some intuition as to why including multiple futures might not be as relevant as expected in a price forecasting model.

\subsection{Short-term regressors}
Short-term regressors refers to explanatory variables that can not be accurately forecasted beyond a few days. These are variables derived from meteorological data such as infeed from solar and wind, but also load since it is partly influenced by temperature and wind speed. Infeed from RES and load have very strong impacts on the day-ahead power price and are crucial for explaining effects such as negative prices, the merit order effect of renewables, and the weekend effect of load (see Figure \ref{fig:price_res_load}). However, since it is only possible to accurately forecast these variables one day in advance, power price models beyond this meteorological forecasting horizon do not benefit from these variables. This is evident in Figure \ref{fig:expert_constr_coefs}a, where the coefficients of renewables infeed and load are essentially only relevant for the day-ahead forecast. This is because the model only includes day-ahead forecasts of these variables. 

\begin{figure}[hbtp]
	\centering
	\scalebox{0.5}{
		\includegraphics[width=0.5\textwidth]{hours_color_legend.pdf}}
	\linebreak
    \subfloat[Renewables actual values, fitted values and forecasts for 360 days ahead by hour.]{
        \includegraphics[width=0.9\textwidth]{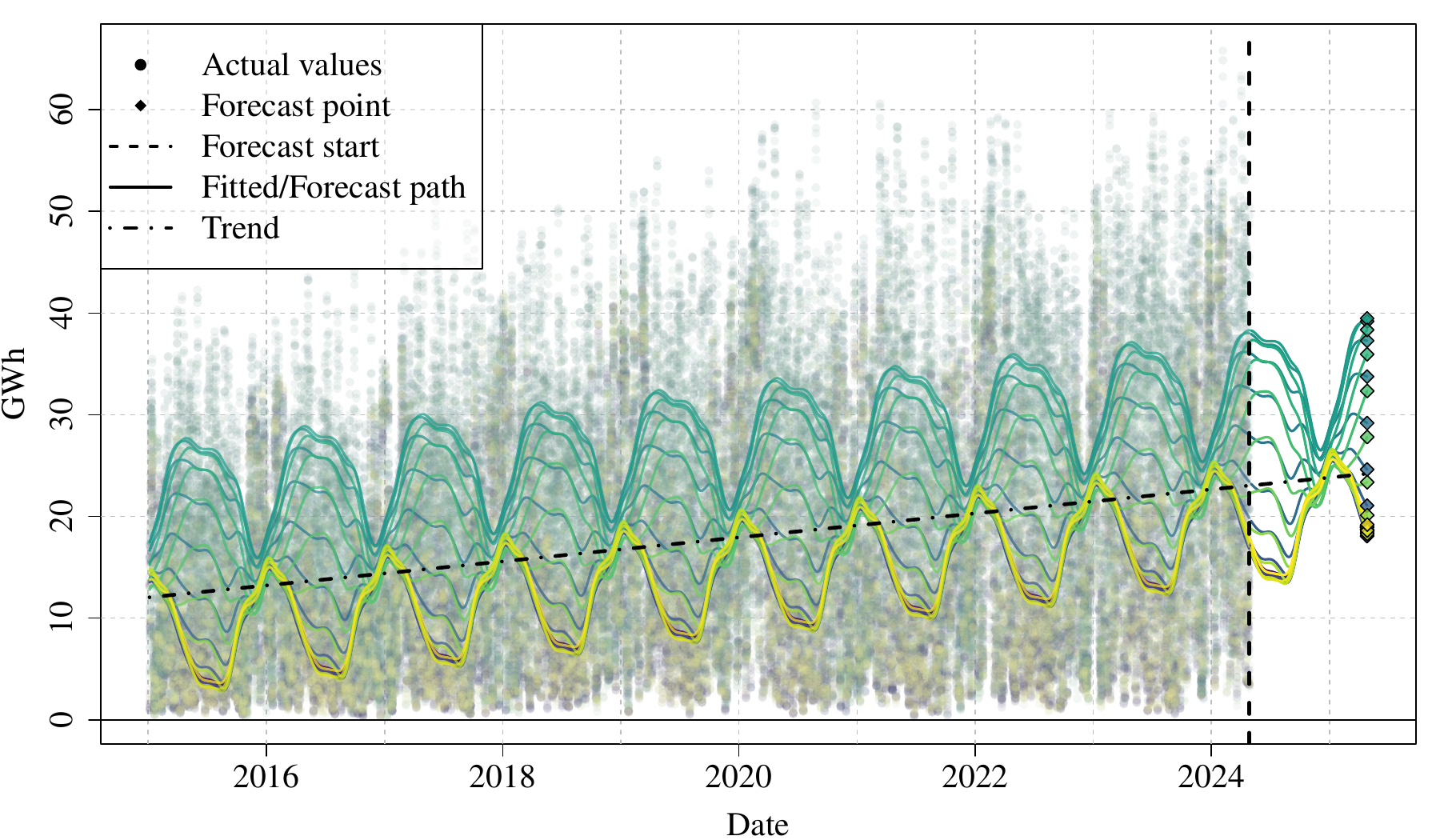}}
	\linebreak
    \subfloat[Load actual values, fitted values and forecasts for 360 days ahead by hour.]{
        \includegraphics[width=0.9\textwidth]{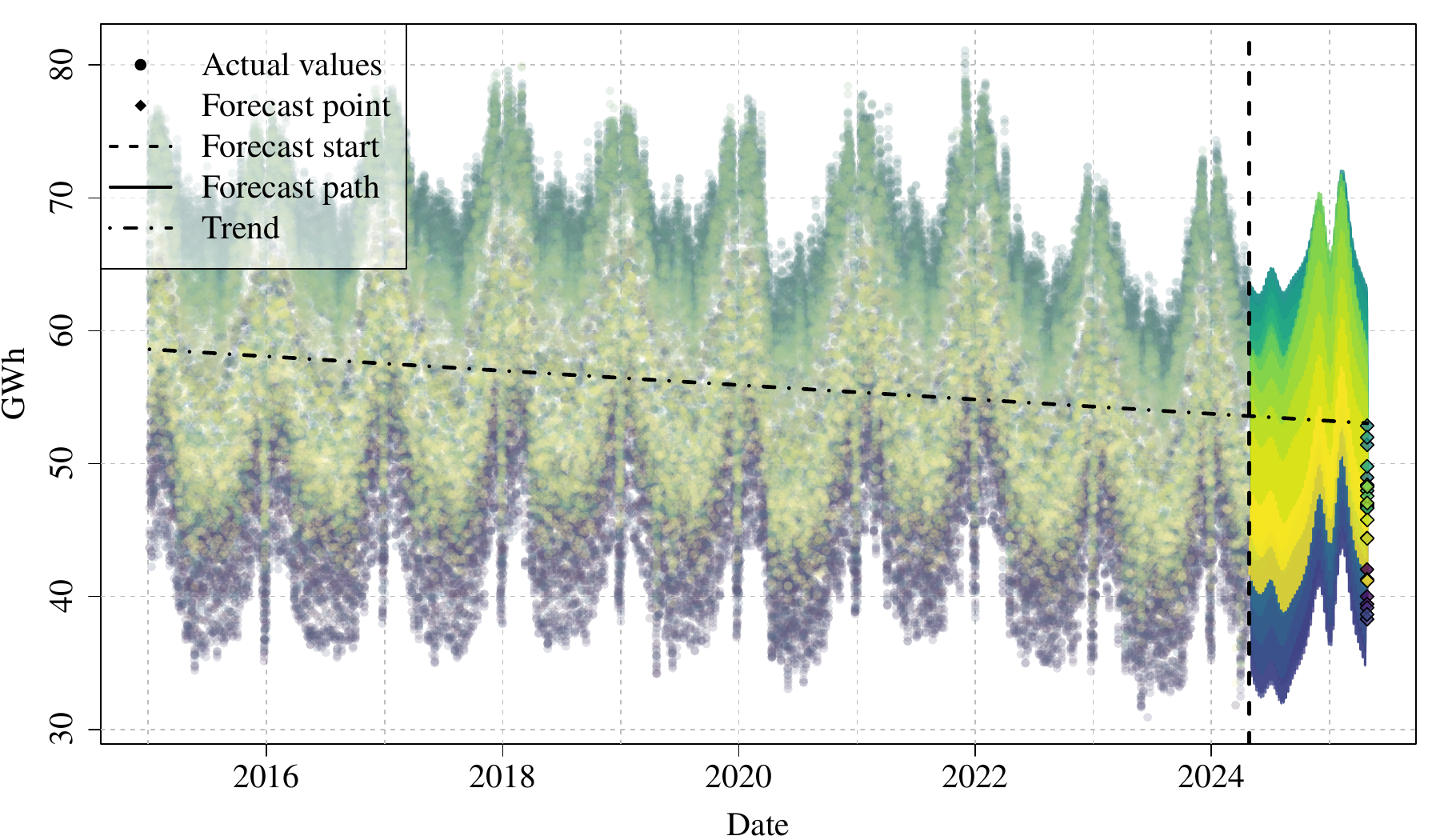}}
    \caption{The forecast was generated using Generalized Additive Models (GAMs).}
    \label{fig:res_load_forecast}
\end{figure}

It is however possible to forecast the daily, weekly and annual seasonalities as shown in Figure \ref{fig:res_load_forecast}. Here, short-term shocks can not be foreseen, but rather only broad observations such as that there is higher solar irradiation around noon and during summer, there is stronger wind during winter and spring, that load is higher on weekdays and in winter, and lower on weekends and in summer. Furthermore, the expected expansion of renewable capacity and development of load can be included in the forecast, which we did in the form of linear trends.
We can now include these forecasts in the \textbf{expert model}. However, simply regressing the power price on these forecasts would still greatly underestimate their contribution as some effects, such as negative price shocks due to sudden high RES infeed, would not be explainable. To address this, the RES and load coefficients can be estimated by regressing the power price on their actual values and then using the seasonal forecasts to generate power price forecasts. We refer to this approach as the \textbf{current model} and introduce it in the next section.

\subsection{Unit root behavior of power prices}

\subsubsection{Commodities as stochastic drivers of power prices}
A widely accepted fact in finance is that stock prices exhibit unit root behavior, meaning that they are non-stationary and possess a stochastic trend. This assumption is evident in the modeling of stocks as Brownian motions, also referred to as Gaussian or Wiener processes, or functions thereof for structuring and valuation purposes, such as pricing derivatives \citep{black1973pricing, schwartz2000short, hull2009optionen}. This unit-root behavior extends to commodities as well \citep{schwartz2000short}. In fact, unit root tests for $\textrm{CO}_2$, gas, coal and oil futures fail to reject the null hypothesis of having a unit root (see Figure \ref{fig:adf_test} in the Appendix) \citep{berrisch2023modeling}. As these commodities represent fuels for power generation, it is reasonable to assume that the power price inherits some unit root behavior from them, which is particularly visible in the long run (see Figure \ref{fig:time_series}). 

Some evidence for this is provided by considering the \textbf{differenced expert model},
where instead of modeling the levels of each component of the \textbf{expert model} we model their one-period differences, which is standard practice for eliminating a stochatic trend. The results are shown in Table \ref{tab:diff_rmse}. There are two important observations. First, differentiation when fuels are not included sometimes improves the model, especially for the energy crisis period, indicating that power price do have a unit root behavior. Second, for the day-ahead case we observe that differentiation does not bring any improvement when fuels are included in the model, indicating that the unit root behavior is captured by the fuel prices. Also, the inclusion of fuels tends to visibly improve the forecast. This effect fades with increasing forecasting horizons since the power price drifts further away from its current value at $t$, which can not be explained anymore by the current values of the fuel prices.

\begin{table}[hbtp]
	\centering
	\scalebox{0.5}{
	\begin{tabular}{|c|rr|rr|rr|rr|rr|rr|rr|rr|}
		\hline
		\textbf{Horizon}  & \multicolumn{4}{|c|}{\textbf{1}} & \multicolumn{4}{|c|}{\textbf{90}} & \multicolumn{4}{|c|}{\textbf{180}} & \multicolumn{4}{|c|}{\textbf{360}}\\ \hline \textbf{Model} & \multicolumn{2}{|c|}{\textbf{No fuels}} & \multicolumn{2}{|c|}{\textbf{Fuels}}& \multicolumn{2}{|c|}{\textbf{No fuels}} & \multicolumn{2}{|c|}{\textbf{Fuels}}& \multicolumn{2}{|c|}{\textbf{No fuels}} & \multicolumn{2}{|c|}{\textbf{Fuels}}& \multicolumn{2}{|c|}{\textbf{No fuels}} & \multicolumn{2}{|c|}{\textbf{Fuels}}\\ \hline  \textbf{Diff}  & \textbf{No} & \textbf{Yes} & \textbf{No} & \textbf{Yes}  & \textbf{No} & \textbf{Yes} & \textbf{No} & \textbf{Yes}  & \textbf{No} & \textbf{Yes} & \textbf{No} & \textbf{Yes}  & \textbf{No} & \textbf{Yes} & \textbf{No} & \textbf{Yes}  \\  \hline  \hline
	  2018 & \cellcolor[rgb]{1,0.6,0.6} $\textbf{8.84}$ & \cellcolor[rgb]{0.6,1,0.6} $\textbf{7.75}$ & \cellcolor[rgb]{0.6,1,0.6} $\textbf{7.17}$ & \cellcolor[rgb]{1,0.6,0.6} $\textbf{7.77}$ & \cellcolor[rgb]{0.6,1,0.6} $\textbf{20.25}$ & \cellcolor[rgb]{1,0.6,0.6} $\textbf{21.72}$ & \cellcolor[rgb]{0.6,1,0.6} $\textbf{17.65}$ & \cellcolor[rgb]{1,0.6,0.6} $\textbf{21.64}$ & \cellcolor[rgb]{0.6,1,0.6} $\textbf{22.34}$ & \cellcolor[rgb]{1,0.6,0.6} $\textbf{26.67}$ & \cellcolor[rgb]{0.6,1,0.6} $\textbf{21.62}$ & \cellcolor[rgb]{1,0.6,0.6} $\textbf{26.55}$ & \cellcolor[rgb]{0.6,1,0.6} $\textbf{23.99}$ & \cellcolor[rgb]{1,0.6,0.6} $\textbf{26.28}$ & \cellcolor[rgb]{0.6,1,0.6} $\textbf{24.00}$ & \cellcolor[rgb]{1,0.6,0.6} $\textbf{26.27}$ \\ 
		2019 & \cellcolor[rgb]{1,0.6,0.6} $\textbf{9.37}$ & \cellcolor[rgb]{0.6,1,0.6} $\textbf{9.32}$ & \cellcolor[rgb]{0.6,1,0.6} $\textbf{8.55}$ & \cellcolor[rgb]{1,0.6,0.6} $\textbf{9.33}$ & \cellcolor[rgb]{0.6,1,0.6} $\textbf{13.73}$ & \cellcolor[rgb]{1,0.6,0.6} $\textbf{21.44}$ & \cellcolor[rgb]{0.6,1,0.6} $\textbf{16.48}$ & \cellcolor[rgb]{1,0.6,0.6} $\textbf{21.35}$ & \cellcolor[rgb]{0.6,1,0.6} $\textbf{13.68}$ & \cellcolor[rgb]{1,0.6,0.6} $\textbf{23.68}$ & \cellcolor[rgb]{0.6,1,0.6} $\textbf{17.43}$ & \cellcolor[rgb]{1,0.6,0.6} $\textbf{23.63}$ & \cellcolor[rgb]{0.6,1,0.6} $\textbf{14.34}$ & \cellcolor[rgb]{1,0.6,0.6} $\textbf{22.79}$ & \cellcolor[rgb]{0.6,1,0.6} $\textbf{14.51}$ & \cellcolor[rgb]{1,0.6,0.6} $\textbf{22.79}$ \\ 
		2020 & \cellcolor[rgb]{0.6,1,0.6} $\textbf{8.38}$ & \cellcolor[rgb]{1,0.6,0.6} $\textbf{8.73}$ & \cellcolor[rgb]{0.6,1,0.6} $\textbf{7.94}$ & \cellcolor[rgb]{1,0.6,0.6} $\textbf{8.72}$ & \cellcolor[rgb]{0.6,1,0.6} $\textbf{16.36}$ & \cellcolor[rgb]{1,0.6,0.6} $\textbf{22.40}$ & \cellcolor[rgb]{0.6,1,0.6} $\textbf{19.97}$ & \cellcolor[rgb]{1,0.6,0.6} $\textbf{22.28}$ & \cellcolor[rgb]{0.6,1,0.6} $\textbf{17.25}$ & \cellcolor[rgb]{1,0.6,0.6} $\textbf{25.25}$ & \cellcolor[rgb]{0.6,1,0.6} $\textbf{20.28}$ & \cellcolor[rgb]{1,0.6,0.6} $\textbf{25.20}$ & \cellcolor[rgb]{0.6,1,0.6} $\textbf{16.15}$ & \cellcolor[rgb]{1,0.6,0.6} $\textbf{22.36}$ & \cellcolor[rgb]{0.6,1,0.6} $\textbf{17.30}$ & \cellcolor[rgb]{1,0.6,0.6} $\textbf{22.37}$ \\ 
		2021 & \cellcolor[rgb]{0.6,1,0.6} $\textbf{29.98}$ & \cellcolor[rgb]{1,0.6,0.6} $\textbf{30.19}$ & \cellcolor[rgb]{0.6,1,0.6} $\textbf{29.11}$ & \cellcolor[rgb]{1,0.6,0.6} $\textbf{30.23}$ & \cellcolor[rgb]{1,0.6,0.6} $\textbf{71.53}$ & \cellcolor[rgb]{0.6,1,0.6} $\textbf{67.79}$ & \cellcolor[rgb]{0.6,1,0.6} $\textbf{62.81}$ & \cellcolor[rgb]{1,0.6,0.6} $\textbf{67.75}$ & \cellcolor[rgb]{1,0.6,0.6} $\textbf{91.38}$ & \cellcolor[rgb]{0.6,1,0.6} $\textbf{83.37}$ & \cellcolor[rgb]{1,0.6,0.6} $\textbf{92.62}$ & \cellcolor[rgb]{0.6,1,0.6} $\textbf{83.35}$ & \cellcolor[rgb]{0.6,1,0.6} $\textbf{92.76}$ & \cellcolor[rgb]{1,0.6,0.6} $\textbf{95.41}$ & \cellcolor[rgb]{1,0.6,0.6} $\textbf{97.60}$ & \cellcolor[rgb]{0.6,1,0.6} $\textbf{95.43}$ \\ 
		2022 & \cellcolor[rgb]{0.6,1,0.6} $\textbf{55.72}$ & \cellcolor[rgb]{1,0.6,0.6} $\textbf{55.96}$ & \cellcolor[rgb]{0.6,1,0.6} $\textbf{54.60}$ & \cellcolor[rgb]{1,0.6,0.6} $\textbf{55.94}$ & \cellcolor[rgb]{1,0.6,0.6} $\textbf{213.56}$ & \cellcolor[rgb]{0.6,1,0.6} $\textbf{201.35}$ & \cellcolor[rgb]{0.6,1,0.6} $\textbf{159.06}$ & \cellcolor[rgb]{1,0.6,0.6} $\textbf{202.02}$ & \cellcolor[rgb]{1,0.6,0.6} $\textbf{186.59}$ & \cellcolor[rgb]{0.6,1,0.6} $\textbf{178.01}$ & \cellcolor[rgb]{0.6,1,0.6} $\textbf{165.02}$ & \cellcolor[rgb]{1,0.6,0.6} $\textbf{178.06}$ & \cellcolor[rgb]{1,0.6,0.6} $\textbf{253.16}$ & \cellcolor[rgb]{0.6,1,0.6} $\textbf{199.91}$ & \cellcolor[rgb]{1,0.6,0.6} $\textbf{245.72}$ & \cellcolor[rgb]{0.6,1,0.6} $\textbf{199.91}$ \\ 
		2023 & \cellcolor[rgb]{1,0.6,0.6} $\textbf{25.80}$ & \cellcolor[rgb]{0.6,1,0.6} $\textbf{25.03}$ & \cellcolor[rgb]{0.6,1,0.6} $\textbf{24.32}$ & \cellcolor[rgb]{1,0.6,0.6} $\textbf{24.94}$ & \cellcolor[rgb]{0.6,1,0.6} $\textbf{73.03}$ & \cellcolor[rgb]{1,0.6,0.6} $\textbf{94.74}$ & \cellcolor[rgb]{1,0.6,0.6} $\textbf{105.49}$ & \cellcolor[rgb]{0.6,1,0.6} $\textbf{94.70}$ & \cellcolor[rgb]{1,0.6,0.6} $\textbf{280.55}$ & \cellcolor[rgb]{0.6,1,0.6} $\textbf{179.53}$ & \cellcolor[rgb]{1,0.6,0.6} $\textbf{247.55}$ & \cellcolor[rgb]{0.6,1,0.6} $\textbf{179.21}$ & \cellcolor[rgb]{1,0.6,0.6} $\textbf{609.41}$ & \cellcolor[rgb]{0.6,1,0.6} $\textbf{197.80}$ & \cellcolor[rgb]{1,0.6,0.6} $\textbf{576.04}$ & \cellcolor[rgb]{0.6,1,0.6} $\textbf{197.93}$ \\ 
		2024 & \cellcolor[rgb]{0.6,1,0.6} $\textbf{22.91}$ & \cellcolor[rgb]{1,0.6,0.6} $\textbf{26.64}$ & \cellcolor[rgb]{0.6,1,0.6} $\textbf{25.67}$ & \cellcolor[rgb]{1,0.6,0.6} $\textbf{26.52}$ & \cellcolor[rgb]{1,0.6,0.6} $\textbf{58.23}$ & \cellcolor[rgb]{0.6,1,0.6} $\textbf{54.84}$ & \cellcolor[rgb]{1,0.6,0.6} $\textbf{77.64}$ & \cellcolor[rgb]{0.6,1,0.6} $\textbf{54.78}$ & \cellcolor[rgb]{1,0.6,0.6} $\textbf{71.60}$ & \cellcolor[rgb]{0.6,1,0.6} $\textbf{58.26}$ & \cellcolor[rgb]{1,0.6,0.6} $\textbf{138.29}$ & \cellcolor[rgb]{0.6,1,0.6} $\textbf{57.98}$ & \cellcolor[rgb]{1,0.6,0.6} $\textbf{119.13}$ & \cellcolor[rgb]{0.6,1,0.6} $\textbf{68.55}$ & \cellcolor[rgb]{1,0.6,0.6} $\textbf{162.18}$ & \cellcolor[rgb]{0.6,1,0.6} $\textbf{68.44}$ \\ 
		 \hline
	\end{tabular}}
	\caption{Yearly RMSE (EUR/MWh) out-of-sample showing effect of differencing with and without fuels as regressors for selected forecasting horizons.} 
	\label{tab:diff_rmse}
\end{table}

Power and fuels prices could be cointegrated for short forecasting horizons, as they do seem to move together in the long run \citep{moutinho2022examining}. However, the cointegration relationship might be complex and might change both over time due to complex effects such as fuel switches and supply-stack non-linearities. Furthermore, when the forecasting horizon is large, it is unlikely that a long-term equilibrium relationship even exists. Lastly, our 24-dimensional model structure, where we model each hour individually, and the usage of external regressors does not fit in the classical framework of cointegration and vector error correction models (VECMs), making standard statistical tests and theory unapplicable.

\subsubsection{Spurious effects}

A well-known effect in time series analysis is that regressing a unit root process on another independent unit root process can lead to spurious regression results \citep{granger1974spurious, lutkepohl2005new}. In our context, we have seen that both the response, which is the power price, and some explanatory variables, such as the fuel prices, exhibit unit root behavior. There is some evidence that load also exhibits long-term unit root behavior \citep{smyth2013fluctuations}. However, these explanatory variables are not independent of the power price. For short forecasting horizons such as for day-ahead they are important variables with a lot of explanatory power, hence there is little risk of spurious effects. However, with increasing forecasting horizons, the model rapidly approaches the case of a spurious regression. 

\begin{figure}[htbp]
    \centering
		\includegraphics[width=0.63\textwidth]{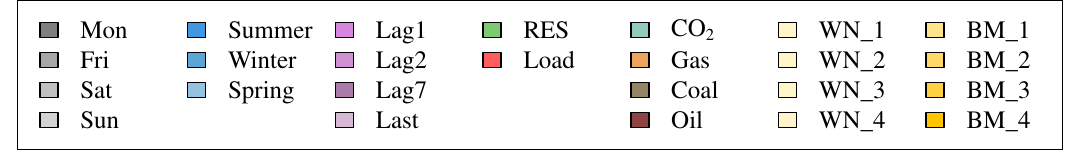}
	\linebreak
    \subfloat[Coefficients 2015-2019]{
        \includegraphics[width=0.49\textwidth]{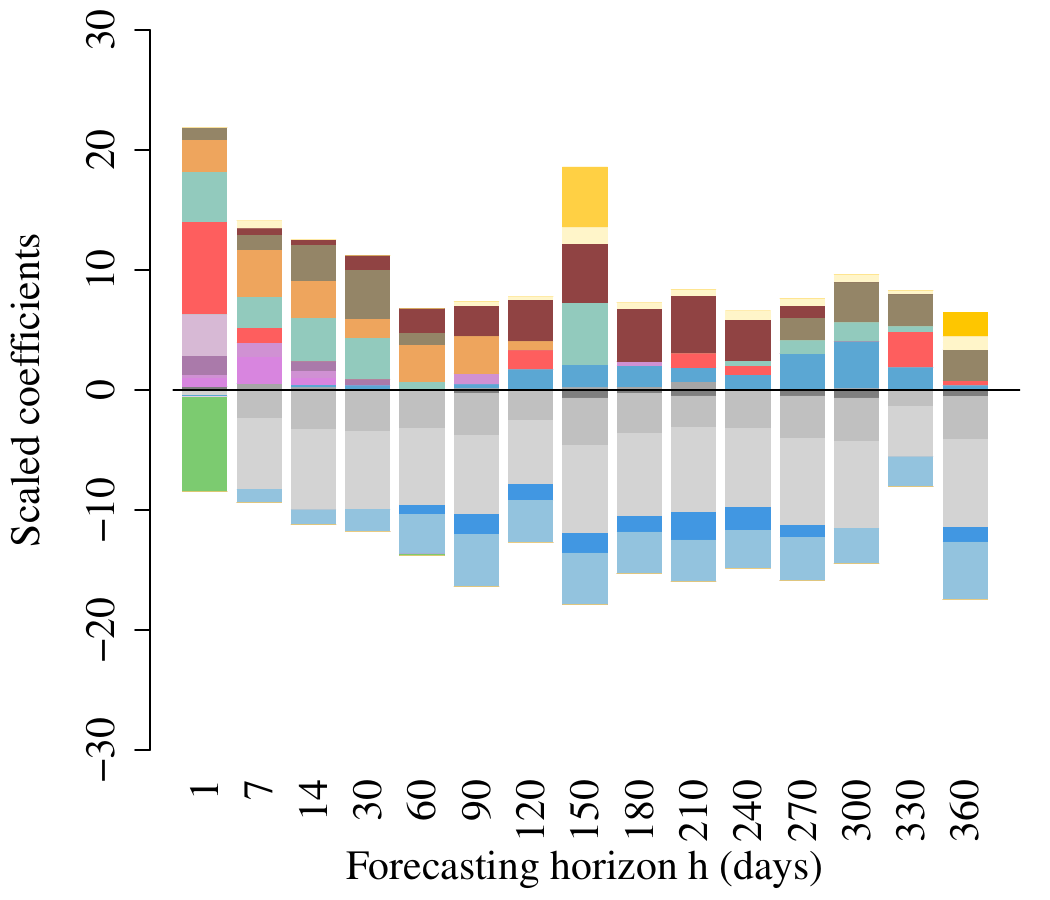}}
    \subfloat[Coefficients 2020-2024]{
        \includegraphics[width=0.49\textwidth]{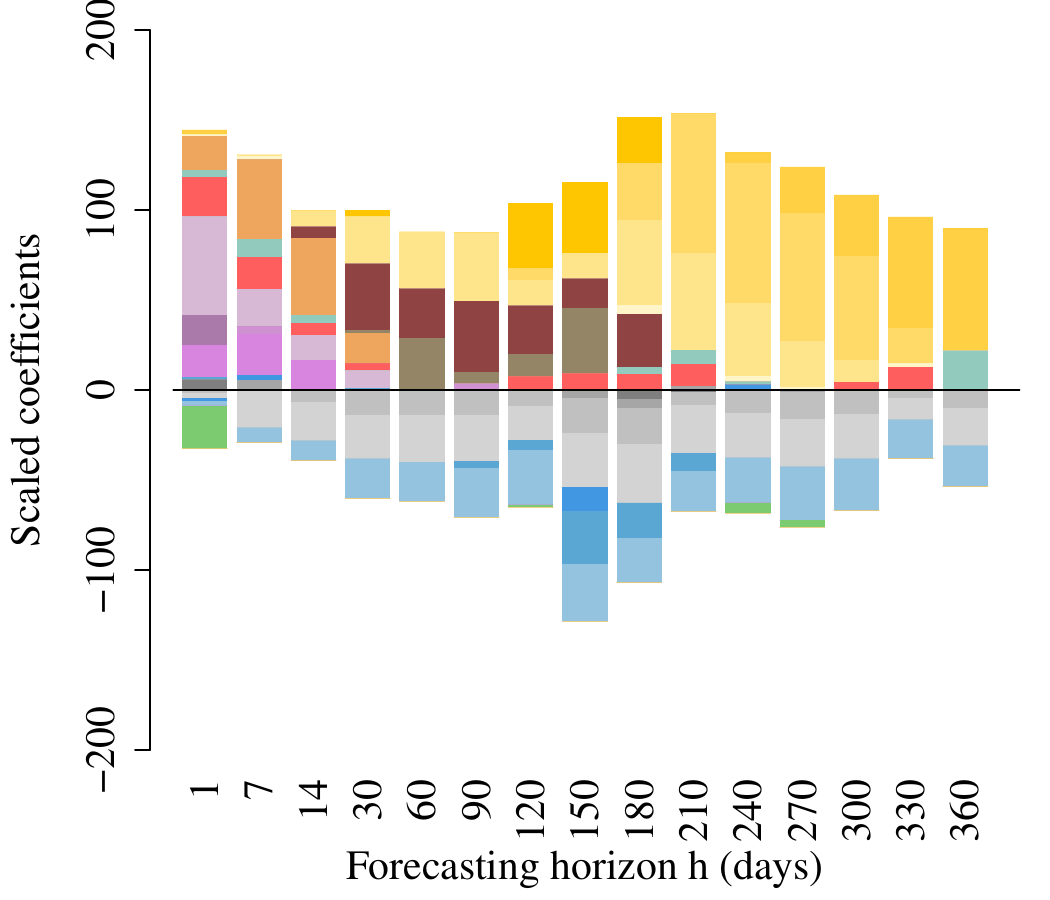}}
	\caption{Scaled coefficients of the constrained model (\ref{eq:expert}) with 4 randomly generated brownian motions and white noises as regressors, for hour 9.}
    \label{fig:expert_rwwn}
\end{figure}

This effect is illustrated in Figure \ref{fig:expert_rwwn} where 4 randomly generated brownian motions and white noises were included as regressors into the \textbf{constrained expert model} with positivity constraints. 
It becomes apparent that for higher horizons the coefficients of the brownian motions are not always shrunk to zero, indicating that the model is struggling to filter out the stochastic trend from the regressors. This effect is especially prominent for the period after the energy crisis, where the coefficients of the brownian motions are at times exclusively chosen. 
This is because as the temporal gap between response and regressor widens the explanatory variables stop being informative. Essentially we experience a loss of explanatory power. For example, the RES infeed of today will have some correlation with the infeed of tomorrow but almost none with the infeed one year from now. The same applies to the fuels, but given that they have a unit root behavior they are prone to strong spurious effects, similar to the brownian motions above.

\subsubsection{Current fit model}

In the previous section we explained how the lag between the response and regressors leads to spurious effects. A natural solution would be to close this gap and estimate the same-day relationships between the response and regressors. This way the model would be able to capture the true relationship between the response and regressors, avoiding spurious effects. However, this approach alone would not be feasible for forecasting, since the regressors at time $t+h$ would be needed, which are not known at time $t$. Therefore, for the forecast we would need to substitute the regressors with their forecasts. The forecasts for RES infeed and load will be their seasonal averages (see Figure \ref{fig:res_load_forecast}). The forecasts for the commodities $\textrm{CO}_2$, gas, coal and oil will be the futures with maturities corresponding to the forecasting horizon. By doing so the current relationship between response and regressors is captured and projected into the future where it is assumed to still hold. This effectively disentagles estimation from forecasting leading to a two-step approach which we refer to as the \textbf{current model}. 
\begin{enumerate}
	\item Estimate the current relationship between the response and regressors.
	\begin{align}
		\text{Price}_{t+h} = & \beta_0 + \beta_1 \text{Mon}_{t+h} + \beta_2 \text{Fri}_{t+h} + \beta_3 \text{Sat}_{t+h} + \beta_4 \text{Sun}_{t+h}  \\
		& + \beta_5 \text{Winter}_{t+h} + \beta_6 \text{Spring}_{t+h} + \beta_7 \text{Summer}_{t+h} \nonumber \\ 
		& + \beta_8 \text{Price}_{t} + \beta_9 \text{Price}_{t-1} + \beta_{10} \text{Price}_{t-6} + \beta_{11} \text{PriceLastHour}_{t} \nonumber  \\
		& + \beta_{12} \text{RES}_{\color{red}{t+h}} + \beta_{13} \text{Load}_{\color{red}{t+h}} \nonumber  \\
		& + \beta_{14} \textrm{CO}^{\text{(01)}}_{2\color{red}{t+h}} + \beta_{15} \text{Gas}^{\text{(01)}}_{\color{red}{t+h}} + \beta_{16} \text{Coal}^{\text{(01)}}_{\color{red}{t+h}} + \beta_{17} \text{Oil}^{\text{(01)}}_{\color{red}{t+h}} + \varepsilon_{\color{red}{t+h}} \nonumber 
		\label{eq:expert_current_est}
	\end{align}
	\item Forecast the response by substituting the regressors with their forecasts.
	\begin{align}
		\what{\text{Price}}_{T+h} = & \beta_0 + \beta_1 \text{Mon}_{T+h} + \beta_2 \text{Fri}_{T+h} + \beta_3 \text{Sat}_{T+h} + \beta_4 \text{Sun}_{T+h}  \\
		& + \beta_5 \text{Winter}_{T+h} + \beta_6 \text{Spring}_{T+h} + \beta_7 \text{Summer}_{T+h} \nonumber \\ 
		& + \beta_8 \text{Price}_{T} + \beta_9 \text{Price}_{T-1} + \beta_{10} \text{Price}_{T-6} + \beta_{11} \text{PriceLastHour}_{T} \nonumber  \\
		& + \beta_{12} \what{\text{RES}}_{\color{red}{T+h}} + \beta_{13} \what{\text{Load}}_{\color{red}{T+h}} \nonumber \\
		& + \beta_{14} \textrm{CO}_{\color{black}{2}}^{\color{red}{\text{(h)}}}_{\color{red}{T}} + \beta_{15} \text{Gas}^{\color{red}{\text{(h)}}}_{\color{red}{T}} + \beta_{16} \text{Coal}^{\color{red}{\text{(h)}}}_{\color{red}{T}} + \beta_{17} \text{Oil}^{\color{red}{\text{(h)}}}_{\color{red}{T}} \nonumber 
		\label{eq:expert_current_for}
	\end{align}
\end{enumerate}
where $(01)$ is the front-month future, which is the most recent available price, and $(h)$ is the future with a maturity corresponding to $h$ as of $T$. Highlighted in red are the main changes from the \textbf{expert model} in (\ref{eq:expert}).

\begin{figure}[htbp]
    \centering
	\includegraphics[width=0.49\textwidth]{expert_win3_coefs_legend.pdf}
	\linebreak
    \subfloat[Coefficients 2015-2019]{
        \includegraphics[width=0.49\textwidth]{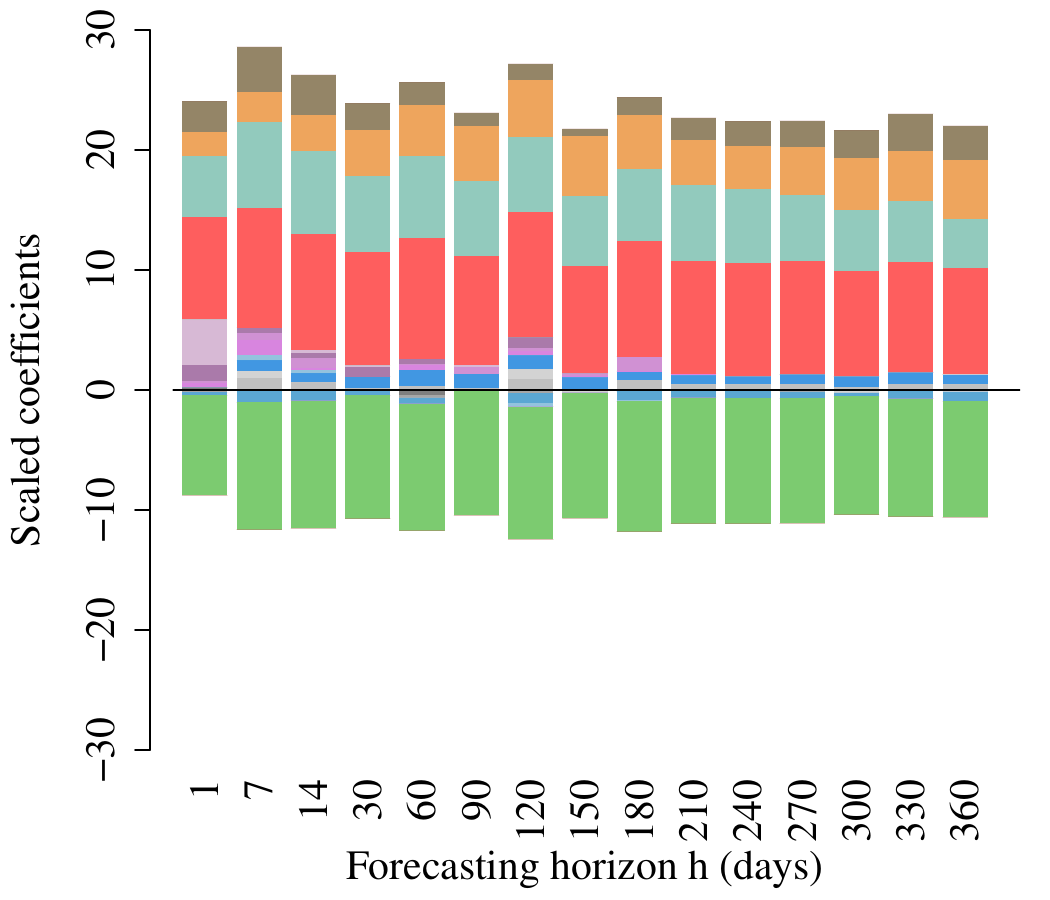}}
    \subfloat[Coefficients 2020-2024]{
        \includegraphics[width=0.49\textwidth]{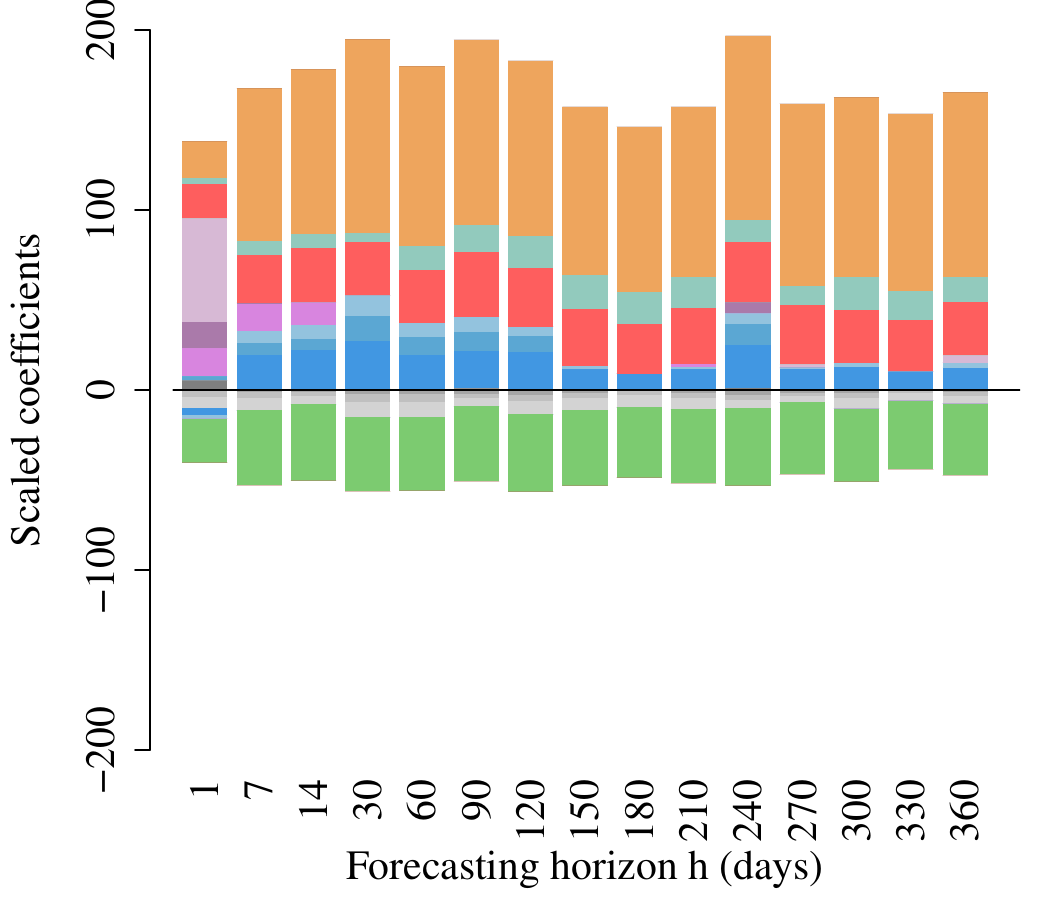}}
	\caption{Scaled coefficients of the \textbf{current model} \ref{eq:expert_current_est}  for hour 9. Similar results were observed for all other hours.}
    \label{fig:expert_current_coefs}
\end{figure}
Figure \ref{fig:expert_current_coefs} shows the coefficients estimated from the \textbf{current model}. These are very stable accross the horizons, with oil completely missing and coal only being active for the pre-crisis period. RES and load become major components for all horizons, similar to the day-ahead case of the \textbf{expert model} (see Figure \ref{fig:expert_noconstr_coefs}). Consequently, the weekday dummies become less relevant, as weekly seasonalities are captured by load. Such results are what would be fundamentally expected, as the price-setting technology remains stable for the same day regardless of when it was forecasted. In the first plot in Figure \ref{fig:expert_current_coefs}a price setting plants are coal and gas, while after the start of the energy crisis gas becomes the only price setting technology for this hour. Contrast this to previous models such as Figure \ref{fig:expert_rwwn}a where the price setting technology switches from coal and gas to oil as the forecasting horizon increases, even though the same day is forecasted every time. For example, according this figure the price setting technology switches between $h=120$ to $h=150$ from less carbon intense oil plants to very high carbon intense oil plants, as the $\textrm{CO}_2$ coefficient is suddenly very high. This is most likely a spurious effect, as also suggested by the high coefficient of the brownian motion at this point. Hence the \textbf{current model} offers much more interpretable results.

\section{Results and discussion}
\label{sec:results}

In the previous section we introduced three different extensions of the original \textbf{merit order model}. To evaluate their performance and pinpoint their contribution to the forecasting accuracy we consider different combinations of these methods. We also include two benchmark models. First, we include the \textbf{naive model}, which implies a random walk, as it simply returns the last available price if the day to be forecasted is Tuesday through Thursday, or the last available price of Monday, Friday, Saturday or Sunday if the day to be forecasted is one of these. The differentiation by days is done to preserve the weekly seasonality. Second, we include and the \textbf{weekday dummies model} (WD) which represents the average weekday prices. As evaluation criteria we use the mean absolute error (MAE) and the root mean squared error (RMSE) which are defined as:
\begin{align*}
	& \text{MAE}(\varepsilon) = \frac{1}{24N} \sum_{i=1}^{N} \sum_{h=1}^{24} |\what \varepsilon_{T+i,h}| \text{  and  } \text{RMSE}(\varepsilon) = \sqrt{\frac{1}{24N} \sum_{i=1}^{N} \sum_{h=1}^{24} \what \varepsilon_{T+i,h}^2}
\end{align*}
where $\what \varepsilon_{t,h}$ is the forecasting error at time $t$ for hour $h$.
Table \ref{tab:rmse_models} shows the RMSE results of the considered models. Table \ref{tab:mae_models} in the Appendix shows similar results for the MAE.

\begin{table}[htbp]
	\centering
	\setlength\extrarowheight{3pt}	
	\setlength{\tabcolsep}{2pt}
	\scalebox{0.58}{\begin{tabular}{|r|ccccc|cccc|rrrrrrrrrrrrrrr|}
	\hline
	\bf Model name & \multicolumn{5}{|c|}{\bf Variables} & \multicolumn{4}{|c|}{\bf Method} & \multicolumn{15}{c|}{\bf Forecasting horizon} \\
	\hline
	 & \rotatebox{90}{Week Dummies} & \rotatebox{90}{Annual Seasons} & \rotatebox{90}{Autoreg} & \rotatebox{90}{RES \& Load} & \rotatebox{90}{Fuels} & \rotatebox{90}{Unconstrained} & \rotatebox{90}{Constrained} & \rotatebox{90}{Differencing} & \rotatebox{90}{Current} & 1 & 7 & 14 & 30 & 60 & 90 & 120 & 150 & 180 & 210 & 240 & 270 & 300 & 330 & 360 \\ 
	\hline
	\hline
	naive &  &  &  &  &  &  &  &  &  & \cellcolor[rgb]{0.832,1,0.6} $39.46$ & \cellcolor[rgb]{0.736,1,0.6} $47.21$ & \cellcolor[rgb]{0.763,1,0.6} $53.12$ & \cellcolor[rgb]{0.855,1,0.6} $61.84$ & \cellcolor[rgb]{0.955,1,0.6} $68.30$ & \cellcolor[rgb]{0.882,1,0.6} $72.62$ & \cellcolor[rgb]{0.986,1,0.6} $81.68$ & \cellcolor[rgb]{0.914,1,0.6} $83.61$ & \cellcolor[rgb]{0.871,1,0.6} $87.25$ & \cellcolor[rgb]{0.804,1,0.6} $89.97$ & \cellcolor[rgb]{0.767,1,0.6} $90.96$ & \cellcolor[rgb]{0.749,1,0.6} $89.96$ & \cellcolor[rgb]{0.722,1,0.6} $94.27$ & \cellcolor[rgb]{0.717,1,0.6} $97.36$ & \cellcolor[rgb]{0.743,1,0.6} $98.38$ \\ 
	WD & $\times$ &  &  &  &  & $\times$ &  &  &  & \cellcolor[rgb]{1,0.6,0.6} $76.95$ & \cellcolor[rgb]{1,0.6,0.6} $76.92$ & \cellcolor[rgb]{1,0.6,0.6} $77.17$ & \cellcolor[rgb]{1,0.6,0.6} $77.79$ & \cellcolor[rgb]{1,0.606,0.6} $78.39$ & \cellcolor[rgb]{1,0.841,0.6} $78.78$ & \cellcolor[rgb]{0.881,1,0.6} $79.27$ & \cellcolor[rgb]{0.742,1,0.6} $79.52$ & \cellcolor[rgb]{0.689,1,0.6} $79.68$ & \cellcolor[rgb]{0.666,1,0.6} $79.84$ & \cellcolor[rgb]{0.638,1,0.6} $80.00$ & \cellcolor[rgb]{0.639,1,0.6} $80.11$ & \cellcolor[rgb]{0.615,1,0.6} $80.08$ & \cellcolor[rgb]{0.611,1,0.6} $80.16$ & \cellcolor[rgb]{0.609,1,0.6} $80.17$ \\ 
	Expert & $\times$ & $\times$ & $\times$ & $\times$ & $\times$ & $\times$ &  &  &  & \cellcolor[rgb]{0.6,1,0.6} $\textbf{24.17}$ & \cellcolor[rgb]{0.615,1,0.6} $41.80$ & \cellcolor[rgb]{0.63,1,0.6} $48.09$ & \cellcolor[rgb]{0.811,1,0.6} $60.54$ & \cellcolor[rgb]{0.666,1,0.6} $61.65$ & \cellcolor[rgb]{0.98,1,0.6} $74.79$ & \cellcolor[rgb]{0.844,1,0.6} $78.44$ & \cellcolor[rgb]{1,0.812,0.6} $90.10$ & \cellcolor[rgb]{1,0.6,0.6} $109.35$ & \cellcolor[rgb]{1,0.6,0.6} $133.67$ & \cellcolor[rgb]{1,0.6,0.6} $144.68$ & \cellcolor[rgb]{1,0.6,0.6} $147.96$ & \cellcolor[rgb]{1,0.6,0.6} $183.76$ & \cellcolor[rgb]{1,0.6,0.6} $208.04$ & \cellcolor[rgb]{1,0.6,0.6} $187.91$ \\ 
	Constr & $\times$ & $\times$ & $\times$ & $\times$ & $\times$ &  & $\times$ &  &  & \cellcolor[rgb]{0.6,1,0.6} $24.18$ & \cellcolor[rgb]{0.6,1,0.6} $\textbf{41.13}$ & \cellcolor[rgb]{0.6,1,0.6} $\textbf{46.96}$ & \cellcolor[rgb]{0.656,1,0.6} $56.00$ & \cellcolor[rgb]{0.6,1,0.6} $\textbf{60.14}$ & \cellcolor[rgb]{0.771,1,0.6} $70.16$ & \cellcolor[rgb]{0.876,1,0.6} $79.17$ & \cellcolor[rgb]{1,0.893,0.6} $88.18$ & \cellcolor[rgb]{1,0.762,0.6} $102.57$ & \cellcolor[rgb]{1,0.878,0.6} $113.28$ & \cellcolor[rgb]{1,0.766,0.6} $130.57$ & \cellcolor[rgb]{1,0.669,0.6} $141.83$ & \cellcolor[rgb]{1,0.816,0.6} $155.23$ & \cellcolor[rgb]{1,0.84,0.6} $169.16$ & \cellcolor[rgb]{1,0.669,0.6} $178.46$ \\ 
	ConstrDiff & $\times$ & $\times$ & $\times$ & $\times$ & $\times$ &  & $\times$ & $\times$ &  & \cellcolor[rgb]{0.638,1,0.6} $26.68$ & \cellcolor[rgb]{0.813,1,0.6} $50.68$ & \cellcolor[rgb]{0.842,1,0.6} $56.10$ & \cellcolor[rgb]{0.974,1,0.6} $65.31$ & \cellcolor[rgb]{1,0.856,0.6} $72.64$ & \cellcolor[rgb]{1,0.911,0.6} $77.20$ & \cellcolor[rgb]{0.943,1,0.6} $80.69$ & \cellcolor[rgb]{1,0.926,0.6} $87.39$ & \cellcolor[rgb]{0.97,1,0.6} $91.38$ & \cellcolor[rgb]{0.831,1,0.6} $91.98$ & \cellcolor[rgb]{0.793,1,0.6} $93.19$ & \cellcolor[rgb]{0.801,1,0.6} $94.56$ & \cellcolor[rgb]{0.752,1,0.6} $98.19$ & \cellcolor[rgb]{0.717,1,0.6} $97.42$ & \cellcolor[rgb]{0.764,1,0.6} $101.30$ \\ 
	Portfolio & $\times$ & $\times$ & $\times$ & $\times$ & $\times$ &  & $\times$ &  &  & \cellcolor[rgb]{0.601,1,0.6} $24.22$ & \cellcolor[rgb]{0.643,1,0.6} $43.05$ & \cellcolor[rgb]{0.736,1,0.6} $52.10$ & \cellcolor[rgb]{0.902,1,0.6} $63.20$ & \cellcolor[rgb]{1,0.93,0.6} $70.94$ & \cellcolor[rgb]{1,0.6,0.6} $84.13$ & \cellcolor[rgb]{1,0.6,0.6} $91.14$ & \cellcolor[rgb]{1,0.622,0.6} $94.61$ & \cellcolor[rgb]{1,0.727,0.6} $104.06$ & \cellcolor[rgb]{1,0.887,0.6} $112.67$ & \cellcolor[rgb]{1,0.839,0.6} $124.37$ & \cellcolor[rgb]{1,0.758,0.6} $133.84$ & \cellcolor[rgb]{1,0.854,0.6} $150.27$ & \cellcolor[rgb]{1,0.942,0.6} $152.68$ & \cellcolor[rgb]{1,0.706,0.6} $173.51$ \\ 
	Short-term & $\times$ & $\times$ & $\times$ & $\times$ & $\times$ &  & $\times$ &  & $\times$* & \cellcolor[rgb]{0.604,1,0.6} $24.41$ & \cellcolor[rgb]{0.657,1,0.6} $43.70$ & \cellcolor[rgb]{0.665,1,0.6} $49.43$ & \cellcolor[rgb]{0.698,1,0.6} $57.22$ & \cellcolor[rgb]{0.65,1,0.6} $61.28$ & \cellcolor[rgb]{0.784,1,0.6} $70.43$ & \cellcolor[rgb]{0.915,1,0.6} $80.05$ & \cellcolor[rgb]{1,0.745,0.6} $91.68$ & \cellcolor[rgb]{1,0.708,0.6} $104.85$ & \cellcolor[rgb]{1,0.84,0.6} $116.07$ & \cellcolor[rgb]{1,0.773,0.6} $129.96$ & \cellcolor[rgb]{1,0.686,0.6} $140.33$ & \cellcolor[rgb]{1,0.802,0.6} $157.15$ & \cellcolor[rgb]{1,0.797,0.6} $176.12$ & \cellcolor[rgb]{1,0.611,0.6} $186.45$ \\ 
	Current & $\times$ & $\times$ & $\times$ & $\times$ & $\times$ &  & $\times$ &  & $\times$ & \cellcolor[rgb]{0.604,1,0.6} $24.41$ & \cellcolor[rgb]{0.655,1,0.6} $43.59$ & \cellcolor[rgb]{0.622,1,0.6} $47.80$ & \cellcolor[rgb]{0.6,1,0.6} $\textbf{54.36}$ & \cellcolor[rgb]{0.61,1,0.6} $60.37$ & \cellcolor[rgb]{0.6,1,0.6} $\textbf{66.34}$ & \cellcolor[rgb]{0.6,1,0.6} $\textbf{72.86}$ & \cellcolor[rgb]{0.6,1,0.6} $\textbf{76.16}$ & \cellcolor[rgb]{0.711,1,0.6} $80.56$ & \cellcolor[rgb]{0.752,1,0.6} $86.16$ & \cellcolor[rgb]{0.757,1,0.6} $90.07$ & \cellcolor[rgb]{0.735,1,0.6} $88.68$ & \cellcolor[rgb]{0.721,1,0.6} $94.17$ & \cellcolor[rgb]{0.711,1,0.6} $96.37$ & \cellcolor[rgb]{0.735,1,0.6} $97.29$ \\ 
	WD+RL & $\times$ &  &  & $\times$ &  &  & $\times$ &  &  & \cellcolor[rgb]{1,0.764,0.6} $66.15$ & \cellcolor[rgb]{1,0.654,0.6} $74.49$ & \cellcolor[rgb]{1,0.63,0.6} $76.02$ & \cellcolor[rgb]{1,0.607,0.6} $77.60$ & \cellcolor[rgb]{1,0.6,0.6} $78.51$ & \cellcolor[rgb]{1,0.822,0.6} $79.20$ & \cellcolor[rgb]{0.823,1,0.6} $77.95$ & \cellcolor[rgb]{0.612,1,0.6} $76.44$ & \cellcolor[rgb]{0.6,1,0.6} $\textbf{75.94}$ & \cellcolor[rgb]{0.6,1,0.6} $\textbf{75.03}$ & \cellcolor[rgb]{0.6,1,0.6} $\textbf{76.77}$ & \cellcolor[rgb]{0.6,1,0.6} $\textbf{76.65}$ & \cellcolor[rgb]{0.604,1,0.6} $78.69$ & \cellcolor[rgb]{0.607,1,0.6} $79.62$ & \cellcolor[rgb]{0.607,1,0.6} $79.89$ \\ 
	WD+RL+C & $\times$ &  &  & $\times$ &  &  & $\times$ &  & $\times$ & \cellcolor[rgb]{1,0.735,0.6} $68.05$ & \cellcolor[rgb]{1,0.657,0.6} $74.39$ & \cellcolor[rgb]{1,0.665,0.6} $74.71$ & \cellcolor[rgb]{1,0.682,0.6} $75.39$ & \cellcolor[rgb]{1,0.702,0.6} $76.17$ & \cellcolor[rgb]{1,0.931,0.6} $76.76$ & \cellcolor[rgb]{0.794,1,0.6} $77.30$ & \cellcolor[rgb]{0.661,1,0.6} $77.60$ & \cellcolor[rgb]{0.639,1,0.6} $77.58$ & \cellcolor[rgb]{0.634,1,0.6} $77.52$ & \cellcolor[rgb]{0.611,1,0.6} $77.69$ & \cellcolor[rgb]{0.615,1,0.6} $77.97$ & \cellcolor[rgb]{0.6,1,0.6} $\textbf{78.15}$ & \cellcolor[rgb]{0.6,1,0.6} $\textbf{78.45}$ & \cellcolor[rgb]{0.6,1,0.6} $\textbf{78.88}$ \\ 
	WD+ARL+C & $\times$ &  & $\times$ & $\times$ &  &  & $\times$ &  & $\times$ & \cellcolor[rgb]{0.611,1,0.6} $24.91$ & \cellcolor[rgb]{0.671,1,0.6} $44.29$ & \cellcolor[rgb]{0.679,1,0.6} $49.94$ & \cellcolor[rgb]{0.7,1,0.6} $57.30$ & \cellcolor[rgb]{0.839,1,0.6} $65.62$ & \cellcolor[rgb]{0.789,1,0.6} $70.54$ & \cellcolor[rgb]{1,0.926,0.6} $83.70$ & \cellcolor[rgb]{1,0.674,0.6} $93.36$ & \cellcolor[rgb]{1,0.757,0.6} $102.80$ & \cellcolor[rgb]{1,0.844,0.6} $115.79$ & \cellcolor[rgb]{1,0.728,0.6} $133.85$ & \cellcolor[rgb]{1,0.627,0.6} $145.52$ & \cellcolor[rgb]{1,0.757,0.6} $163.03$ & \cellcolor[rgb]{1,0.776,0.6} $179.61$ & \cellcolor[rgb]{1,0.63,0.6} $183.76$ \\ 
	WD+F & $\times$ &  &  &  & $\times$ &  & $\times$ &  &  & \cellcolor[rgb]{0.816,1,0.6} $38.40$ & \cellcolor[rgb]{0.646,1,0.6} $43.20$ & \cellcolor[rgb]{0.613,1,0.6} $47.44$ & \cellcolor[rgb]{0.674,1,0.6} $56.52$ & \cellcolor[rgb]{0.656,1,0.6} $61.43$ & \cellcolor[rgb]{0.861,1,0.6} $72.16$ & \cellcolor[rgb]{1,0.946,0.6} $83.24$ & \cellcolor[rgb]{1,0.643,0.6} $94.11$ & \cellcolor[rgb]{1,0.759,0.6} $102.69$ & \cellcolor[rgb]{1,0.88,0.6} $113.17$ & \cellcolor[rgb]{1,0.845,0.6} $123.86$ & \cellcolor[rgb]{1,0.728,0.6} $136.53$ & \cellcolor[rgb]{1,0.855,0.6} $150.04$ & \cellcolor[rgb]{1,0.899,0.6} $159.67$ & \cellcolor[rgb]{1,0.798,0.6} $160.87$ \\ 
	WD+F+C & $\times$ &  &  &  & $\times$ &  & $\times$ &  & $\times$ & \cellcolor[rgb]{0.855,1,0.6} $41.01$ & \cellcolor[rgb]{0.712,1,0.6} $46.16$ & \cellcolor[rgb]{0.699,1,0.6} $50.70$ & \cellcolor[rgb]{0.797,1,0.6} $60.13$ & \cellcolor[rgb]{0.788,1,0.6} $64.45$ & \cellcolor[rgb]{0.957,1,0.6} $74.27$ & \cellcolor[rgb]{1,0.899,0.6} $84.31$ & \cellcolor[rgb]{1,0.6,0.6} $95.12$ & \cellcolor[rgb]{1,0.707,0.6} $104.88$ & \cellcolor[rgb]{1,0.826,0.6} $117.07$ & \cellcolor[rgb]{1,0.813,0.6} $126.63$ & \cellcolor[rgb]{1,0.716,0.6} $137.58$ & \cellcolor[rgb]{1,0.87,0.6} $148.14$ & \cellcolor[rgb]{1,0.91,0.6} $157.85$ & \cellcolor[rgb]{1,0.803,0.6} $160.27$ \\ 
	 \hline
\end{tabular}}
\tiny{*For the short-term model the method as used in the \textbf{current model} is only applied to RES and Load. The fuels are regressed on at their corresponding lags, not at the current value.}
\caption{RMSE (EUR/MWh) for the evaluation period April/2018-April/2024 for all considered models and forecasting horizons over all 24 hours. Abbreviation of the models are defined in the first column and correspond to the marked variables and methods.} 
\label{tab:rmse_models}
\end{table}

\begin{figure}[htbp]
    \centering
	\scalebox{1}{
		\includegraphics[width=1\textwidth]{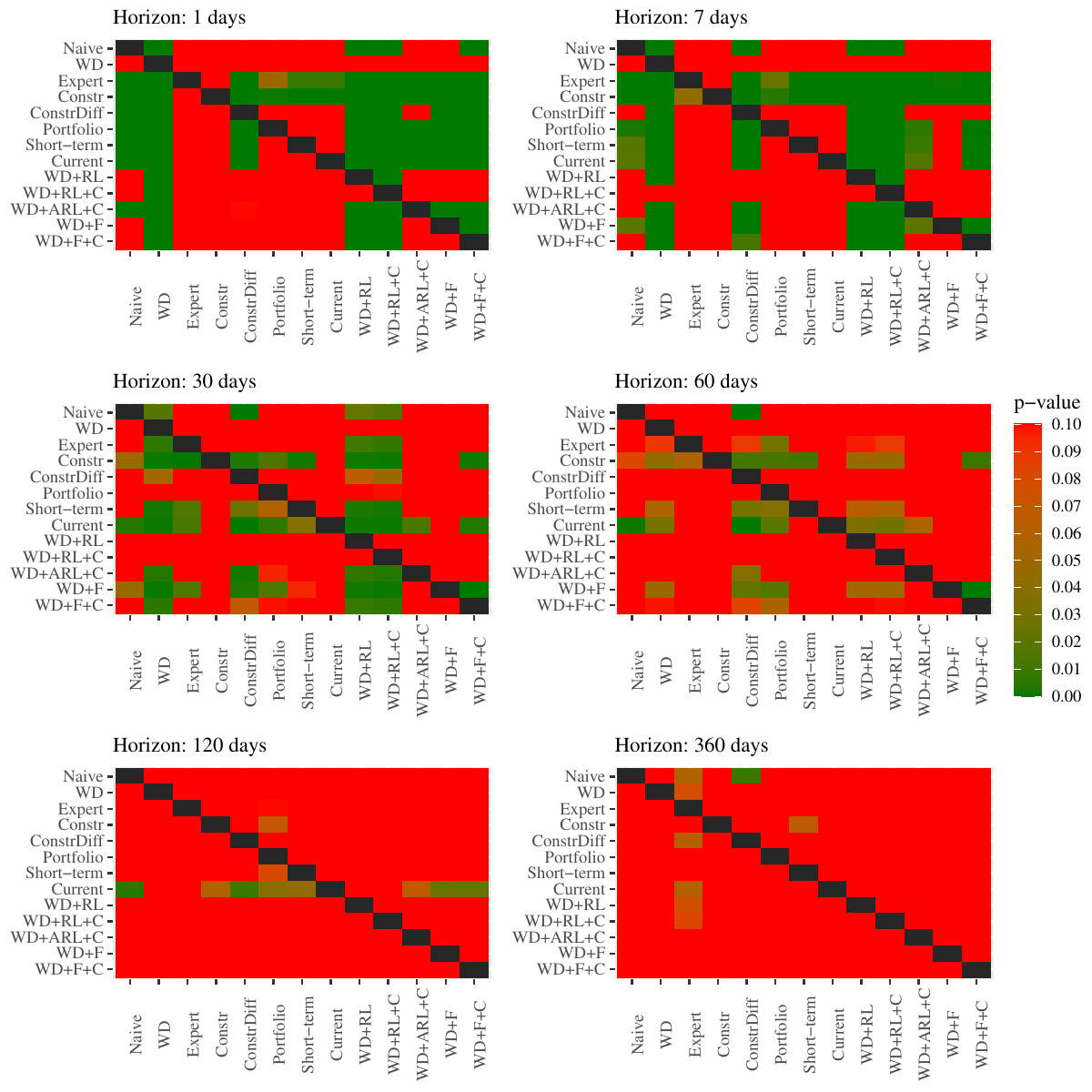}}	
	\caption{DM test results over all considered models and selected horizons.}
    \label{fig:dmtest}
\end{figure}

We observe that while the \textbf{expert model} performs best for the day-ahead setting, for which it was originally designed, other models outperform it for higher horizons. To disentangle the drivers of what makes some models better or worse for differen horizons we looked at different combinations, referred to in Table \ref{tab:rmse_models} as the \textbf{WD+} models. 
There are important conclusions that we can draw from the results in Table \ref{tab:rmse_models}. The corresponding Diebold-Mariano (DM) tests are shown in Figure \ref{fig:dmtest}. We summarize these findings in Figure X as a decision guideline on what regressors and methods to use depending on the forecasting horizon. 
For ease of description we split the considered forecasting horizons into three groups: short horizons up to 30 days or 1 month ahead, mid horizons for 1 month up to 180 days or 6 months ahead, and long horizons for 6 months up to 1 year ahead. 

First, for \emph{short horizons} fuels and autoregressive effects are highly important. Models that do not include either perform consistently worse such as \textbf{WD}, \textbf{WD+RL} and \textbf{WD+RL+C}. On the other hand, models \textbf{WD+F} and \textbf{WD+F+C} which only use fuels also do not perform well, but much better in comparison to those that do not. It appears that a combination of RES, load and fuels are important for short horizons. Using fundamental constraints leads to significant improvements compared to the unconstrained model starting from 2 weeks onward. 

Second, for \emph{mid horizons} the \textbf{current model} seems to outperform the rest on average. This is an important inflection point as this is the period when the models that do not include fuels start to outperform models that do. This can be interpreted as the critical horizon after which the unit root behavior of the commodities becomes dominant making them exhibit spurious effects as they become uninformative for future power prices. The \textbf{current model} mitigates this effect by forcing current relationships between the commodities and the power price and projecting them into the future. In essence, at this horizon modelling the power price structure becomes more important than forecasting it. For this horizon all regressors seem to be relevant, but the modelling method is different.

Third, for \emph{long horizons} models without fuels and autoregressive terms perform best on average. This is in contrast to current literature, where all models include autoregressive terms (see Tables \ref{tab:papers1} and \ref{tab:papers2}). In fact, all models that simply include fuels as lags perform worse than the \textbf{WD} model. This is a clear indication that the commodities are not informative. The \textbf{current model} performs slightly better than the \textbf{WD} model for two main reasons: the large RES and load coefficients act as model stabilizers (see Figure \ref{fig:expert_current_coefs}) and the fuels coefficients are regressed on actual values thus avoiding spurious estimations. The models \textbf{WD+RL} and \textbf{WD+RL+C} which include only weekday dummies and forecasts for renewables and load. It is interesting to see that simply adding an autoregressive effect as in \textbf{WD+ARL+C} makes the model much worse and has very similar effects to adding fuels (see \textbf{WD+F} and \textbf{WD+F+C}). 

\begin{figure}[htbp]
    \centering
	\scalebox{0.75}{
		\includegraphics[width=0.63\textwidth]{expert_win3_coefs_legend.pdf}}
	\scalebox{0.26}{
	\includegraphics[width=0.63\textwidth]{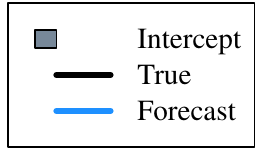}}
	\linebreak
    \subfloat{
        \includegraphics[width=0.49\textwidth]{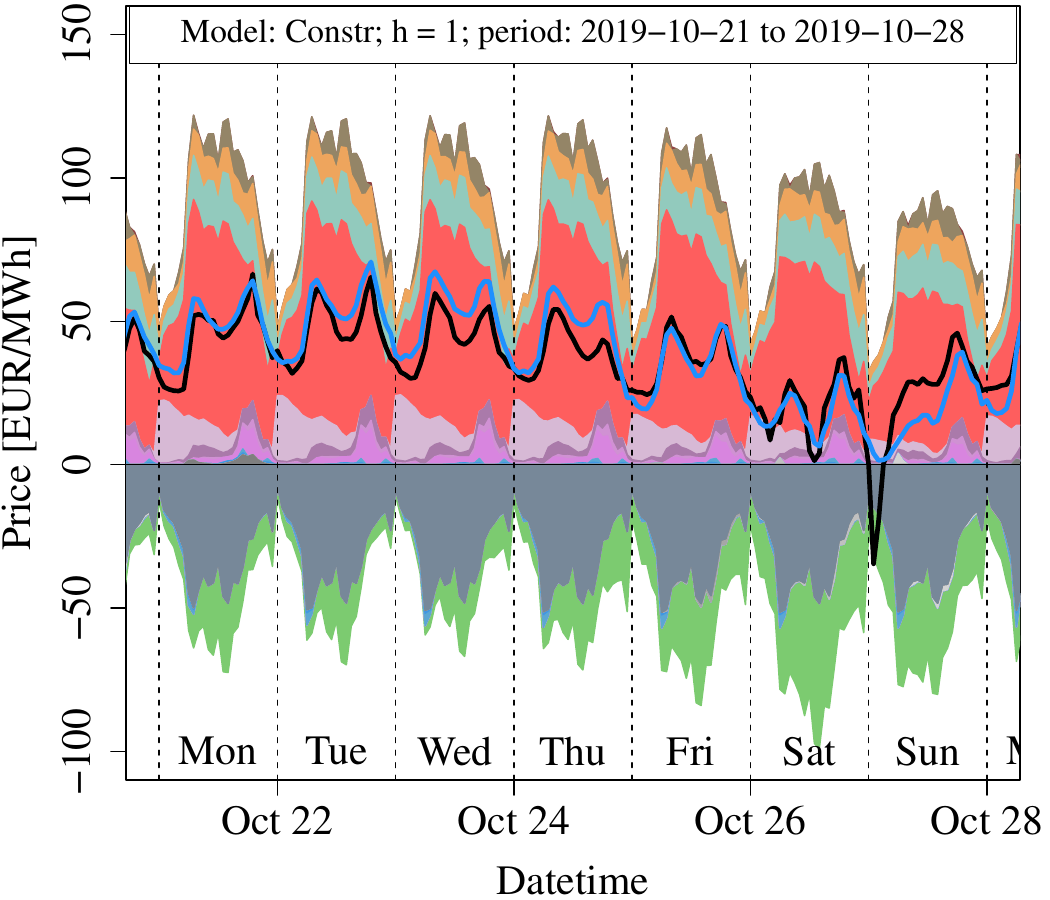}}
    \subfloat{
        \includegraphics[width=0.49\textwidth]{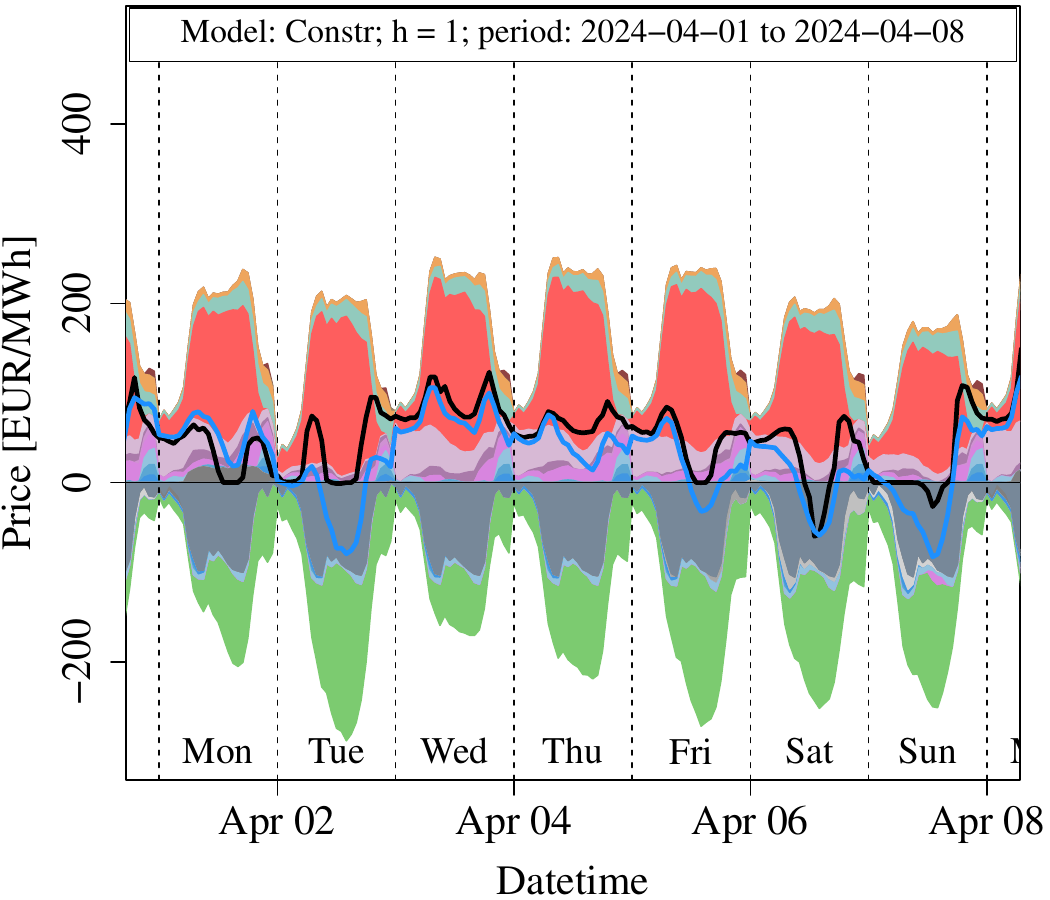}}
	\linebreak
	\subfloat{
		\includegraphics[width=0.49\textwidth]{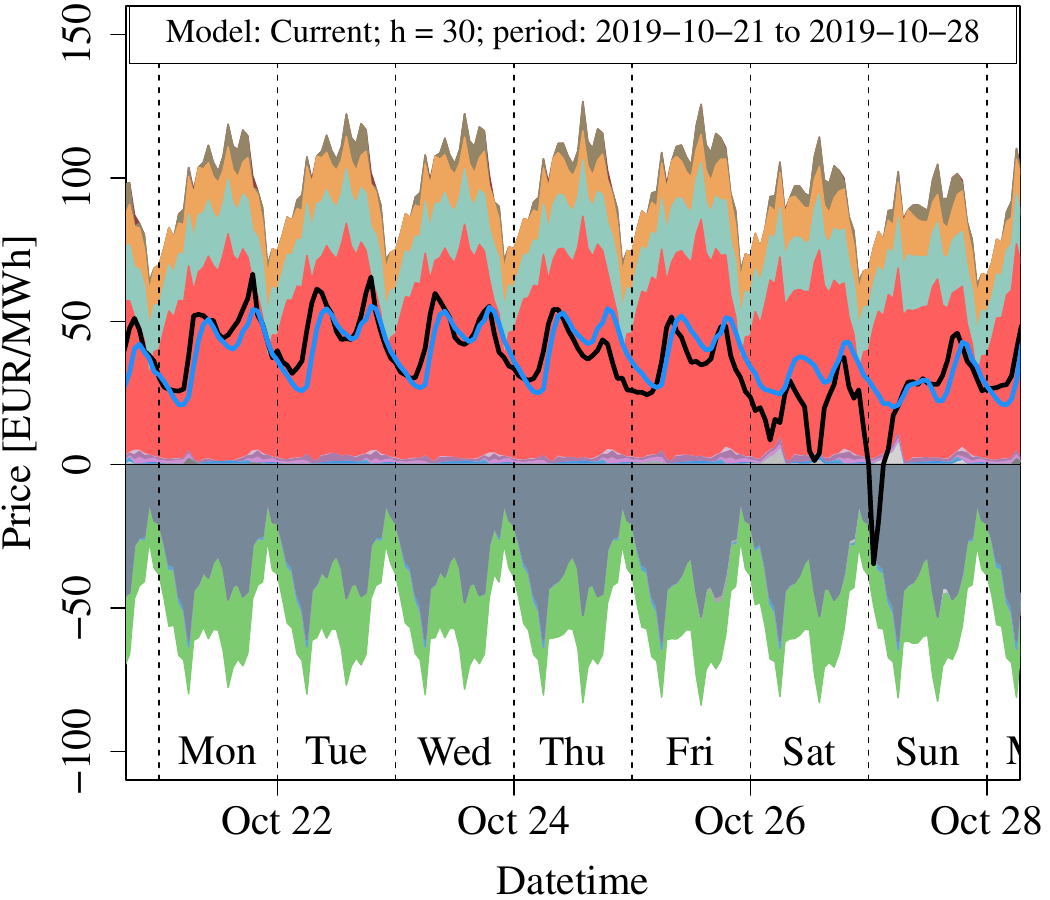}}
	\subfloat{
		\includegraphics[width=0.49\textwidth]{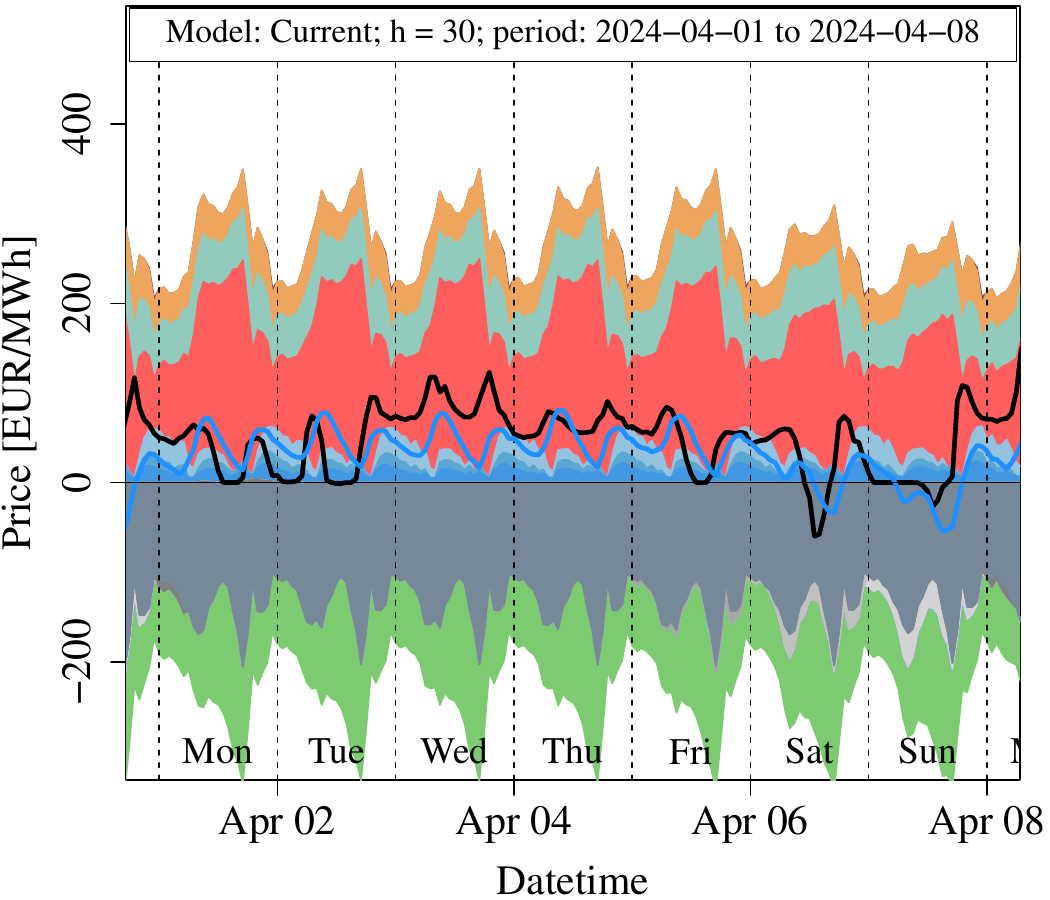}}		
	\linebreak
	\subfloat{
		\includegraphics[width=0.49\textwidth]{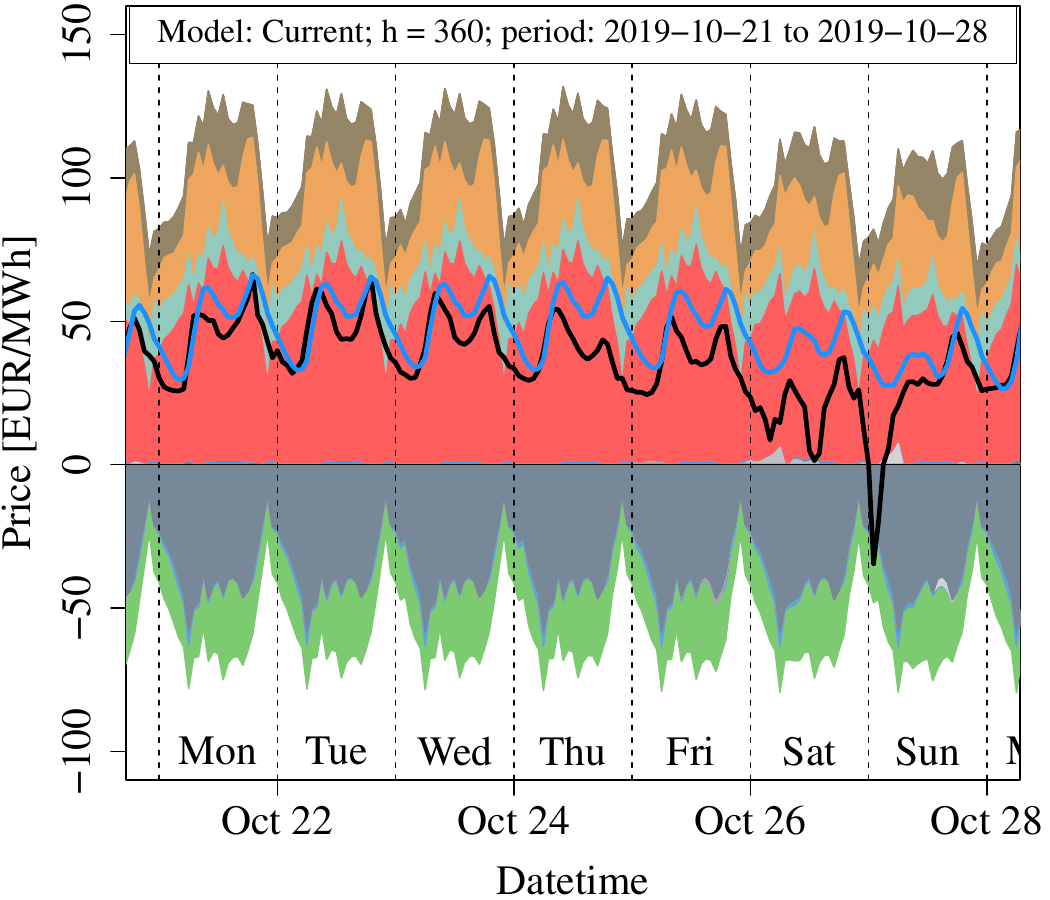}}
	\subfloat{
		\includegraphics[width=0.49\textwidth]{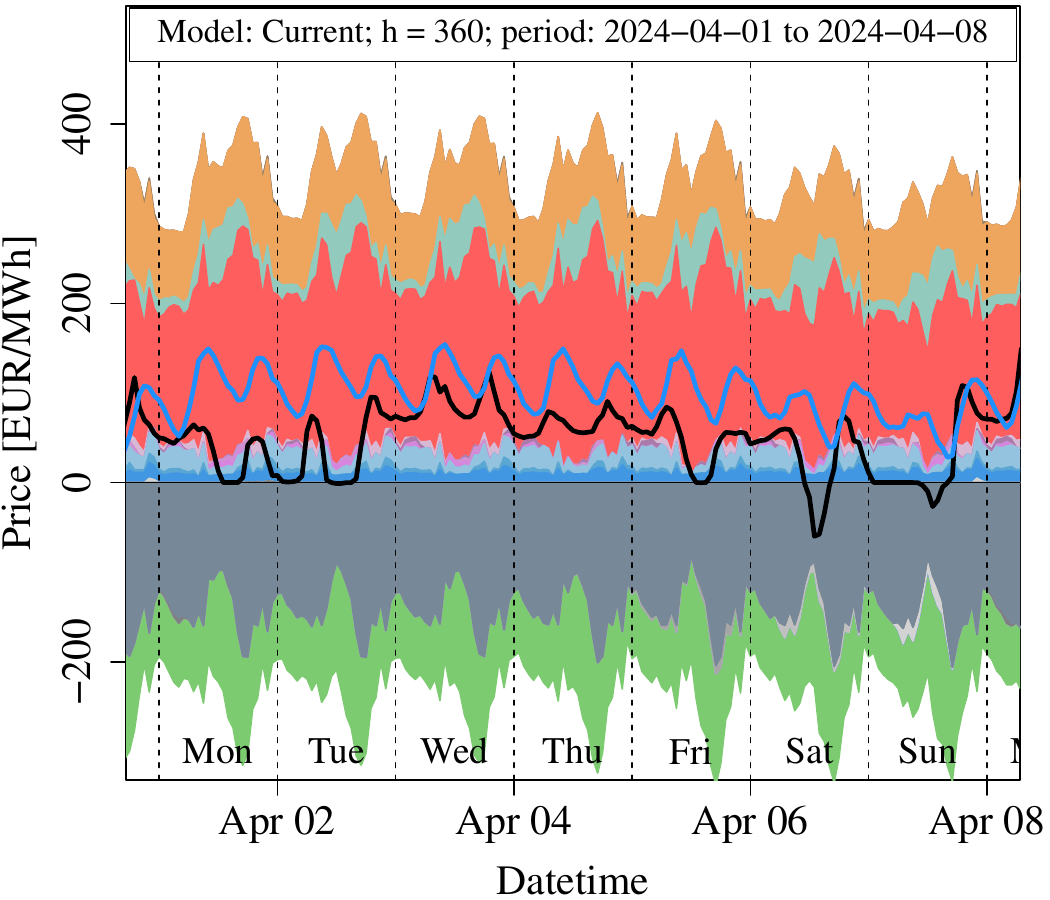}}		
	\caption{Price forecasts and their components for selected models and periods.}
    \label{fig:components}
\end{figure}

In addition to this, we can also inspect the components of the forecasts. Figure \ref{fig:components} displays the contributions of the regressors to the forecasts for two selected weeks. The sum of the components at each time point equals the forecast. As expected, in the short term we are able to capture negative price shocks due to strong RES infeed while in the longer term we predict average seasonal patterns of the power price.

\begin{figure}[htb]
	\centering
	\scalebox{0.7}{	
	\begin{tikzpicture}[
		neutralnode/.style={rectangle, draw=black, very thick, minimum width = 40mm,  minimum height = 10mm},
		varnode/.style={rectangle, draw=black, very thick, minimum width = 40mm,  minimum height = 35mm, align = center, anchor = base north, text depth = 0cm, valign = top},
		varnodesub/.style={rectangle, node distance = 1mm, minimum width = 40mm,  draw=black, very thick, align = center, anchor = base north, text depth = 0cm, valign = top},
		metnode/.style={rectangle, draw=black, very thick, minimum width = 40mm,  minimum height = 35mm, align = center, anchor = north, text depth = 0cm, valign = top},
		metnodesub/.style={rectangle, node distance = 1mm, draw=black, very thick, minimum width = 40mm, minimum height = 15mm, align = center, anchor = north, text depth = 0cm, valign = top},
		legendnode/.style={rectangle, draw=black, very thick, minimum width = 40mm, align = center, node distance = 5mm}
		]

		\node[neutralnode, minimum height = 7mm, minimum width = 140mm]      (horizon)                              {\bf Forecasting horizon};
		\node[neutralnode]      (horizon_m)      [below=of horizon]    {\scalebox{1}{ 
			\renewcommand{\arraystretch}{0.5}
			\begin{tabular}{c}\textbf{Mid} \\ (1 month - 6 months)\end{tabular}}};
		\node[neutralnode]      (horizon_s)      [left=of horizon_m]   {\renewcommand{\arraystretch}{0.5} \begin{tabular}{c}\textbf{Short} \\ (1 day - 1 month)\end{tabular}};
		\node[neutralnode]      (horizon_l)      [right=of horizon_m]  {\renewcommand{\arraystretch}{0.5} \begin{tabular}{c}\textbf{Long} \\ (6 months - 1 year)\end{tabular}};
		
		\node[varnodesub, node distance = 12mm, diagonal fill={yellowsignif}{greenpos}]     (var_s1)      [below= of horizon_s]  {				
			\scalebox{0.9}{
			\renewcommand{\arraystretch}{0.5}
			\begin{tabular}{p{3cm}}
				\multicolumn{1}{c}{\bf ($\boldsymbol{++}$)} \\
				Autoreg  \\
				RES \& Load \\
				Fuels \\
			\end{tabular}
			} \\	
		};
		\node[varnodesub, diagonal fill={blueinsignif}{greenpos}]      (var_s2)      [below=of var_s1]  {
			\scalebox{0.9}{
			\renewcommand{\arraystretch}{0.5}	
			\begin{tabular}{p{3cm}}
				\multicolumn{1}{c}{\bf ($\boldsymbol{+}$)} \\
				Week Dummies  \\
				Annual Seasons \\
			\end{tabular}} \\
		}; 
		\node [varnode, draw, fit={(var_s1) (var_s2)}] (var_s) [below=of horizon_s] {};

		\node[metnodesub, node distance = 12mm, diagonal fill={yellowsignif}{greenpos}]      (method_s1)      [below=of var_s2]  {
			\scalebox{0.9}{
			\renewcommand{\arraystretch}{0.5}
			\begin{tabular}{p{3cm}}
				\multicolumn{1}{c}{\bf ($\boldsymbol{+}$)} \\
				Constrained \\
			\end{tabular}} \\
		};
		\node[metnodesub, diagonal fill={yellowsignif}{redneg}]      (method_s2)      [below=of method_s1]  {
			\scalebox{0.9}{
				\renewcommand{\arraystretch}{0.5}
				\begin{tabular}{p{3cm}}
					\multicolumn{1}{c}{\bf ($\boldsymbol{-}$)} \\
					Unconstrained  \\
					Differencing \\
					Current \\
				\end{tabular}} \\
			};

		\node [metnode, draw, fit={(method_s1) (method_s2)}] (method_s) [below=of var_s] {};

		\node[varnodesub, node distance = 12mm, diagonal fill={yellowsignif}{greenpos}]     (var_m1)      [below= of horizon_m]  {				
			\scalebox{0.9}{
			\renewcommand{\arraystretch}{0.5}
			\begin{tabular}{p{3cm}}
				\multicolumn{1}{c}{\bf ($\boldsymbol{++}$)} \\
				RES \& Load \\
				Fuels \\
			\end{tabular}} \\
		};
		\node[varnodesub, diagonal fill={blueinsignif}{greenpos}]     (var_m2)      [below= of var_m1]  {				
			\scalebox{0.9}{
			\renewcommand{\arraystretch}{0.5}
			\begin{tabular}{p{3cm}}
				\multicolumn{1}{c}{\bf ($\boldsymbol{+}$)} \\
				Week Dummies  \\
				Annual Seasons \\
				Autoreg \\
			\end{tabular}} \\
		};	
		\node [varnode, draw, fit={(var_m1) (var_m2)}] (var_m) [below=of horizon_m] {};
	
		\node[metnodesub, node distance = 12mm, diagonal fill={yellowsignif}{greenpos}]      (method_m1)      [below=of var_m2]  {
			\scalebox{0.9}{
				\renewcommand{\arraystretch}{0.5}
				\centering
				\begin{tabular}{p{3cm}}
					\multicolumn{1}{c}{\bf ($\boldsymbol{+}$)} \\
					Current  \\
				\end{tabular}} \\
		};
		\node[metnodesub, diagonal fill={yellowsignif}{redneg}]      (method_m2)      [below=of method_m1]  {
			\scalebox{0.9}{
				\renewcommand{\arraystretch}{0.5}
			\begin{tabular}{p{3cm}}
				\multicolumn{1}{c}{\bf ($\boldsymbol{-}$)} \\
				Unconstrained  \\
				Differencing \\
			\end{tabular}} 
		};

		\node [metnode, draw, fit={(method_m1) (method_m2)}] (method_m) [below=of var_m] {};

		\node[varnodesub, diagonal fill={blueinsignif}{greenpos}, node distance = 12mm]     (var_l2)      [below= of horizon_l]  {				
			\scalebox{0.9}{
				\renewcommand{\arraystretch}{0.5}
				\begin{tabular}{p{3cm}}
					\multicolumn{1}{c}{\bf ($\boldsymbol{+}$)} \\
					RES \& Load \\
					Week Dummies  \\
					Annual Seasons \\
				\end{tabular}} \\
		};	
		\node[varnodesub, diagonal fill={blueinsignif}{redneg}]     (var_l3)      [below= of var_l2]  {				
			\scalebox{0.9}{
				\renewcommand{\arraystretch}{0.5}
				\begin{tabular}{p{3cm}}
					\multicolumn{1}{c}{\bf ($\boldsymbol{-}$)} \\
					Fuels \\
					Autoreg \\
				\end{tabular}} \\
		};	
		\node [varnode, draw, fit={(var_l2) (var_l3)}] (var_l) [below=of horizon_l] {};
			
		\node[metnodesub, node distance = 12mm, diagonal fill={blueinsignif}{greenpos}]      (method_l1)      [below=of var_l3]  {
			\scalebox{0.9}{
				\renewcommand{\arraystretch}{0.5}
				\begin{tabular}{p{3cm}}
					\multicolumn{1}{c}{\bf ($\boldsymbol{+}$)} \\
					Current \\
					Constrained  \\
				\end{tabular}} \\
		};
		\node[metnodesub, diagonal fill={blueinsignif}{redneg}]      (method_l2)      [below=of method_l1]  {
			\scalebox{0.9}{
			\renewcommand{\arraystretch}{0.5}
			\begin{tabular}{p{3cm}}
				\multicolumn{1}{c}{\bf ($\boldsymbol{-}$)} \\
				Unconstrained  \\
				Differentiation \\
			\end{tabular}} \\
		};

		\node [metnode, draw, fit={(method_l1) (method_l2)}] (method_l) [below=of var_l] {};

	\node[neutralnode, minimum width = 5mm, minimum height = 35mm]      (var)      [left=of var_s]   {\rotatebox{90}{\bf Variables}};
	\node[neutralnode, minimum width = 5mm, minimum height = 35mm]      (method)      [left=of method_s]   {\rotatebox{90}{\bf Methods}};
	\node[legendnode, fill = greenpos, xshift = 30mm]   (pos)   [below=of method_s] {Positive effect on accuracy};
	\node[legendnode, fill = redneg]   (neg)   [right=of pos] {Negative effect on accuracy};
	\node[legendnode, fill = yellowsignif, node distance = 2mm]   (signif)   [below=of pos] {Significant};
	\node[legendnode, fill = blueinsignif, node distance = 2mm]   (insignif)   [below=of neg] {Not significant};
		\draw[->] (horizon.south) -- (horizon_m.north);
		\draw[->] (horizon.south) -- (horizon_s.north);
		\draw[->] (horizon.south) -- (horizon_l.north);
		\draw[->] (horizon_s.south) -- (var_s.north);
		\draw[->] (horizon_m.south) -- (var_m.north);
		\draw[->] (horizon_l.south) -- (var_l.north);
		\draw[->] (var_s.south) -- (method_s.north);
		\draw[->] (var_m.south) -- (method_m.north);
		\draw[->] (var_l.south) -- (method_l.north);
	\end{tikzpicture}}
	\tiny{\\($\boldsymbol{++}$) = strong impact on accuracy, ($\boldsymbol{+}$) = positive impact on accuracy, ($\boldsymbol{-}$) = negative impact on accuracy}
	\caption{Guideline on choosing the inputs, methods and models for a forecasting horizon.}
\end{figure}
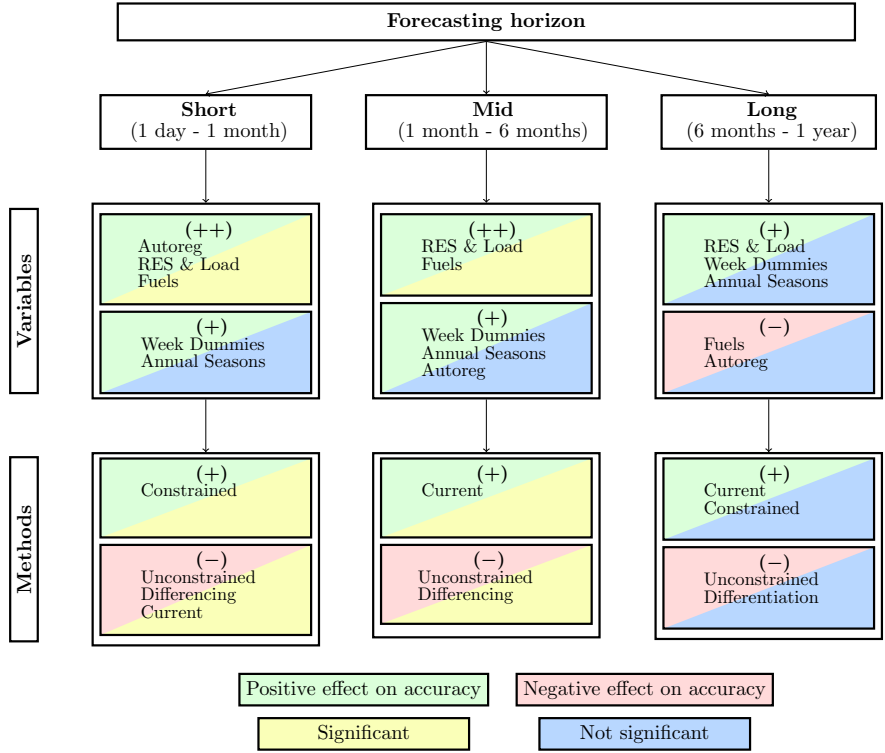

\section{Summary and conclusion}
\label{sec:conclusion}

This study showcases the challenges associated with forecasting beyond short-term horizons in the power market using data-driven econometric models. First, it shows that fundamentally-derived bounds for the coefficients are needed for ensuring model stability and alignment with energy economic principles. It also illustrate how econometric models relate to fundamental models such as the merit order model and that some coefficients correspond to certain technical power plant characteristics. Second, it explains how regressors that are meteorologically limited in their forecasting accuracy to less than a few days ahead, such as renewables, load and temperature, can still be used in a long-term forecasting framework. This is done by generating seasonal renewables infeed and load forecasts which take into account buildout of renewable capacity and other macroeconomic trends. Third, it demonstrates that power prices exhibit a unit root behavior in the long run which is driven by the commodities prices $\textrm{CO}_2$ gas, coal and oil. It lays open the issue of spurious effects becoming highly relevant with increasing forecasting horizons especially since the onset of the energy crisis in 2021. Lastly, it also discusses the idea of portfolio effects where the power price is influenced by a mix of different futures of the same commodity.

To mitigate these issues we propose the \textbf{current model} together with fundamental constraints which is a highly interpretable model robust to spurious effects during estimation. This model can be used for forecasting power prices starting one month ahead. We also provide a guideline as to which explanatory variables are important and which methods perform the best for different forecasting horizons.

Further research could focus on extending the model to include probabilistic forecasts. It would require simulating future values of the regressors, particularly commodity prices, while accounting for their correlation or cointegration relationships. Finding a distribution that would accurately represent the power prices would be challening as it would most likely be multi-modal, with one mode around zero and a right skeweness that drastically changes over the years.
Another aspect would be incorporating more complex relationships between the regressors and power prices, such as interactions between fuels. This could account for effects like fuel switches, where two or more fuels change order in the supply stack model. Furthermore, inspection of extending standard cointegration theory and vector error correction to include a 24-hour price model would be a nother area of extension. This could allow better capturing of the relationship between power price and fuel commodities, and potentially offer deeper insight into unit root behavior for longer forecasting horizons.

\clearpage
\section*{Declaration of Generative AI and AI-assisted technologies}
During the preparation of this work the authors used GitHub Copilot and OpenAI ChatGPT-3.5 in order to generate rephrasing suggestions to improve the language and readability of this paper. After using this tool/service, the authors reviewed and edited the content as needed and take full responsibility for the content of the published article.

\section*{Acknowledgements}
Funding: This work was supported by the German Research Foundation (DFG, Deutsche Forschungsgemeinschaft) project number 505565850 to P.G. and F.Z.

\section*{Data Availability}
The raw data required to reproduce the findings in this work include freely available resources such as day-ahead power prices, actual values and day-ahead forecasts for load, solar, onshore wind, and offshore wind infeed. This data can be downloaded from the ENTSOE Transparency Platform (https://transparency.entsoe.eu/).
Data not freely available include daily closing prices for futures of gas, coal, oil, and European Emission Allowances (EUA), accessible via the information platform Refinitiv Eikon at a cost (https://eikon.refinitiv.com/).

\clearpage
\section{Appendix}
\label{sec:appendix}

\subsection{Scopus Queries}
\label{sec:appendix_scopus}
\begin{itemize}
	\item Scopus query for papers covering mid-term forecasting: 
\end{itemize}
(TITLE((((("electric*" OR "energy market" OR "power price" OR "power market" OR "power system" OR pool OR"market clearing" OR "energy clearing") AND (price OR prices OR pricing)) OR lmp OR "locational marginal price") AND (forecast OR forecasts OR forecasting OR prediction OR predicting OR predictability OR "predictive densit*")) OR ("price forecasting" AND "smart grid*")) OR TITLE-ABS("electricity price forecasting" OR "forecasting electricity price" OR "day-ahead price forecasting" OR "day-ahead mar* price forecasting" OR (gefcom2014 AND price) OR (("electricity market" OR "electric energy market") AND "price forecasting") OR ("electricity price" AND ("prediction interval" OR "interval forecast" OR "density forecast" OR "probabilistic forecast"))) AND NOT TITLE ("unit commitment")) AND ( TITLE("mid-term") OR TITLE("mid term") OR TITLE("medium-term") OR TITLE("medium term") OR TITLE("long-term") OR TITLE("long term") OR TITLE-ABS-KEY("mid-term electric*") OR TITLE-ABS-KEY("mid term electric*") OR TITLE-ABS-KEY("medium-term electric*") OR TITLE-ABS-KEY("medium term electric*") OR TITLE-ABS-KEY("long-term electric*") OR TITLE-ABS-KEY("long term electric*") OR TITLE-ABS-KEY("mid-term price*") OR TITLE-ABS-KEY("mid term price*") OR TITLE-ABS-KEY("medium- term price*") OR TITLE-ABS-KEY("medium term price*") OR TITLE-ABS-KEY("weeks-ahead") OR TITLE-ABS-KEY("weeks ahead") OR TITLE-ABS-KEY("month-ahead") OR TITLE-ABS-KEY("month ahead") OR TITLE-ABS-KEY("mid term* horizon") OR TITLE-ABS-KEY("mid-term* horizon") OR TITLE-ABS-KEY("medium term* horizon") OR TITLE-ABS-KEY("medium-term* horizon"))

\begin{itemize}
	\item Scopus query for papers covering long-term forecasting:
\end{itemize}
 (TITLE((((("electric*" OR "energy market" OR "power price" OR "power market" OR "power system" OR pool OR"market clearing" OR "energy clearing") AND (price OR prices OR pricing)) OR lmp OR "locational marginal price") AND (forecast OR forecasts OR forecasting OR prediction OR predicting OR predictability OR "predictive densit*")) OR ("price forecasting" AND "smart grid*")) OR TITLE-ABS("electricity price forecasting" OR "forecasting electricity price" OR "day-ahead price forecasting" OR "day-ahead mar* price forecasting" OR (gefcom2014 AND price) OR (("electricity market" OR "electric energy market") AND "price forecasting") OR ("electricity price" AND ("prediction interval" OR "interval forecast" OR "density forecast" OR "probabilistic forecast"))) AND NOT TITLE ("unit commitment")) AND ( TITLE("long-term") OR TITLE("long term") OR TITLE-ABS-KEY("long-term electric*") OR TITLE-ABS-KEY("long term electric*") OR TITLE-ABS-KEY("long-term price*") OR TITLE-ABS-KEY("long term price*") OR TITLE-ABS-KEY("year-ahead") OR TITLE-ABS-KEY("year ahead") OR TITLE-ABS-KEY("long term* horizon") OR TITLE-ABS-KEY("long-term* horizon") ) 

\subsection{Derivation of fundamental bounds for coefficients}
\label{sec:appendix_bounds}
The coefficients $\beta$ for $\textrm{CO}_2$, Gas, Coal and Oil from Table \ref{tab:fundamentals} are derived in such a way as to get the units using power plant efficiencies, CO2 intensity factors, and fuel conversion rates as follows:
\begin{align*}
	\beta_{\textrm{CO}_2}  \left[ \frac{tCO2}{MWh_{el}} \right]     & \le \max \left(\frac{\varepsilon}{\eta} \right) = \frac{0.4}{0.3} = 1.33 \\
	\beta_{gas}  \left[ \frac{MWh_{th}}{MWh_{el}} \right] & \le \max \left( \frac{1}{\eta_{gas}} \right) = \frac{1}{0.25} = 4 \\
	\beta_{coal} \left[ \frac{t}{MWh_{el}} \right]        & \le \max \left(\frac{1}{\eta_{coal}} \right) \frac{1}{8.141} \left[ \frac{t}{MWh_{th}} \right]  = \frac{0.123}{0.35} = 0.35 \\
	\beta_{oil}  \left[ \frac{bbl}{MWh_{el}} \right]      & \le \max \left(\frac{1}{\eta_{oil}} \right) \frac{1000}{1700} \left[ \frac{bbl}{MWh_{th}} \right]  = \frac{0.588}{0.24} = 2.45 \\
\end{align*}

\subsection{Model formulas for renewables and load forecasts}
The hourly forecasts for renewable infeed and load, and for horizons ranging from 1 day to 365 days ahead as shown in Figure \ref{fig:res_load_forecast} were generated using Generalized Additive Models (GAMs) as implemented in the mgcv R package \citep{mgcv4}. The regressors consist of only deterministic terms including a linear trend component, daily, weekly and annual seasonalities as well as interactions. The corresponding models are:

\begin{align}
	\text{Load}_{t} & \sim t + ps(\text{HoD}_{t},24) + ps(\text{DoW}_{t},7) + cp(\text{SoY}_{t},12)  \nonumber \\ 
	& + ti(ps(\text{HoD}_{t},12),cp(\text{SoY}_{t},6)) + ti(ps(\text{HoD},12),ps(\text{DoW}_{t},6))  \nonumber \\
	\text{RES}_{t} & \sim t + ps(\text{HoD}_{t},24) + cp(\text{SoY}_{t},12)  + ti(ps(\text{HoD}_{t},12),cp(\text{SoY}_{t},6)) \nonumber 
\end{align}
where $\text{HoD}$ stands for Hour of the Day, $\text{DoW}$ for Day of the Week, and $\text{SoY}$ for Season of the Year which refers to an index covering the meteorological year of $365.24$ days on an hourly basis, i.e. $\text{SoY} \in \{0, 1, \dots, 8765, 0.24, 1.24, \dots,\}$. $h \in \{0, \dots, 23 \}$ is the hour of the day and $t$ represents the hourly time index. $ps(\cdot,k)$ represents a fit using penalized cubic splines using $k$ knots, $cp(\cdot,k)$ represents a periodic cubic spline with $k$ knots and $ti(\cdot, \cdot)$ represents a tensor interaction, which models the interaction effects between two variables. The model is fitted using an expanding window with actual data from 2015 to 2024. At least 3 years of data is used for every model. Hence, the first window is from 2015 to 2018, to produce forecasts for 2018.

\subsection{Plots and tables}

\begin{table}[hbtp]
	\centering
	\scalebox{0.7}{
	\setlength\extrarowheight{-3pt}		
\begin{tabular}{|c|ccccc|ccccc|ccccc|}
		\hline
	   \textbf{Horizon (days)}  & 1 & 30 & 120 & 180 & 360 & 1 & 30 & 120 & 180 & 360 & 1 & 30 & 120 & 180 & 360 \\ \hline
	   \textbf{Horizon (months)} & 0 & 1 & 4 & 6 & 12 & 0 & 1 & 4 & 6 & 12 & 0 & 1 & 4 & 6 & 12 \\\hline
	Lag (months) &\multicolumn{5}{|c|}{\text{Delivery D0 maturity (months)}} & \multicolumn{5}{|c|}{\text{Delivery D-1 maturity (months)}} &\multicolumn{5}{|c|}{\text{Delivery D+1 maturity (months)}} \\ 
	  \hline
	   0  &    &     &    &    &    &    &    &    &    &    & 1  &    &    &    &  \\ 
	   1  & 1  & 1   &    &    &    &    &    &    &    &    & 2  & 2  &    &    &  \\ 
	   2  & 2  & 2   &    &    &    & 1  & 1  &    &    &    & 3  & 3  &    &    &  \\  
	   3  & 3  & 3   &    &    &    & 2  & 2  &    &    &    & 4  & 4  &    &    &  \\  
	   4  & 4  & 4   & 4  &    &    & 3  & 3  & 3  &    &    & 5  & 5  & 5  &    &  \\
	   5  & 5  & 5   & 5  &    &    & 4  & 4  & 4  &    &    & 6  & 6  & 6  &    &  \\   
	   6  & 6  & 6   & 6  & 6  &    & 5  & 5  & 5  & 5  &    & 7  & 7  & 7  & 7  &  \\   
	   7  & 7  & 7   & 7  & 7  &    & 6  & 6  & 6  & 6  &    & 8  & 8  & 8  & 8  &  \\  
	   8  & 8  & 8   & 8  & 8  &    & 7  & 7  & 7  & 7  &    & 9  & 9  & 9  & 9  &  \\ 
	   9  & 9  & 9   & 9  & 9  &    & 8  & 8  & 8  & 8  &    & 10 & 10 & 10 & 10 &  \\  
	   10 & 10 & 10  & 10 & 10 &    & 9  & 9  & 9  & 9  &    & 11 & 11 & 11 & 11 &  \\ 
	   11 & 11 & 11  & 11 & 11 &    & 10 & 10 & 10 & 10 &    & 12 & 12 & 12 & 12 &  \\  
	   12 & 12 & 12  & 12 & 12 & 12 & 11 & 11 & 11 & 11 & 11 &    &    &    &    &  \\ 
	  \hline
	\end{tabular}}
	\caption{Futures and their lag structure used as regressors for the portfolio model. They include the actual delivery period $D0$ in months, which corresponds to the forecast horizon, trailing delivery period $D-1$ and leading delivery period $D+1$.} 
	\label{tab:portfolio_lag_struct}
\end{table}

\begin{table}[hbtp]
	\centering
	\scalebox{0.5}{
	\setlength\extrarowheight{-5pt}	
	\begin{tabular}{|>{\centering\arraybackslash}p{2cm}>{\centering\arraybackslash}p{2cm}>{\centering\arraybackslash}p{1cm}|p{0.7cm}p{0.7cm}p{0.7cm}p{0.7cm}|p{0.7cm}p{0.7cm}p{0.7cm}p{0.7cm}|p{0.7cm}p{0.7cm}p{0.7cm}p{0.7cm}|p{0.7cm}p{0.7cm}p{0.7cm}p{0.7cm}|p{0.7cm}p{0.7cm}p{0.7cm}p{0.7cm}|}
	\hline
	\multicolumn{3}{|c|}{\textbf{Horizon (days)}}  & \multicolumn{4}{|c|}{\textbf{1}} & \multicolumn{4}{|c|}{\textbf{30}} & \multicolumn{4}{|c|}{\textbf{120}} & \multicolumn{4}{|c|}{\textbf{180}} & \multicolumn{4}{|c|}{\textbf{360}}\\ 
	\multicolumn{3}{|c|}{\textbf{Horizon (months)}}  & \multicolumn{4}{|c|}{\textbf{0}} & \multicolumn{4}{|c|}{\textbf{1}} & \multicolumn{4}{|c|}{\textbf{4}} & \multicolumn{4}{|c|}{\textbf{6}} & \multicolumn{4}{|c|}{\textbf{12}} \\
	\hline
	Period & Delivery & Lag & Gas & Coal & EUA & Oil & Gas & Coal & EUA & Oil & Gas & Coal & EUA & Oil & Gas & Coal & EUA & Oil & Gas & Coal & EUA & Oil \\ 
	\hline \hline
	\multirow{39}{*}{\rotatebox[origin=c]{90}{2015-2019}} & D0 & 0 & \cellcolor[rgb]{0.79,0.79,0.79} - & \cellcolor[rgb]{0.79,0.79,0.79} - & \cellcolor[rgb]{0.79,0.79,0.79} - & \cellcolor[rgb]{0.79,0.79,0.79} - & \cellcolor[rgb]{0.79,0.79,0.79} - & \cellcolor[rgb]{0.79,0.79,0.79} - & \cellcolor[rgb]{0.79,0.79,0.79} - & \cellcolor[rgb]{0.79,0.79,0.79} - & \cellcolor[rgb]{0.79,0.79,0.79} - & \cellcolor[rgb]{0.79,0.79,0.79} - & \cellcolor[rgb]{0.79,0.79,0.79} - & \cellcolor[rgb]{0.79,0.79,0.79} - & \cellcolor[rgb]{0.79,0.79,0.79} - & \cellcolor[rgb]{0.79,0.79,0.79} - & \cellcolor[rgb]{0.79,0.79,0.79} - & \cellcolor[rgb]{0.79,0.79,0.79} - & \cellcolor[rgb]{0.79,0.79,0.79} - & \cellcolor[rgb]{0.79,0.79,0.79} - & \cellcolor[rgb]{0.79,0.79,0.79} - & \cellcolor[rgb]{0.79,0.79,0.79} - \\
\hline
	\end{tabular}}
	\caption{Average fuels coefficients of the portfolio model for two periods. $D0$, $D-1$, and $D+1$ are the actual, trailing and leading delivery periods.} 
	\label{tab:portfolio}
\end{table}

\begin{figure}[hbt]
    \centering
	\includegraphics[width=0.7\textwidth]{fuels_color_legend.pdf}
	\scalebox{0.5}{\includegraphics[width=0.49\textwidth]{hours_color_legend.pdf}}
	\linebreak
	\subfloat[p-values of ADF test for fuels]{
        \includegraphics[width=0.49\textwidth]{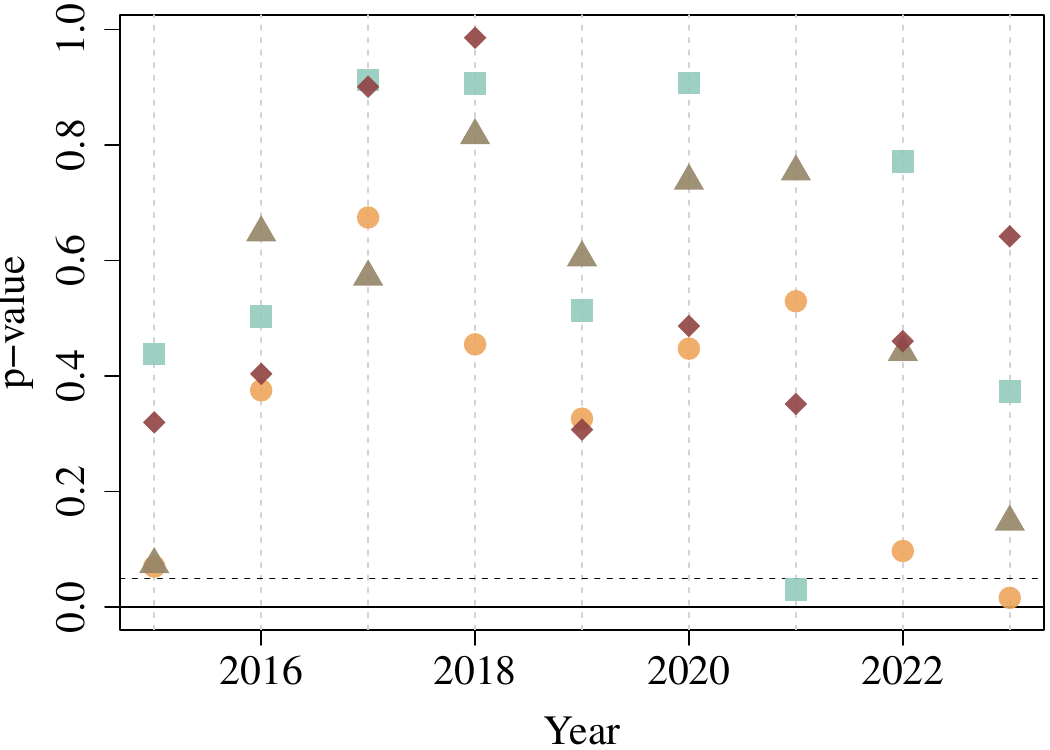}}
    \subfloat[p-values of ADF test for power prices]{
		\includegraphics[width=0.49\textwidth]{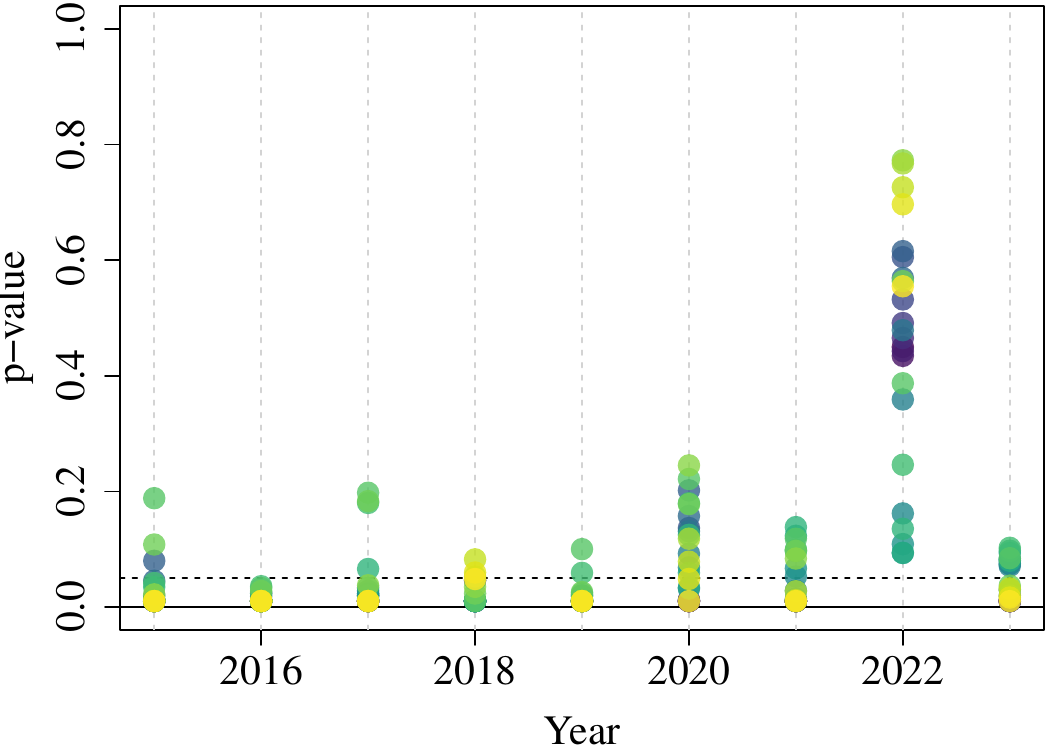}}
    \caption{p-values of augmented Dickey-Fuller (ADF) unit root test with the alternative hypothesis "stationarity", for (a) EUA, gas, coal and oil front month futures prices by year and for (b) German day-ahead power prices by hour and year. The dashed horizontal line represents the 0.05 significance level. The results must be interpreted with care as the underlying assumptions of the ADF test might be questionable for the data at hand.}
    \label{fig:adf_test}
\end{figure}

\begin{table}[htbp]
	\centering
	\setlength\extrarowheight{3pt}	
	\setlength{\tabcolsep}{2pt}
	\scalebox{0.58}{\begin{tabular}{|r|ccccc|cccc|rrrrrrrrrrrrrrr|}
	\hline
	\bf Model name & \multicolumn{5}{|c|}{\bf Variables} & \multicolumn{4}{|c|}{\bf Method} & \multicolumn{15}{c|}{\bf Forecasting horizon} \\
	\hline
	\hline
	 & \rotatebox{90}{Week Dummies} & \rotatebox{90}{Annual Seasons} & \rotatebox{90}{Autoreg} & \rotatebox{90}{RES \& Load} & \rotatebox{90}{Fuels} & \rotatebox{90}{Unconstrained} & \rotatebox{90}{Constrained} & \rotatebox{90}{Differencing} & \rotatebox{90}{Current} & 1 & 7 & 14 & 30 & 60 & 90 & 120 & 150 & 180 & 210 & 240 & 270 & 300 & 330 & 360 \\ 
	\hline  \hline
	Naive &  &  &  &  &  &  &  &  &  & \cellcolor[rgb]{0.782,1,0.6} $26.57$ & \cellcolor[rgb]{0.692,1,0.6} $33.02$ & \cellcolor[rgb]{0.71,1,0.6} $37.52$ & \cellcolor[rgb]{0.738,1,0.6} $43.89$ & \cellcolor[rgb]{0.815,1,0.6} $49.26$ & \cellcolor[rgb]{0.767,1,0.6} $53.16$ & \cellcolor[rgb]{0.787,1,0.6} $58.08$ & \cellcolor[rgb]{0.715,1,0.6} $59.51$ & \cellcolor[rgb]{0.699,1,0.6} $63.50$ & \cellcolor[rgb]{0.692,1,0.6} $65.20$ & \cellcolor[rgb]{0.685,1,0.6} $67.43$ & \cellcolor[rgb]{0.687,1,0.6} $67.31$ & \cellcolor[rgb]{0.686,1,0.6} $72.39$ & \cellcolor[rgb]{0.7,1,0.6} $76.25$ & \cellcolor[rgb]{0.711,1,0.6} $76.58$ \\ 
	  WD & $\times$ &  &  &  &  & $\times$ &  &  &  & \cellcolor[rgb]{1,0.6,0.6} $60.46$ & \cellcolor[rgb]{1,0.6,0.6} $60.41$ & \cellcolor[rgb]{1,0.6,0.6} $60.59$ & \cellcolor[rgb]{1,0.6,0.6} $61.12$ & \cellcolor[rgb]{1,0.6,0.6} $61.66$ & \cellcolor[rgb]{1,0.864,0.6} $62.08$ & \cellcolor[rgb]{0.943,1,0.6} $62.62$ & \cellcolor[rgb]{0.825,1,0.6} $62.86$ & \cellcolor[rgb]{0.687,1,0.6} $62.97$ & \cellcolor[rgb]{0.665,1,0.6} $63.15$ & \cellcolor[rgb]{0.637,1,0.6} $63.33$ & \cellcolor[rgb]{0.641,1,0.6} $63.41$ & \cellcolor[rgb]{0.613,1,0.6} $63.36$ & \cellcolor[rgb]{0.609,1,0.6} $63.42$ & \cellcolor[rgb]{0.607,1,0.6} $63.41$ \\ 
	  Expert & $\times$ & $\times$ & $\times$ & $\times$ & $\times$ & $\times$ &  &  &  & \cellcolor[rgb]{0.6,1,0.6} $16.60$ & \cellcolor[rgb]{0.603,1,0.6} $29.57$ & \cellcolor[rgb]{0.622,1,0.6} $34.58$ & \cellcolor[rgb]{0.811,1,0.6} $45.79$ & \cellcolor[rgb]{0.815,1,0.6} $49.25$ & \cellcolor[rgb]{1,0.889,0.6} $61.48$ & \cellcolor[rgb]{1,0.948,0.6} $65.82$ & \cellcolor[rgb]{1,0.739,0.6} $76.08$ & \cellcolor[rgb]{1,0.6,0.6} $95.87$ & \cellcolor[rgb]{1,0.6,0.6} $117.70$ & \cellcolor[rgb]{1,0.6,0.6} $128.18$ & \cellcolor[rgb]{1,0.6,0.6} $127.45$ & \cellcolor[rgb]{1,0.6,0.6} $160.40$ & \cellcolor[rgb]{1,0.6,0.6} $175.30$ & \cellcolor[rgb]{1,0.6,0.6} $163.87$ \\ 
	  Constr & $\times$ & $\times$ & $\times$ & $\times$ & $\times$ &  & $\times$ &  &  & \cellcolor[rgb]{0.6,1,0.6} $\textbf{16.58}$ & \cellcolor[rgb]{0.6,1,0.6} $\textbf{29.47}$ & \cellcolor[rgb]{0.6,1,0.6} $\textbf{33.85}$ & \cellcolor[rgb]{0.664,1,0.6} $41.98$ & \cellcolor[rgb]{0.721,1,0.6} $47.26$ & \cellcolor[rgb]{0.933,1,0.6} $57.16$ & \cellcolor[rgb]{1,0.992,0.6} $64.53$ & \cellcolor[rgb]{1,0.903,0.6} $71.10$ & \cellcolor[rgb]{1,0.856,0.6} $84.07$ & \cellcolor[rgb]{1,0.959,0.6} $91.07$ & \cellcolor[rgb]{1,0.862,0.6} $105.95$ & \cellcolor[rgb]{1,0.727,0.6} $116.76$ & \cellcolor[rgb]{1,0.883,0.6} $125.44$ & \cellcolor[rgb]{1,0.916,0.6} $130.59$ & \cellcolor[rgb]{1,0.864,0.6} $130.45$ \\ 
	  ConstrDiff & $\times$ & $\times$ & $\times$ & $\times$ & $\times$ &  & $\times$ & $\times$ &  & \cellcolor[rgb]{0.623,1,0.6} $17.82$ & \cellcolor[rgb]{0.762,1,0.6} $35.74$ & \cellcolor[rgb]{0.782,1,0.6} $39.93$ & \cellcolor[rgb]{0.842,1,0.6} $46.62$ & \cellcolor[rgb]{0.968,1,0.6} $52.51$ & \cellcolor[rgb]{0.927,1,0.6} $57.02$ & \cellcolor[rgb]{0.785,1,0.6} $58.02$ & \cellcolor[rgb]{0.801,1,0.6} $62.13$ & \cellcolor[rgb]{0.768,1,0.6} $66.70$ & \cellcolor[rgb]{0.712,1,0.6} $66.68$ & \cellcolor[rgb]{0.702,1,0.6} $68.86$ & \cellcolor[rgb]{0.728,1,0.6} $70.79$ & \cellcolor[rgb]{0.707,1,0.6} $74.93$ & \cellcolor[rgb]{0.698,1,0.6} $75.93$ & \cellcolor[rgb]{0.725,1,0.6} $78.41$ \\ 
	  Portfolio & $\times$ & $\times$ & $\times$ & $\times$ & $\times$ &  & $\times$ &  &  & \cellcolor[rgb]{0.602,1,0.6} $16.69$ & \cellcolor[rgb]{0.638,1,0.6} $30.93$ & \cellcolor[rgb]{0.705,1,0.6} $37.36$ & \cellcolor[rgb]{0.902,1,0.6} $48.17$ & \cellcolor[rgb]{1,0.879,0.6} $55.76$ & \cellcolor[rgb]{1,0.6,0.6} $68.45$ & \cellcolor[rgb]{1,0.6,0.6} $75.96$ & \cellcolor[rgb]{1,0.6,0.6} $80.28$ & \cellcolor[rgb]{1,0.745,0.6} $89.19$ & \cellcolor[rgb]{1,0.888,0.6} $96.36$ & \cellcolor[rgb]{1,0.835,0.6} $108.23$ & \cellcolor[rgb]{1,0.721,0.6} $117.24$ & \cellcolor[rgb]{1,0.831,0.6} $131.88$ & \cellcolor[rgb]{1,0.887,0.6} $134.63$ & \cellcolor[rgb]{1,0.806,0.6} $137.72$ \\ 
	  Short-term & $\times$ & $\times$ & $\times$ & $\times$ & $\times$ &  & $\times$ &  & $\times$* & \cellcolor[rgb]{0.603,1,0.6} $16.74$ & \cellcolor[rgb]{0.667,1,0.6} $32.08$ & \cellcolor[rgb]{0.678,1,0.6} $36.47$ & \cellcolor[rgb]{0.707,1,0.6} $43.09$ & \cellcolor[rgb]{0.759,1,0.6} $48.07$ & \cellcolor[rgb]{0.909,1,0.6} $56.58$ & \cellcolor[rgb]{0.999,1,0.6} $64.25$ & \cellcolor[rgb]{1,0.797,0.6} $74.31$ & \cellcolor[rgb]{1,0.818,0.6} $85.83$ & \cellcolor[rgb]{1,0.929,0.6} $93.29$ & \cellcolor[rgb]{1,0.879,0.6} $104.45$ & \cellcolor[rgb]{1,0.749,0.6} $114.91$ & \cellcolor[rgb]{1,0.877,0.6} $126.28$ & \cellcolor[rgb]{1,0.881,0.6} $135.53$ & \cellcolor[rgb]{1,0.819,0.6} $136.15$ \\ 
	  Current & $\times$ & $\times$ & $\times$ & $\times$ & $\times$ &  & $\times$ &  & $\times$ & \cellcolor[rgb]{0.603,1,0.6} $16.74$ & \cellcolor[rgb]{0.671,1,0.6} $32.23$ & \cellcolor[rgb]{0.652,1,0.6} $35.61$ & \cellcolor[rgb]{0.6,1,0.6} $\textbf{40.31}$ & \cellcolor[rgb]{0.6,1,0.6} $\textbf{44.70}$ & \cellcolor[rgb]{0.6,1,0.6} $\textbf{49.12}$ & \cellcolor[rgb]{0.6,1,0.6} $\textbf{52.62}$ & \cellcolor[rgb]{0.6,1,0.6} $\textbf{56.02}$ & \cellcolor[rgb]{0.645,1,0.6} $61.02$ & \cellcolor[rgb]{0.697,1,0.6} $65.55$ & \cellcolor[rgb]{0.703,1,0.6} $69.00$ & \cellcolor[rgb]{0.693,1,0.6} $67.83$ & \cellcolor[rgb]{0.686,1,0.6} $72.38$ & \cellcolor[rgb]{0.681,1,0.6} $73.58$ & \cellcolor[rgb]{0.699,1,0.6} $75.05$ \\ 
	  WD+RL & $\times$ &  &  & $\times$ &  &  & $\times$ &  &  & \cellcolor[rgb]{1,0.777,0.6} $50.74$ & \cellcolor[rgb]{1,0.665,0.6} $57.88$ & \cellcolor[rgb]{1,0.64,0.6} $59.27$ & \cellcolor[rgb]{1,0.615,0.6} $60.73$ & \cellcolor[rgb]{1,0.607,0.6} $61.51$ & \cellcolor[rgb]{1,0.847,0.6} $62.48$ & \cellcolor[rgb]{0.891,1,0.6} $61.10$ & \cellcolor[rgb]{0.709,1,0.6} $59.34$ & \cellcolor[rgb]{0.6,1,0.6} $\textbf{58.95}$ & \cellcolor[rgb]{0.6,1,0.6} $\textbf{58.34}$ & \cellcolor[rgb]{0.6,1,0.6} $\textbf{60.23}$ & \cellcolor[rgb]{0.6,1,0.6} $\textbf{59.99}$ & \cellcolor[rgb]{0.6,1,0.6} $\textbf{61.73}$ & \cellcolor[rgb]{0.605,1,0.6} $62.86$ & \cellcolor[rgb]{0.606,1,0.6} $63.29$ \\ 
	  WD+RL+C & $\times$ &  &  & $\times$ &  &  & $\times$ &  & $\times$ & \cellcolor[rgb]{1,0.733,0.6} $53.19$ & \cellcolor[rgb]{1,0.664,0.6} $57.93$ & \cellcolor[rgb]{1,0.671,0.6} $58.21$ & \cellcolor[rgb]{1,0.687,0.6} $58.86$ & \cellcolor[rgb]{1,0.699,0.6} $59.57$ & \cellcolor[rgb]{1,0.944,0.6} $60.13$ & \cellcolor[rgb]{0.874,1,0.6} $60.62$ & \cellcolor[rgb]{0.761,1,0.6} $60.91$ & \cellcolor[rgb]{0.642,1,0.6} $60.91$ & \cellcolor[rgb]{0.635,1,0.6} $60.97$ & \cellcolor[rgb]{0.612,1,0.6} $61.25$ & \cellcolor[rgb]{0.619,1,0.6} $61.57$ & \cellcolor[rgb]{0.601,1,0.6} $61.80$ & \cellcolor[rgb]{0.6,1,0.6} $\textbf{62.12}$ & \cellcolor[rgb]{0.6,1,0.6} $\textbf{62.52}$ \\ 
	  WD+ARL+C & $\times$ &  & $\times$ & $\times$ &  &  & $\times$ &  & $\times$ & \cellcolor[rgb]{0.611,1,0.6} $17.18$ & \cellcolor[rgb]{0.686,1,0.6} $32.81$ & \cellcolor[rgb]{0.697,1,0.6} $37.10$ & \cellcolor[rgb]{0.686,1,0.6} $42.56$ & \cellcolor[rgb]{0.778,1,0.6} $48.47$ & \cellcolor[rgb]{0.741,1,0.6} $52.52$ & \cellcolor[rgb]{0.786,1,0.6} $58.06$ & \cellcolor[rgb]{0.898,1,0.6} $65.06$ & \cellcolor[rgb]{0.924,1,0.6} $73.92$ & \cellcolor[rgb]{0.965,1,0.6} $85.39$ & \cellcolor[rgb]{1,0.925,0.6} $100.60$ & \cellcolor[rgb]{1,0.803,0.6} $110.33$ & \cellcolor[rgb]{1,0.91,0.6} $122.11$ & \cellcolor[rgb]{1,0.919,0.6} $130.11$ & \cellcolor[rgb]{1,0.874,0.6} $129.15$ \\ 
	  WD+F & $\times$ &  &  &  & $\times$ &  & $\times$ &  &  & \cellcolor[rgb]{0.803,1,0.6} $27.72$ & \cellcolor[rgb]{0.632,1,0.6} $30.71$ & \cellcolor[rgb]{0.6,1,0.6} $33.86$ & \cellcolor[rgb]{0.689,1,0.6} $42.64$ & \cellcolor[rgb]{0.787,1,0.6} $48.66$ & \cellcolor[rgb]{1,0.998,0.6} $58.82$ & \cellcolor[rgb]{1,0.851,0.6} $68.62$ & \cellcolor[rgb]{1,0.683,0.6} $77.75$ & \cellcolor[rgb]{1,0.819,0.6} $85.77$ & \cellcolor[rgb]{1,0.926,0.6} $93.48$ & \cellcolor[rgb]{1,0.895,0.6} $103.09$ & \cellcolor[rgb]{1,0.769,0.6} $113.21$ & \cellcolor[rgb]{1,0.909,0.6} $122.24$ & \cellcolor[rgb]{1,0.948,0.6} $126.13$ & \cellcolor[rgb]{1,0.92,0.6} $123.31$ \\ 
	  WD+F+C & $\times$ &  &  &  & $\times$ &  & $\times$ &  & $\times$ & \cellcolor[rgb]{0.837,1,0.6} $29.56$ & \cellcolor[rgb]{0.691,1,0.6} $33.00$ & \cellcolor[rgb]{0.682,1,0.6} $36.60$ & \cellcolor[rgb]{0.812,1,0.6} $45.84$ & \cellcolor[rgb]{0.903,1,0.6} $51.13$ & \cellcolor[rgb]{1,0.927,0.6} $60.54$ & \cellcolor[rgb]{1,0.827,0.6} $69.34$ & \cellcolor[rgb]{1,0.649,0.6} $78.80$ & \cellcolor[rgb]{1,0.765,0.6} $88.25$ & \cellcolor[rgb]{1,0.862,0.6} $98.27$ & \cellcolor[rgb]{1,0.846,0.6} $107.25$ & \cellcolor[rgb]{1,0.731,0.6} $116.38$ & \cellcolor[rgb]{1,0.904,0.6} $122.89$ & \cellcolor[rgb]{1,0.941,0.6} $127.03$ & \cellcolor[rgb]{1,0.901,0.6} $125.79$ \\ 
	   \hline
\end{tabular}}
\tiny{*For the short-term model the method as used in the \textbf{current model} is only applied to RES and Load. The fuels are regressed on at their corresponding lags, not at the current value.}
\caption{MAE (EUR/MWh) for the evaluation period April/2018-April/2024 for all considered models and forecasting horizons over all 24 hours. Abbreviation of the models are defined in the first column and correspond to the marked variables and methods.} 
\label{tab:mae_models}
\end{table}

\clearpage
\vspace{-5mm} 

 \bibliographystyle{unsrtnat1.bst}

\clearpage
\spacingset{0.8} 
\bibliography{bibliography}

\end{document}